\documentclass[aps,prb,reprint,twocolumn,superscriptaddress,amsmath,amssymb,amsfonts]{revtex4-2}

\AtBeginDocument{\usepackage{booktabs}}               
\makeatletter
\g@addto@macro\bfseries{\boldmath}
\makeatother
\usepackage[utf8]{inputenc}
\usepackage[varg]{txfonts}
\usepackage{chngcntr}
\usepackage[usenames,dvipsnames]{xcolor}
\usepackage{hyperref}
\usepackage{physics}
\usepackage{gensymb}
\definecolor{cL}{RGB}{0,0,255}\hypersetup{colorlinks=true,linkcolor=cL,citecolor=cL,urlcolor=cL} 
\usepackage{amsmath}
\usepackage{graphicx}
\usepackage[percent]{overpic}
\usepackage[newcommand]{ragged2e}
\usepackage{mathrsfs}
\usepackage{bm}
\usepackage{braket}
\usepackage[nolist,nohyperlinks]{acronym}
\usepackage{natbib}
\usepackage{microtype}
\usepackage{mathtools}
\usepackage{graphicx}
\usepackage{caption}
\DeclareCaptionJustification{justified}{\justifying}
\usepackage{subcaption}
\usepackage{tikz}
\usepackage{soul}
\usepackage{hyperref}
\usepackage{comment}
\hypersetup{colorlinks,urlcolor=blue}
\usepackage[percent]{overpic}
\usepackage{dcolumn}
\usepackage{bm}
\usepackage{float}

\makeatletter
\newcommand*{\addFileDependency}[1]{
  \typeout{(#1)}
  \@addtofilelist{#1}
  \IfFileExists{#1}{}{\typeout{No file #1.}}
}
\makeatother

\begin{document}
\preprint{APS/123-QED}
\title{Order-by-disorder without quantum zero-point fluctuations in the pyrochlore Heisenberg ferromagnet with Dzyaloshinskii-Moriya interactions}

\author{Alexander Hickey}
\altaffiliation[]
{These authors contributed equally to the project.}
\affiliation{Department of Physics and Astronomy, University of Waterloo, Waterloo, Ontario N2L 3G1, Canada}

\author{Daniel Lozano-G\'omez}
\altaffiliation[]
{These authors contributed equally to the project.}

\affiliation{Department of Physics and Astronomy, University of Waterloo, Waterloo, Ontario N2L 3G1, Canada}
\affiliation{Institut f\"ur Theoretische Physik and W\"urzburg-Dresden Cluster of Excellence ct.qmat, Technische Universit\"at Dresden, 01062 Dresden, Germany}	

\author{Michel J. P. Gingras}
\affiliation{Department of Physics and Astronomy, University of Waterloo, Waterloo, Ontario N2L 3G1, Canada}

\begin{abstract}
Order-by-disorder, whereby fluctuations lift an accidental classical ground state degeneracy to stabilize a subset of ordered states, is a recurrent and prominent theme in the field of frustrated magnetism where magnetic moments, or spins, are subject to competing spin-spin interactions. 
Thus far, such a phenomenon has been discussed in systems where the quantum ground state is not a ``classical'' product state. 
In such a case, both thermal and quantum fluctuations act to lift the accidental classical degeneracy, begging the question of whether one mechanism of order-by-disorder is possible without the other. 
In this paper, we present results exposing an uncharted route to order-by-disorder, one without quantum zero-point fluctuations, in the ferromagnetic pyrochlore Heisenberg system with the Dzyaloshinskii-Moriya (DM) interaction as the leading perturbation. 
We prove that any colinear ferromagnetic state is an \emph{exact} eigenstate even in the presence of the anisotropic DM interaction while thermal fluctuations give rise to a preference in the magnetization direction.
Using linear spin wave theory, we find that the anisotropy appears at lowest order as a sub-leading term in the low-temperature expansion of the free energy, proportional to $T^{7/2}$. 
Our results thus show that the phenomenon of thermal order-by-disorder can in principle occur even in the absence of quantum zero-point fluctuations driving quantum order-by-disorder -- this being so in particular when the accidentally degenerate ground state of the classical model turns out to be an exact eigenstate of the quantum version of the model.
However, and in addition,  we find that when the DM interaction is large, the fully polarized ferromagnetic ground state becomes unstable for a spin-$\frac{1}{2}$ system within the framework of non-linear spin wave theory, a result that is presumably closely related to the recent report of a quantum spin liquid in this spin-$\frac{1}{2}$ model at $D/J \approx 2$ reported in [Lozano-G\'omez {\it et al.}, 
 PNAS {\bf 121}, e2403487121 (2024)].

\end{abstract}

\date{\today}
\maketitle
\section{Introduction}

In correlated many-body systems with competing or frustrated interactions, thermal and quantum fluctuations may suppress 
the development of a symmetry-breaking transition to long-range order down to extremely low temperatures and, in some cases, completely prevent 
it~\cite{villain1979,moessner1998,Canals1998,kitaev2006a,Balents2010,gingras2014,Imai2016,Savary2017,Knolle2019}.
This observation, however, does not mean that the presence of such fluctuations only has the effect of undermining the development of long-range ordered phases.
Indeed, a contrary and at first sight seemingly paradoxical scenario may arise where the long-range order can be \emph{assisted} by the presence of quantum or thermal fluctuations, when the level of frustration is ``not too high'' \cite{moessner1998}, 
a phenomenon known as \emph{order-by-disorder}~\footnotetext[9]{Since frozen random disorder can also lift the accidental classical degeneracy of the parent disorder-free system~\cite{villain1979,henley1989},  
the term order-by-disorder may be confusing in this context and should perhaps be referred to as ``order-by-fluctuations''.}\cite{Note9,henley1989,villain1980,belorizky1980}. 

The study of frustrated magnetic systems has proven to be fruitful ground in the study and realization of this phenomenon, where competing exchange interactions result in a set of ``accidentally'' degenerate ground-state configurations at the classical level that are not protected by symmetries of the Hamiltonian \cite{Shender1982,henley1989,prakash1990}. 
Thanks to the lack of any symmetry protecting this ground state degeneracy, the system becomes prone to selecting a subset of configurations within the degenerate manifold of classical ground states due to spin fluctuations at temperature $T$.
These fluctuations can be both thermal ($T>0$) or quantum ($T = 0$) in origin. 
In the thermal case, first discussed in Ref.~\cite{villain1980}, a subset of the degenerate states are selected at $T=0^+$ by maximizing the entropy associated with fluctuations about the ultimately ``selected'' ordered magnetic moment direction \cite{Henley1987}. 
Conversely, the lifting of degeneracy at the quantum level~\cite{tessman1954,Shender1982} occurs as a result of perturbative corrections to the ground state energy, relative to the classical model, from an inherently semi-classical description of a quantum many-body system \cite{kittel1991} --- the true ground state of the quantum model not having any accidental degeneracy to lift in the first place
\footnotetext[19]{From that perspective, and as the present work illustrates through the specific model considered, quantum order-by-disorder ($T=0$) is not \textit{per se} a ``real'' physical phenomenon. 
Rather, it is a statement reflecting a mathematical procedure that  perturbatively corrects the approximation of a classical (direct product) ground state for a quantum spin model. 
Conversely, thermal order-by-disorder ($T>0$) being driven by ``real'' thermal fluctuations, can be viewed as a genuine physical phenomenon.}\cite{Note19}.

Despite the distinction between thermal and quantum fluctuations, both of these routes to fluctuation-induced selection are referred to colloquially as order-by-disorder (ObD).
Over the last five decades, there has been a large body of theoretical work focusing on models which exhibit thermal ObD \cite{villain1980,Henley1987,prakash1990,reimers1993,bramwell1994,Elhajal2005,bergman2007,chern2010,Gvozdikova2011,oitmaa2013,zhitomirsky2014,mcclarty2014,Francini2024nematicR2} or quantum ObD \cite{tessman1954,Shender1982,belorizky1980,chandra1988,kubo1991,sachdev1991,Chubukov_1991,chubukov1992,henley1994,Lecheminant1995,sobral1997,champion2003,baskaran2008,bernier2008,mulder2010,savary2012,zhitomirsky2012,Chernyshev2014,Rousochatzakis2015,rau2018,Placke2020,Chen2020,Liu2020,Khatua2021}. 
In this context, it has been commonplace in the literature to focus on a single mechanism of fluctuations, either thermal or quantum. While it is often the case that both quantum and thermal fluctuations select the same configurations \cite{henley1989,oitmaa2013,KHATUA2023,khatua2024}, these two mechanisms of fluctuations may instead favor \emph{distinct} long-range order, resulting in an additional transition between two ordered states at $T < T_{\rm c}$, where $T_{\rm c}$ is the paramagnetic critical temperature~\cite{Lee2014,danu2016,schick2020,noculak2023}. 
This implies that, in the most general case, both thermal and quantum fluctuations ought to be accounted for to obtain a comprehensive understanding of the selected ordered phases at finite temperature. We note that the lifting of degeneracy due to quenched random disorder can also be referred to as ObD \cite{villain1979,Fyodorov1991,Savary2011,Maryasin2013,Andreanov, Maryasin2015,Smirnov2017,AndradePRL2018,consoli2024}. 
This is, however, a distinct physical mechanism that is not considered in this paper.

While there is an abundance of theoretical work about ObD in spin models, the prevalence of material realizations of ObD is still not fully appreciated. 
So far, there is only a handful of real materials in which experimental evidence for ObD exists.
Examples of magnetic materials where ObD selection is somewhat compelling include$\text{Er}_2\text{Ti}_2\text{O}_7$~\cite{zhitomirsky2012,savary2012,ross2014,
Note64,mcclarty2009a,Petit2014,Rau2016},~\footnotetext[64]{Although selection, at least in part, due to virtual crystal field fluctuations has not been ruled out \cite{mcclarty2009a,Petit2014,Rau2016}} $\text{Co}\text{Ti}\text{O}_3$ \cite{elliot2021}, $\text{Sr}_2\text{Cu}_3\text{O}_4\text{Cl}_2$ \cite{kim1999}, $\text{Fe}_2\text{Ca}_3\left(\text{GeO}_4\right)_3$ \cite{brueckel1988}, $\text{Yb}_2\text{Ge}_2\text{O}_7$ \cite{sarkis2020},
$\text{RbFe(MoO}_4\text{)}_2$ \cite{Inami1996,Smirnov2007}, $\text{Ba}_3\text{CoSb}_2\text{O}_9$ \cite{Susuki2013}, $\text{NaYbSe}_2$ \cite{Ranjith2019}
and $\text{BaCoS}_2$ \cite{lenz2024}. 
However, it is perhaps fair to say that in no case has ObD being at play in these systems been definitively established. 
This is because there is no agreed-upon ``smoking gun'' 
evidence that ObD is operating in a material without proceeding to quantitatively relate experimental results with theoretical calculations predicting ObD for the spin model argued to faithfully describe the material under consideration~\cite{KHATUA2023}.

To the best of our knowledge, all known examples of ObD selection in magnetic systems thus far studied have been found to exhibit both quantum ($T=0$) and thermal ($T>0$) ObD. 
Naturally, this leads to a fundamental question surrounding our current framework for understanding ObD: does the existence of thermal ObD \textit{necessarily} imply the existence of quantum ObD? Or, can we have one form of ObD without the other? 
In the present work, we address this question by studying the \emph{ferromagnetic} pyrochlore Heisenberg model with an additional anisotropic Dzyaloshinskii–Moriya (DM) interaction as the leading perturbation. 
In particular, we present the completely worked out example of ObD on the pyrochlore lattice with nearest-neighbor DM interactions. 
For a wide range of parameters, the classical ground state of this Hamiltonian is a uniform colinear ferromagnet.
As our principal result, we show that this system displays a \emph{purely} thermal ObD selection, while quantum ObD selection is \emph{absent} at zero temperature. 
This illustrates that thermal ObD can in principle exist independently of quantum ObD in a quantum magnetic system.
We investigate how the ground state degeneracy is lifted due to thermal fluctuations, using both classical and quantum calculation methods, to provide an insight on the order parameter selection at all temperatures within the ordered phase.
The results pertaining to this spin model therefore represents a rather rare realization of a system exhibiting ObD in the absence of zero-point fluctuations. 
Additionally, we analyze the stability of the ferromagnetic order in the model considered in the quantum limit for an $S=\frac{1}{2}$ system using non-linear spin wave theory. 
This allows us to identify a region of instability of the ferromagnetic order which was first alluded to in very recent work~\cite{noculak2023,lozano-gomez2023}. While the results presented here are specifically for the pyrochlore lattice, we propose that the same qualitative effects should arise in any Heisenberg ferromagnet with DM interactions as long as the sum of the DM vectors along the bonds surrounding each site vanishes.

\begin{figure*}[t!]
    \centering
    \begin{overpic}[width=1.9\columnwidth]{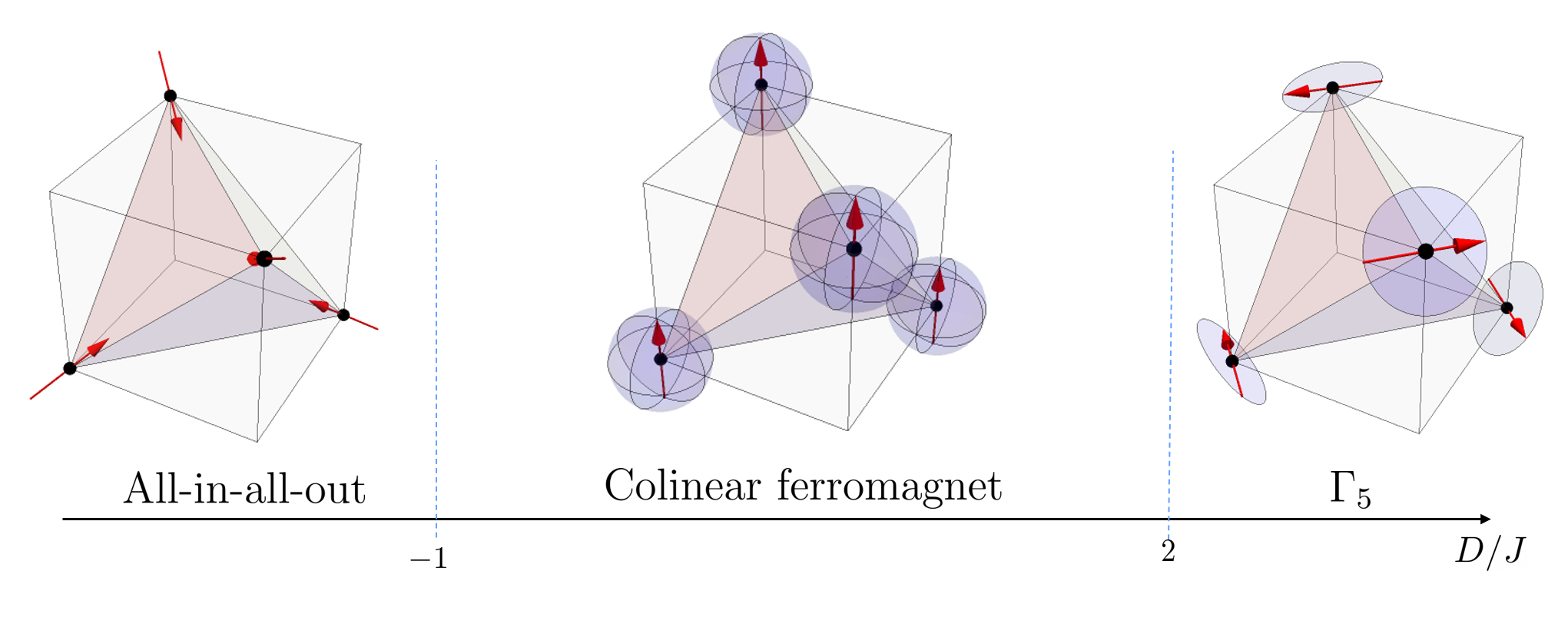}
    \end{overpic}
    \caption{Classical ground state configurations of the Hamiltonian in Eq.~(\ref{eq:hamiltonian}) with ferromagnetic $J>0$, depicted on a single tetrahedron of the pyrochlore lattice.
    The ground state configuration corresponds to colinear ferromagnet for $-1 < D/J < 2$, as well as the all-in-all-out and $\Gamma_5$ antiferromagnetic configurations for $D/J<-1$ and $D/J > 2$ respectively.
    The spheres and discs surrounding the spins illustrate manifolds of degenerate spin configurations. The point where $D/J=0$ corresponds to the Heisenberg ferromagnet.}
    \label{fig:classical_phases}
\end{figure*}

\section{Model}
\label{sec:Model}

We consider the Heisenberg and Dzyaloshinskii–Moriya (DM) Hamiltonian on the pyrochlore lattice, described by the Hamiltonian
\begin{equation}
\mathcal{H} = -J\sum_{\langle i,j\rangle}\bm S_i\cdot \bm S_j - D \sum_{\langle i,j\rangle} \vb*{d}_{ij} \cdot (\bm S_i\times \bm S_j),  \label{eq:hamiltonian}
\end{equation}
where $\langle i,j \rangle$ refers to the sum over all nearest-neighbor sites, $J>0$ is the ferromagnetic Heisenberg coupling~\footnote{In this work we do not consider the possibility of antiferromagnetic $J<0$~\cite{noculak2023}.}, and $\{\bm d_{ij}\}$ are the nearest-neighbor DM vectors determined by the Moriya rules~\cite{Moriya1960} and listed in Appendix~\ref{appendix:Pyrochlore}. 
This choice of the DM vectors, along with sign convention of the DM interaction in Eq.~\eqref{eq:hamiltonian}, is referred to as being \emph{indirect} (\emph{direct})  when $D>0 \; (D<0)$ \cite{Elhajal2005}.

Classically, the ground states of this model are magnetically ordered phases with $\bm q= \bm 0$ ordering wavevector~\cite{Yan2017}, i.e. the magnetic order is fully determined by the spin configuration on a single (basis cell) tetrahedron. 
The classical ground state configurations are determined by expressing Eq.~\eqref{eq:hamiltonian} as a sum over all individual tetrahedra,
each labeled $\boxtimes$, $\mathcal{H}=\sum_\boxtimes \mathcal{H}^{\boxtimes}$, and then minimizing the resulting single-tetrahedron Hamiltonian $\mathcal{H}^{\boxtimes}$. 
The result of this calculation is pictorially illustrated in Fig.~\ref{fig:classical_phases}, where three distinct sets of classical spin configurations are identified: the ``all-in-all-out'' antiferromagnetic states for $D/J<-1$~\footnote{Although at $D/J=-1$ there is a degeneracy between the internal energy of the all-in-all-out and colinear ferromagnetic phases, we note that, according to our classical Monte Carlo simulation, the colinear ferromagnetic phase is selected at $D/J=-1$ at $T_c$ down to the $T\to 0^+$ limit.}, colinear ferromagnetic configurations for $-1 < D/J < 2$, and the antiferromagnetic ``$\Gamma_5$'' configurations for $D/J>2$~\cite{Elhajal2005,chern2010,noculak2023}. 

It is well known that the $\Gamma_5$ region of the phase diagram exhibits an accidental ground state degeneracy at the classical level, parameterized by the global rotation group $O(2)$~\cite{savary2012,Yan2017}. 
This continuous symmetry group is reduced by both thermal and quantum fluctuations, resulting in six degenerate ordered configurations that are related by local discrete rotation about a local $z$ axis (the local cubic $[111]$ direction)~\cite{Elhajal2005,chern2010,noculak2023}. 
In the present work, we point out that the ferromagnetic regime ($-1 < D/J < 2$) of Eq.~\eqref{eq:hamiltonian} also exhibits an accidental ground state degeneracy at the classical level whenever $D\neq 0$. 
Inspecting each term in Eq.~\eqref{eq:hamiltonian}, we note that the Heisenberg exchange is completely isotropic, while the cross product appearing in the DM interaction term obviously vanishes for any perfectly colinear configuration. This implies that the different ferromagnetic ground states are related by rotations in three-dimensional space, which can be parametrized by the continuous rotation group $O(3)$. 
This degeneracy of the classical ground state is accidental, as Eq.~\eqref{eq:hamiltonian} has no continuous symmetry for an arbitrary spin state \footnote{There is an exception to this statement when $D=0$. In this case, Eq.~\eqref{eq:hamiltonian} is simply the ferromagnetic Heisenberg model which has an exact $O(3)$ symmetry, and therefore the ground state degeneracy is no longer accidental}. 
Naturally, one may ask whether this accidental degeneracy is lifted by $\cramped{T=0}$ quantum or $T>0$ thermal fluctuations, selecting a subset of the ferromagnetic configurations.

Physically, the DM interaction arises as the leading order exchange interaction generated by spin-orbit interactions~\cite{Moriya1960,Coffey1991a},  while, in general, the spin-orbit interaction will give rise to additional anisotropic exchange interactions that are typically weaker than the DM interaction~\cite{Moriya1960,Riedl2016a}. 
In particular, for the pyrochlore lattice, there are two additional symmetry-allowed nearest-neighbor exchange couplings in the bilinear spin Hamiltonian~\cite{mcclarty2009a,Ross2011}. 
These two couplings correspond to a bond-dependent diagonal ``Kitaev'' interaction $K$~\footnotetext[80]{
Here, we refer to the Kitaev interaction for the pyrochlore lattice in the same way that it is used in Ref.~\cite{Rau2018B}. 
The colinear ferromagnet is still the classical ground state configuration when a small but non-zero Kitaev interaction is present. 
However, the ferromagnetic product state will no longer be an exact eigenstate of the quantum Hamiltonian. 
This, in turn, will generate zero-point fluctuations proportional to $K$, and there will be quantum ObD present in addition to thermal ObD, as is expected in a typical quantum ObD scenario.}~\cite{Note80}, 
and a symmetric off-diagonal ``pseudo-dipolar'' interaction
$\Gamma$~\cite{mcclarty2009a,Rau2018B}. 
When spin-orbit coupling is one of the dominant energy scales, these additional couplings cannot be ignored in general, as is the case of rare-earth pyrochlore magnets~\cite{rau2016b,Rau2019Review}. 
The effect of a small but nonzero $\Gamma/J$ is to induce a finite canting angle to the spins, these splaying away from a perfect colinear ferromagnetic configuration~\cite{Yan2017}. 

In this paper, we only consider the Heisenberg and DM interactions, and ignore the remaining two symmetry-allowed $K$ and $\Gamma$ couplings. 
This model was shown to display colinear ferromagnetic order in both the classical and quantum scenarios~\cite{noculak2023}, resulting in an accidental O(3) degeneracy in the ground state manifold. 
From a material perspective, this model is justified in the regime where spin-orbit interactions can be treated perturbatively, in which case the isotropic Heisenberg coupling is expected to be the dominant exchange interaction, and the DM coupling is the leading order anisotropic exchange \cite{Moriya1960,Coffey1991a,noculak2023,Riedl2016a}. 
This situation is relevant to $3d$ transition metal ions, where spin-orbit interactions are expected to be perturbatively small. 
This model is of particular relevance for the ferromagnetic materials $\text{Lu}_2\text{V}_2\text{O}_7$ and $\text{Y}_2\text{V}_2\text{O}_7$ which crystallize into a pyrochlore lattice, with a magnetic $\text{V}^{4+}$ ion ($S=\frac{1}{2}$) located at each pyrochlore lattice site~\cite{Shamoto2002,AliBiswas2013,Nazipov2016,Nazipov2016a}.
Indeed, the realization of a colinear ferromagnet in $\text{Lu}_2\text{V}_2\text{O}_7$ is supported by neutron scattering measurements \cite{Mena2014}, as well as density-functional theory calculations \cite{Riedl2016a}. 
In passing, we note that this material has garnered significant interest due to the observation of a thermal Hall effect of magnons \cite{onose2010}.

The remainder of this paper focuses on ObD in the ferromagnetic regime of Eq.~\eqref{eq:hamiltonian} which, to the best of our knowledge, has not been explored previously. 
As discussed just above, the case of ``weak DM'' ($|D| < J$) is motivated by material realizations of this model in the context of transition metal pyrochlores. 
However, out of theoretical interest, we extend our analysis to the entirety of the ferromagnetic phase as depicted in Fig.~\ref{fig:classical_phases} and which ranges from $D/|J|=-1$ to 
$D/J=+2$~\cite{noculak2023,lozano-gomez2023}.
While the region where $|D|>J$ may not have any particular relevance to real known regular pyrochlore 
materials~\footnotetext[91]{
Interestingly, the breathing pyrochlore (BP) material Ba$_3$Yb$_2$Zn$_5$O$_{11}$ seems well-described by the Hamiltonian of Eq.~\eqref{eq:hamiltonian} over the small tetrahedra of the BP structure~\cite{rau2016b}.}~\cite{Note91,rau2016b}, it is of theoretical interest in the broader context the mechanism of ObD in the \emph{absence} of quantum fluctuations. 
In the following section, we discuss the methods used to investigate ObD in the ferromagnetic regime of Eq.~\eqref{eq:hamiltonian}. centering

\section{Methods}
In the present work, we use a number of methods to determine the direction of the global (bulk) magnetization per spin order parameter, denoted as $\vb*{m}$, in the ferromagnetic regime of Eq.\eqref{eq:hamiltonian}. 
This section briefly discusses each of these methods and their implementation.

\subsection{Classical low-temperature expansion}

For the classical version of the Hamiltonian in Eq.~\eqref{eq:hamiltonian}, the low-temperature selection of the magnetization direction as a function of the reduced interaction parameter $D/J$ is studied using a classical low-temperature expansion (CLTE), describing Gaussian spin fluctuations $\delta n^\alpha_{i}$ about an ordered ground state configuration, where $\alpha=1,2$ labels the two transverse directions. 
At low temperatures, only small transverse spin deviations  away from the local spin direction occur, i.e. $|\delta n_i^\alpha | \ll S$.
Under this assumption, the spins take the form 
\begin{equation}
    \bm{S}_{i}\simeq  \delta n^1_{i} \vu*{e}_1 + \delta n^2_{i}\vu*{e}_2 + S\left(1-\frac{(\delta n^1_{i})^2}{2S^2}-\frac{(\delta n^2_{i})^2}{2S^2}\right)\vu*{m} \label{eq:fluctuations_CLTE},
\end{equation}
where $\vu*{m} = \vb*{m}/|\vb*{m}|$ is the unit vector in the direction of the bulk magnetization, and $\vu*{e}_1,\vu*{e}_2$ are two unit vectors that form an orthonormal triad with the magnetization direction, i.e. $\vu*{e}_\alpha \cdot \vu*{e}_{\alpha'} = \delta_{\alpha \alpha'}$ and $\vu*{m} = \vu*{e}_1 \times \vu*{e}_2$. 
Substituting Eq.~\eqref{eq:fluctuations_CLTE} into Eq.~\eqref{eq:hamiltonian} and keeping only terms quadratic in $\delta n_i^\alpha $ gives a quadratic form 
\begin{equation}
    \mathcal{H}^{(\vu*{m})}_{\rm CLTE}= S^2 E_0+\frac{1}{2}\sum_{\bm q} \sum_{\mu,\nu=0}^3\sum_{\alpha,\beta=1}^2 \delta n_\mu^\alpha(-\bm q){\sf H}_{\alpha\beta,\mu\nu}^{(\vu*{m})}(\bm q)\delta n_\nu^\beta(\bm q),
\end{equation}
where $S^2 E_0 = -12JNS^2$ is the classical ground state energy, $N$ is the number of FCC primitive unit cells, each containing four sublattices defining a tetrahedron, where $\mu, \nu$ label the sublattices, and  
\begin{align}
    {\sf H}_{\alpha\beta,\mu\nu}^{(\vu*{m})}(\bm q) \equiv & \; 6J \delta_{\alpha \beta}\delta_{\mu \nu} -2J\delta_{\alpha\beta} \left(1-\delta_{\mu \nu}\right) \cos \left( \vb*{q} \cdot \vb*{r}_{\mu \nu}\right) \nonumber \\
    & \; +  2D \vb*{d}_{\mu \nu}\cdot \vu*{m} (-1)^\alpha\left(1-\delta_{\alpha \beta}\right) \left(1-\delta_{\mu \nu}\right) \cos \left( \vb*{q} \cdot \vb*{r}_{\mu \nu}\right)
\end{align}
is the Hessian matrix expressed in momentum (${\bm q}$) space. 
The entropy associated with the fluctuations about a given polarized configuration $\vu*{m}$ is then given by the expression~\cite{mcclarty2014,Yan2017,KHATUA2023}
\begin{equation}
\label{eq:entropy}
\mathcal{S}_{\hat{\bm m}}=\mathrm{constant} - \frac{1}{2}\sum_{\bm q}\ln
\left(\det {{\sf H}^{(\vu*{m})}(\bm q )}\right).
\end{equation}
The thermal selection among the fully polarized states can be exposed by maximizing the entropy in Eq.~\eqref{eq:entropy} with respect to the direction of the magnetization, $\vu*{m}$. 
Note that throughout this paper, calculations are carried out in units where $\hbar = k_\text{B} = 1$.

\subsection{Classical Monte Carlo}

The finite temperature properties of the ferromagnetic phase of Eq.~\eqref{eq:hamiltonian} are explored using classical Monte Carlo (MC) simulations. 
Simulations were carried out on a finite pyrochlore lattice with $4L^3$ spins with $L=10$ and periodic boundaries, using both Metropolis importance sampling and over-relaxation updates \cite{ALZATE-CARDONA2019,BINDER2010,CREUTZ1987}, and averaged over 10 independent MC simulations.
For each simulation, the system was initialized in a random configuration at high temperature $T>T_{\text{c}}$, and then gradually cooled down to the target temperature, thermalizing with $5\times 10^4$ sweeps between each step in temperature. 
Each of the thermodynamic observables measured is averaged over $10^5$ independent MC sweeps and then averaged over the independent simulations. 

For the colinear ferromagnetic phase, the onset of long-range order is tracked by measuring the magnetization per spin, ${\bm m}$
 \begin{equation}\label{eq:MagnetizationPerSpin}
    \vb*{m} \equiv \frac{1}{4N}\sum_{i}\vb*{S}_i,
\end{equation}
where $4N$ is the number of spins in a system consisting of $N=L^3$ FCC unit cells. 
To investigate the evolution and selection of the orientation of the magnetization, $\vb*{m}$, as a function of temperature, it is not enough to study the thermodynamic average of Eq.~\eqref{eq:MagnetizationPerSpin}. Indeed, a generic Ginzburg-Landau (GL) expansion of the free energy for a system with cubic symmetry yields~\cite{bouchaud1993,CHAIKIN2000} 
\begin{eqnarray}
\label{eq:GL-theory}
    \mathcal{F} =&& \frac{r}{2} |\bm m|^2 + \frac{u}{4}|\bm m|^4 + v\left( m_x^4 +m_y^4+m_z^4\right)\nonumber\\ 
    &&+\frac{w}{6} |\bm m|^6 + \gamma \left( m_x^2m_y^2m_z^2\right) + \cdots 
\end{eqnarray}
where the coefficients $\{r,u,v,w,\gamma\}$ are temperature dependent. The resulting terms can be broadly separated into isotropic terms, i.e. those involving powers of the magnitude of magnetization, $|\bm m|$,  in Eq.~\eqref{eq:GL-theory}, which do not distinguish between different magnetization orientations and the anisotropic terms, which do. 
While the GL free energy in Eq.~\eqref{eq:GL-theory} is not necessarily valid  away from the critical temperature, it motivates the lowest order symmetry allowed functions of the order parameter, which can distinguish between different orientations of the order parameter. 
In our MC simulations, we study the temperature evolution of the average magnetization per spin $\langle |\bm m|  \rangle$, as well as the anisotropic observables 
\begin{align}
M_{4} &\equiv \langle m_x^4 +m_y^4+m_z^4 \rangle , \label{eq:M_4}\\
\delta M_4 &\equiv \frac{3}{2}\left( \langle|\bm m|^4\rangle - M_4 \right), \label{eq:dM_4} \\
M_6 &\equiv 27 \langle m_x^2m_y^2m_z^2\rangle, \label{eq:M6}
\end{align}
where $\expval{\cdots}$ denotes thermal averaging, and the factors of $3/2$ and $27$ are introduced to normalize the range of expectation values between 0 and 1. 
We refer to the anisotropic terms defined in Eqs.~(\ref{eq:M_4}-\ref{eq:M6}) as the \emph{cubic parameters}.
The values of the cubic parameters for the magnetization  ${\bm m}$ along various high symmetry directions are listed in Table~\ref{table:CubicParams} in Appendix \ref{appendix:MC}. 
Note that throughout this work, we use the notation $\langle hkl \rangle$ to denote the direct space vector $h\vu*{x}+k\vu*{y}+l\vu*{z}$, along with all other symmetry-related directions.

To further investigate the evolution of the magnetization, we sample the \emph{instantaneous} magnetization direction of the system, $\vu*{m} \equiv\sum_i \vb{S}_i / |\sum_i \vb{S}_i|$, along each Cartesian direction $2\times 10^4$ times for each temperature, using a set of temperatures ranging from $T\leq T_{\rm c}$ to $T\ll T_{\rm c}$. 
The resulting distribution, $p(\vu*{m})$, is visualized as a function of the polar and azimuthal angles, $\theta$ and $\phi$, of the global magnetization $\vb*{m}$ restricted to the first octant where $\phi\in [0,\pi/2]$ and $\cos(\theta)\in [0,1]$. 
Clustering of the magnetization distribution about a given magnetization orientation is therefore considered as evidence for a thermal selection of the corresponding magnetization direction.

\subsection{Quantum spin waves} \label{sec:LSW}

To investigate the putative thermal ObD selection in the low temperature quantum version of Eq.~\eqref{eq:hamiltonian}, we perform an Holstein-Primakoff spin wave expansion. 
While the results presented in the remainder of this paper focus on the quantum limit $S=\frac{1}{2}$, we first leave the spin quantum number $S$ as a free parameter in our equations for organizing the $1/S$ Holstein-Primakoff expansion.
In particular, spin operators are recast as Holstein-Primakoff bosons \cite{kittel1991,Holstein1940} 
\begin{eqnarray} 
S_{\vb*{R},\mu}^{(\hat{\vb*{m}})} &=& S- a_{\vb*{R},\mu}^\dagger a_{\vb*{R},\mu}^{}, \nonumber \\
S_{\vb*{R},\mu}^{+}&=&(2S- a_{\vb*{R},\mu}^\dagger a_{\vb*{R},\mu}^{})^{1/2} a_{\vb*{R},\mu}^{} , \nonumber \\
S_{\vb*{R},\mu}^{-}&=&a_{\vb*{R},\mu}^\dagger(2S-a_{\vb*{R},\mu}^\dagger a_{\vb*{R},\mu}^{})^{1/2},\label{eq:HP_transform}
\end{eqnarray}
where $S_{\vb*{R},\mu}^{(\hat{\vb*{m}})} = \vb*{S}_{\vb*{R},\mu} \cdot \vu*{m}$ is the component of the spin operator aligned along the global polarization axis $\hat{\vb*{m}}$ of the ferromagnetic configuration, and $\vb*{R}$ is a FCC lattice vector.
In this picture, each boson is interpreted as a quantum of spin fluctuation about the ordered ground state. Plugging this transformation into the spin Hamiltonian in Eq.~\eqref{eq:hamiltonian} and expanding the square root in powers of $a_{\vb*{R},\mu}^\dagger a_{\vb*{R},\mu}^{}/2S$ yields a semi-classical expansion
\begin{equation}
 \mathcal{H} = \sum_{n=0}^\infty S^{2-n/2} \mathcal{H}^{(\hat{\vb*{m}})}_{n},
\label{eq:Boson_Hamiltonian}
\end{equation} 
where each order of the expansion is now dependent on the global polarization axis. Truncating at $\mathcal{O}(S)$ gives the non-interacting magnon Hamiltonian 
\begin{align}
\mathcal{H}^{(\hat{\vb*{m}})}_{\text{QSW}} = S^{2} \mathcal{H}^{(\hat{\vb*{m}})}_{0} + S^{3/2} \mathcal{H}^{(\hat{\vb*{m}})}_{1} + S \mathcal{H}^{(\hat{\vb*{m}})}_{2}.  \label{eq:QSW_expansion}
\end{align}
Note that $\mathcal{H}^{(\hat{\vb*{m}})}_1 = 0$ whenever the spin configuration is classically metastable \cite{kittel1991}, in which case Eq.~\eqref{eq:QSW_expansion} simplifies to
\begin{align}
\mathcal{H}^{(\hat{\vb*{m}})}_{\mathrm{QSW}} = &S(S+1)E_0 + \nonumber \\ &+\frac{S}{2} \sum_{\bm q}\sum_{\mu,\nu=0}^3 \left[ X^{(\hat{\vb*{m}})}_{\mu\nu}(\vb*{q}) a_{\mu,\bm q}^\dagger a^{}_{\nu,\bm q} + X^{(\hat{\vb*{m}})}_{\nu \mu}(-\vb*{q}) a_{\mu,-\bm q}^{} a^{\dagger}_{\nu,-\bm q} \right],
\label{eq:LSW_HamiltonianA}
\end{align}
where
\begin{equation}\label{eq:LSWMatrixElements}
X^{(\hat{\vb*{m}})}_{\mu\nu}(\vb*{q}) = 6J\delta_{\mu\nu} - 2\left( J - i D  \vb*{d}_{\mu \nu} \cdot \hat{\vb*{m}}  \right) (1-\delta_{\mu\nu}) \cos \left( \vb*{q} \cdot \vb*{r}_{\mu \nu} \right).
\end{equation}
An appropriate unitary transformation of the magnon operators $a_{\mu,\bm q}^{} = \sum_{\alpha} U^{(\hat{\vb*{m}})}_{\mu,\alpha} (\bm q) b_{\alpha,\bm q}^{} $ brings Eq.~\eqref{eq:LSW_HamiltonianA} to the diagonal form 
\begin{equation}
\mathcal{H}^{(\hat{\vb*{m}})}_{\mathrm{QSW}} = S^2 E_0+S E^{(\hat{\vb*{m}})}_2 + S \sum_{\bm q}\sum_{\mu=0}^3  \omega^{(\hat{\vb*{m}})}_{\mu,\vb*{q}} b_{\mu,\bm q}^\dagger b^{}_{\mu,\bm q} ,
\label{eq:LSW_HamiltonianB}
\end{equation}
where $\omega^{(\hat{\vb*{m}})}_{\mu,\vb*{q}}$ is the magnon dispersion and
\begin{equation}
S E^{(\hat{\vb*{m}})}_2 = S E_0 + \frac{S}{2}\sum_{\bm q}\sum_{\mu=0}^3\omega^{(\hat{\vb*{m}})}_{\mu,\vb*{q}} \label{eq:zeropoint}
\end{equation}
is the $\mathcal{O}(S)$ correction to the ground state energy from quantum zero-point fluctuations~\cite{kittel1991}. The free energy, $\mathcal{F}^{(\hat{\vb*{m}})}$, in linear spin wave theory is then calculated as 
\begin{equation}\label{eq:FullFreeEnergy}
\mathcal{F}^{(\hat{\vb*{m}})} = S^2 E_0+S E^{(\hat{\vb*{m}})}_2 + T \sum_{\vb*{q}}\sum_{\mu=0}^3 \ln \left( 1-e^{-S\omega^{(\hat{\vb*{m}})}_{\mu,\vb*{q}}/T}\right).
\end{equation}  
The selection of certain configurations over others is understood at this level by minimizing Eq.~\eqref{eq:FullFreeEnergy} with respect to the various classical ground state configurations characterized by 
${\hat{\vb*{m}}}$. At $T=0$, this amounts to minimizing the zero-point energy in Eq.~\eqref{eq:zeropoint}, which is the typical approach taken to expose quantum ObD \cite{Yan2017,ross2014,savary2012,sarkis2020,noculak2023}. 
At nonzero temperature, the thermal population of quasiparticles leads to an entropic contribution to Eq.~\eqref{eq:FullFreeEnergy} as well as a nonzero temperature contribution to the total internal energy which may or may not select the same configurations as the zero-point contribution \cite{schick2020,noculak2023}.

 \begin{figure}[t!]
    \centering
   \begin{overpic}[width=\columnwidth]{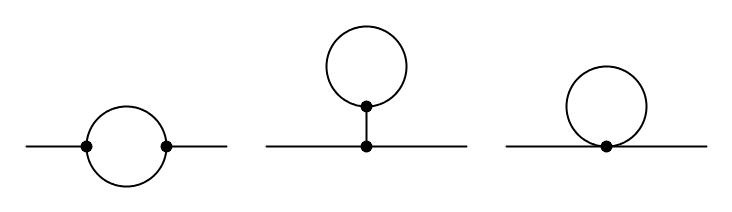}
    \end{overpic}
    \caption{Diagrams contributing to the magnon self-energy at $\mathcal{O}(S^0)$ in the Holstein-Primakoff expansion. At $T=0$, only the first diagram, connecting a pair of three-magnon interaction vertices, has a non-zero contribution. The explicit evaluation of these diagrams is discussed in Appendix~\ref{appendix:SelfEnergy}.}
    \label{fig:FeynmanBubble}
\end{figure}

\subsection{Non-linear spin waves} \label{sec:NLSW}

To analyze the stability of the ferromagnetic phase, we extend our spin wave theory to include magnon-magnon interactions. 
Corrections to the linear spin wave theory can be included perturbatively using the leading order magnon-magnon interactions in the $1/S$ expansion in Eq.~\eqref{eq:Boson_Hamiltonian}, given by the symmetrized three-magnon and four-magnon interaction terms \cite{Rau2019a}
\begin{equation}\label{eq:ThreeMagnonHamiltonian}
\mathcal{H}^{(\hat{\vb*{m}})}_{3} = \frac{1}{2! \sqrt{N}} \sum_{\vb*{k} \vb*{q}} \sum_{\mu \nu \lambda} \left[ Y_{\mu\nu\lambda}^{(\vu*{m})}(\vb*{k} ,\vb*{q}) a^\dagger_{\vb*{k},\mu} a^\dagger_{\vb*{q},\nu} a^{}_{\vb*{k}+\vb*{q},\lambda} + \text{H.c.}\right] 
\end{equation}
and
\begin{equation}\label{eq:FourMagnonHamiltonian}
\mathcal{H}^{(\hat{\vb*{m}})}_{4} = \frac{1}{N(2!)^2} \sum_{\vb*{k} \vb*{q} \vb*{Q}} \sum_{\mu \nu \lambda \rho} V_{\mu\nu\lambda \rho}^{(\vu*{m})}(\vb*{k} ,\vb*{q},\vb*{Q}) a^\dagger_{\vb*{k}+\vb*{Q},\mu} a^\dagger_{\vb*{q}-\vb*{Q},\nu} a^{}_{\vb*{q},\lambda} a^{}_{\vb*{k},\rho} ,
\end{equation}
where the interaction vertices $Y_{\mu\nu\lambda}^{(\vu*{m})}(\vb*{k} ,\vb*{q}) $ and $V_{\mu\nu\lambda \rho}^{(\vu*{m})}(\vb*{k} ,\vb*{q},\vb*{Q})$, are defined in Appendix~\ref{appendix:SelfEnergy}. 
To probe the stability of the ferromagnetic phase, we analyze the single-magnon spectrum at $T=0$ in the presence of three and four-magnon interactions by numerically computing the magnon spectral function, $\mathcal{A}(\vb*{k},\omega)$, which can be understood as the momentum-resolved single-magnon density of states. 
This quantity is closely related to observables such as the dynamical structure factor~\cite{MOURIGAL2013}, which is experimentally accessible from inelastic neutron scattering.

The single-magnon spectral function is given by \cite{mahan2000}
\begin{equation}\label{eq:SpectralFunction}
\mathcal{A} (\vb*{k},\omega) = -\frac{1}{\pi} \text{Im} \left\{ \text{Tr} \left[ \vb*{G}^{(\hat{\vb*{m}})}_{\text{ret}}(\vb*{k},\omega) \right] \right\} ,
\end{equation}
where
\begin{equation}\label{eq:GreensFunction}
\vb*{G}^{(\hat{\vb*{m}})}_{\text{ret}} (\vb{k},\omega) = \left[ \omega - S \vb{\Omega}^{(\hat{\vb*{m}})}_{\vb*{k}} - \vb{\Sigma}^{(\hat{\vb*{m}})} (\vb*{k},\omega) + i0^+ \right]^{-1},
\end{equation}
is the retarded magnon Green's function and $\left[ \vb{\Omega}^{(\hat{\vb*{m}})}_{\vb*{k}}\right]_{\mu\nu} =  \delta_{\mu\nu} \omega^{(\hat{\vb*{m}})}_{\mu,\vb*{k}}$ is the diagonalized linear-spin wave Hamiltonian. 
The (retarded) self-energy $\vb{\Sigma}^{(\hat{\vb*{m}})} (\vb*{k},\omega)$ arises from magnon-magnon interactions and can be included perturbatively in the $1/S$ expansion.
At $\mathcal{O}(S^0)$, the self-energy is calculated by evaluating the one-loop Feynman diagrams depicted in Fig.~\ref{fig:FeynmanBubble}.
At $T=0$, we find that only the leftmost diagram has a non-zero contribution (see Appendix~\ref{appendix:SelfEnergy} for more details). In this case, the components of the self-energy matrix are evaluated as
\begin{equation} \label{eqn:SelfEnergy}
\Sigma_{\mu\nu}^{(\hat{\vb*{m}})}(\vb*{k},\omega) = \frac{S}{2N} \sum_{\vb*{q}} \sum_{\lambda \rho}  \frac{ T_{\lambda \rho \mu}^{(\vu*{m})}(\vb*{q} ,\vb*{k}-\vb*{q}) \overline{T}_{\lambda \rho \nu}^{(\vu*{m})}(\vb*{q} ,\vb*{k}-\vb*{q})}{ \omega + i0^+ - S \omega^{(\hat{\vb*{m}})}_{\lambda,\vb*{k}-\vb*{q}} - S\omega^{(\hat{\vb*{m}})}_{\rho,\vb*{q}}  },
\end{equation}
where the $T_{\alpha \beta \mu}^{(\vu*{m})}(\vb*{q} ,\vb*{k}-\vb*{q})$ coefficients are the three-magnon vertices in the basis which diagonalizes the linear spin wave theory, and the overbar represents complex conjugation. 
A more detailed derivation of Eq.~\eqref{eqn:SelfEnergy} is given in Appendix~\ref{appendix:SelfEnergy}.

In the absence of magnon-magnon interactions, the single-magnon excitations have an infinite lifetime and appear as sharp delta function peaks. 
In general, the self-energy will consist of both real and imaginary parts, which will renormalize and broaden the single-magnon spectrum respectively, the latter of which is interpreted as the magnons acquiring a finite (quasiparticle) lifetime. 
In practice, the self-energy real and imaginary components are calculated by numerically computing the sum over wavevectors in Eq.~\eqref{eqn:SelfEnergy} with a finite broadening factor $0^+ \approx 10^{-3} J$.
The self-energy is then used to calculate the leading $1/S$ correction to the bare Green's function using Eq.~\eqref{eq:GreensFunction}.

\section{Results}\label{sec:Results}

In this section, we present results regarding ObD in the ferromagnetic regime of Eq.~\eqref{eq:hamiltonian}, for both the classical and quantum versions of the model.
We begin our discussion of ObD with a purely classical outlook on the entropic selection of the magnetization order parameter, using the classical low-temperature expansion (CLTE) supplemented with classical Monte Carlo (MC) simulations.
Next, we study the ObD selection in the quantum case by using linear spin wave theory. 
Finally, we extend this analysis to the non-linear spin wave theory to assess the stability of the ferromagnetic ground state in the purely quantum model.

\subsection{Classical selection} 

\begin{figure}[t!]
    \centering
    \begin{overpic}[width=\columnwidth]{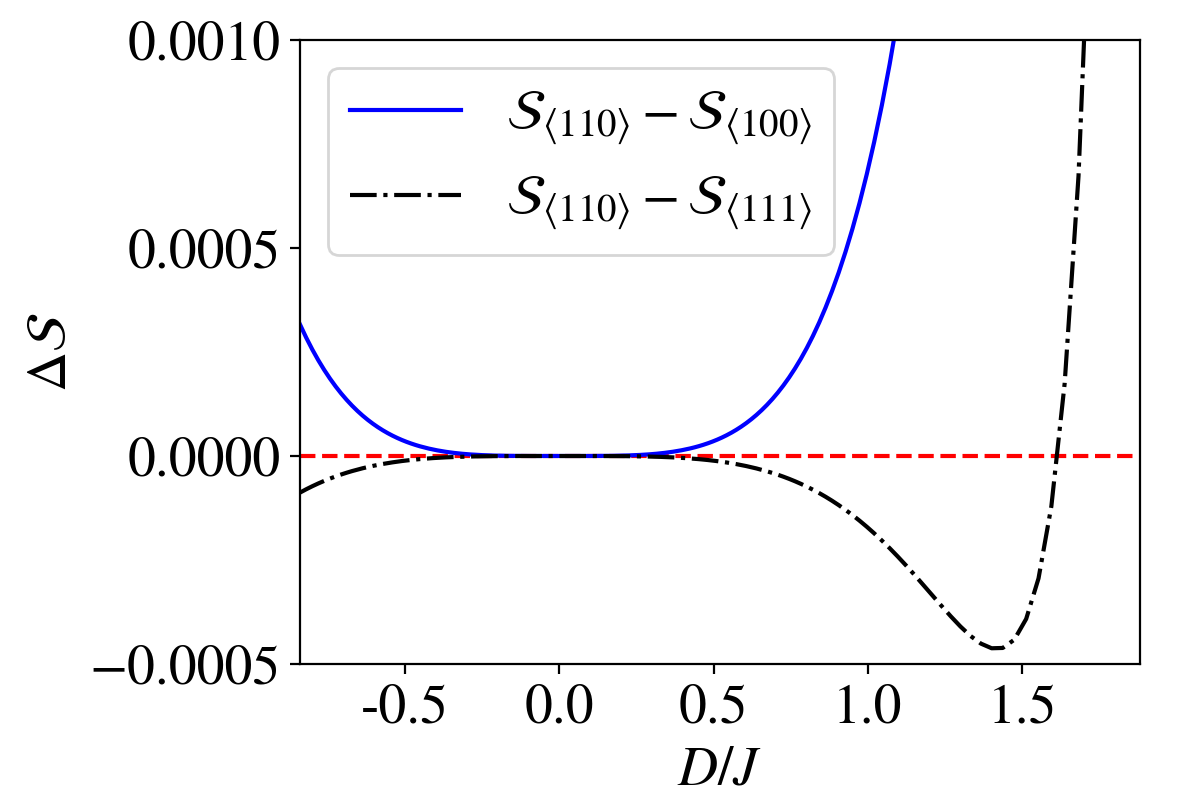}
    \end{overpic}
\caption{Thermodynamic entropy from the CLTE about a ferromagnetic ground state. 
The entropy difference between the $\expval{110}$ and the $\expval{100}$ (blue line), and the $\expval{110}$ and the $\expval{111}$ (black line) cubic directions is calculated using Eq.~\eqref{eq:entropy}. 
We note that the entropy is maximized for the magnetization along the $\expval{111}$ directions for $-1 <D/J \lesssim 1.66$ and along the $\expval{110}$ directions for $1.66 \lesssim D/J < 2$, except at $D/J = 0$ when the magnetization is isotropic. We note that the two curves never cross each other, i.e. $\mathcal{S}_{\expval{110}}-\mathcal{S}_{\expval{100}}\geq \mathcal{S}_{\expval{110}}-\mathcal{S}_{\expval{111}}$.}
    \label{fig:CLTE_entropic_weight}
\end{figure}

As a first approach to expose ObD thermal selection, we use the CLTE to study the preferred orientation of the net magnetization at low temperature ($\cramped{0 < T \ll T_{\rm c}}$). 
We note that in the classical low-temperature limit, the equipartition theorem guarantees that the free energy is minimized by the maximal entropy configurations (further discussion regarding this point can be found in Appendix~\ref{appendix:EnergyEntropy}). 
We calculate the entropy using Eq.~\eqref{eq:entropy} for magnetization along the $\expval{100}$, $\expval{110}$, and $\expval{111}$ cubic directions~\footnote{Note that we also considered the possibility that $\vu*{m}$ is not along one of the high-symmetry directions, but this does not change any of our results.}. 
The results of this calculation are shown in Fig.~\ref{fig:CLTE_entropic_weight} where the difference in entropy between distinct magnetization directions is shown. 
First, we find that the configuration for which the magnetization is polarized along the $\expval{100}$ has the lowest entropy, as exemplified by the blue curve in Fig.~\ref{fig:CLTE_entropic_weight} illustrating the entropy difference between the $\expval{110}$ and the $\expval{100}$ magnetization directions, i.e. $\mathcal{S}_{\expval{110}}-\mathcal{S}_{\expval{100}}>0$~\footnote{Similar results where obtained for the entropy difference between the $\expval{111}$ and the $\expval{100}$ magnetization directions, i.e. $\mathcal{S}_{\expval{111}}-\mathcal{S}_{\expval{100}}>0$ (not shown).}. 
A similar comparison of the entropy associated with the $\expval{111}$ and the $\expval{110}$ directions allows us to identify two distinct regimes, one for $D/J\in (-1,1.66)$ and another for $D/J\in (1.66,2)$, where a magnetization along the $\expval{111}$ and $\expval{110}$ directions respectively maximizes the entropy. 
We recall here the results of Fig.~\ref{fig:classical_phases} showing that the colinear ferromagnet transitions to the all-in-all-out (antiferromagnetic) phase at $D/J=-1$ while the former transitions to the $\Gamma_5$ (planar antiferromagnet) at $D/J=2$.
Hence, the entropic selection of the $\expval{111}$ magnetization direction for the colinear ferromagnet persists all the way to its transition point with
the all-in-all-out phase at $D/J=-1$.
We also note that the entropy difference grows as the ratio $D/J$ approaches either one of the phase boundaries with the other ordered phases (see Fig.~\ref{fig:classical_phases}).

Although the CLTE calculation of the entropy already predicts a selection of the magnetization at low temperatures, we cannot assume that the same state selection remains as the temperature increases upon approaching $T_c$ from below. 
Indeed, taking the example of the $\Gamma_5$ phase depicted in Fig.~\ref{fig:classical_phases}, it was previously found that different states may be selected near the critical temperature $T \lesssim T_{\rm c}$ compared to the low-temperature selection at $T \ll T_{\rm c}$~\cite{noculak2023,chern2010}. 
We use classical MC simulations to follow the evolution of the magnetization direction~\cite{AndradePRL2018,noculak2023}. 
We record the magnetization direction, $\vu*{m} = \sum_i \vb{S}_i / |\sum_i \vb{S}_i|$, to generate a magnetization orientation distribution function, $p(\vu*{m})$, as a function of temperature. 
This distribution function allows one to probe the free energy difference between configurations, given by
\begin{equation}
\label{eqn:magnetizationdistribution}
\frac{p(\vu*{m})}{p(\vu*{n})} = \exp \left(- \frac{\mathcal{F}^{(\vu*{m})}-\mathcal{F}^{(\vu*{n})}}{T} \right).
\end{equation}
In other words, the configuration minimizing the free energy occurs when $p(\vu*{m})$ is maximized. 
We therefore refer to a magnetization direction as being ``selected" if it is the most probable orientation to occur at the specified temperature.
For clarity, and without loss of generality, we display the distribution $p(\vu*{m})$ solely in the first octant and use spherical coordinates $\vu*{m} = (\cos \phi \sin \theta,\sin \phi \sin \theta,\cos \theta)$ where  $\theta$ and $\phi$ are the polar and azimuthal angles, respectively.

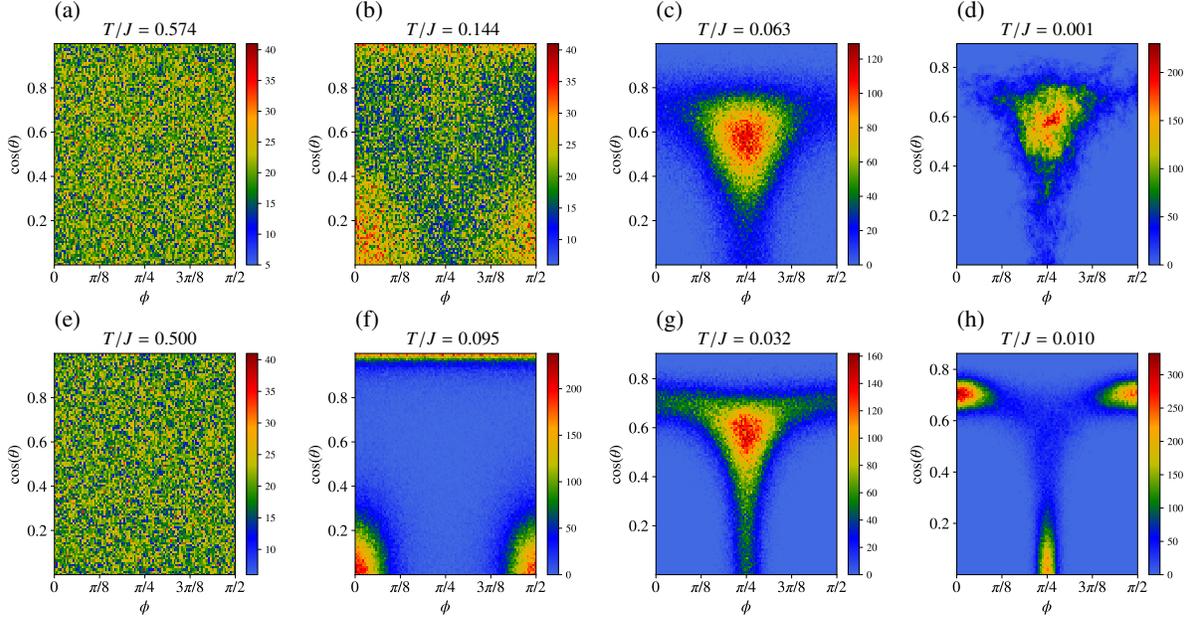
\begin{figure*}[t!]
    \centering
    \begin{overpic}[width=0.9\textwidth]{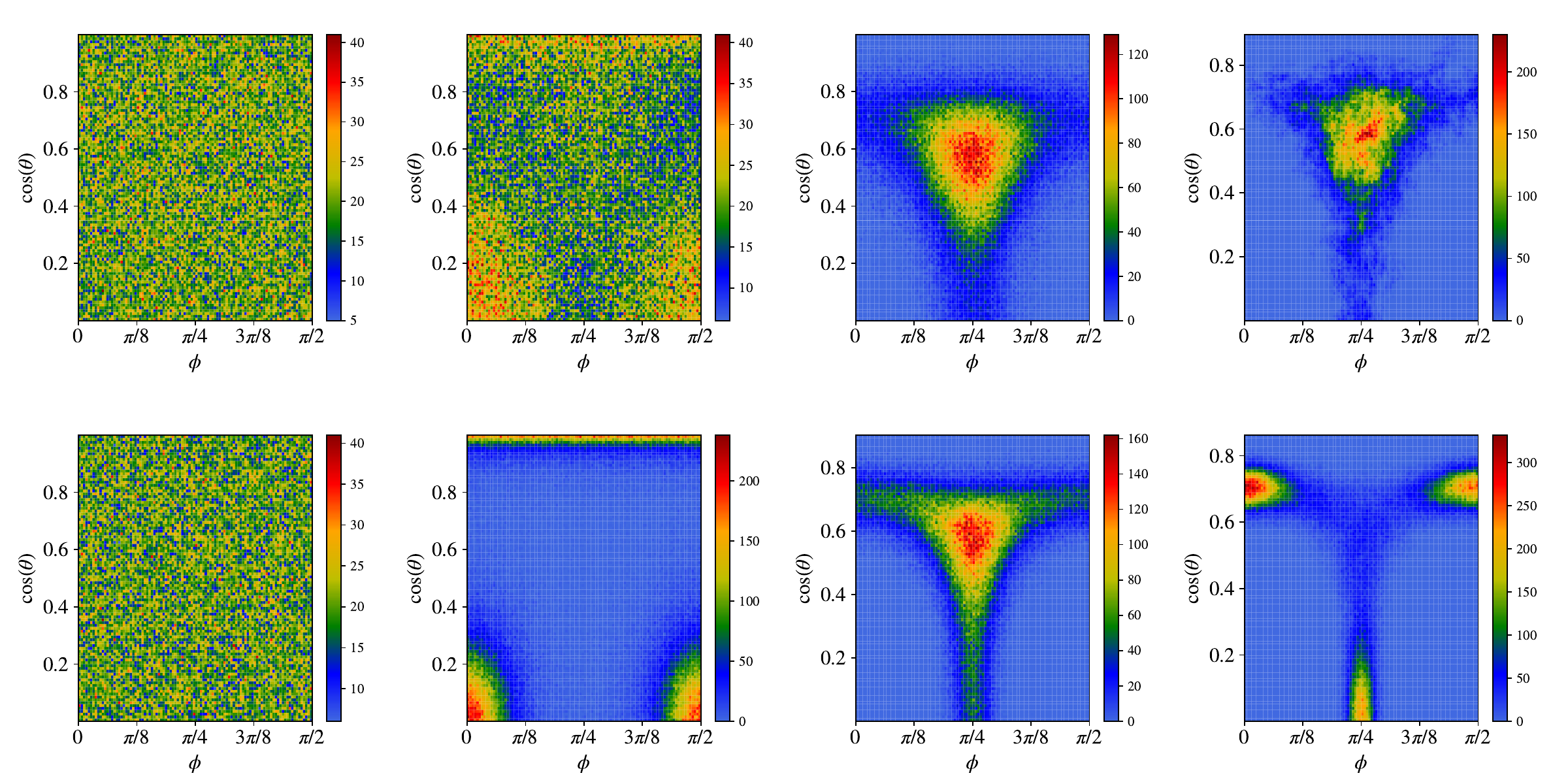}

    \put(5,50) {\textcolor{black}{(a)}}
    \put(29.8,50) {\textcolor{black}{(b)}}
    \put(54.6,50) {\textcolor{black}{(c)}}
    \put(79.4,50) {\textcolor{black}{(d)}}

    \put(5,24.5) {\textcolor{black}{(e)}}
    \put(29.8,24.5) {\textcolor{black}{(f)}}
    \put(54.6,24.5) {\textcolor{black}{(g)}}
    \put(79.4,24.5) {\textcolor{black}{(h)}}
    
    \put(9,48.5) {\scriptsize{$T/J=0.574$}}
    \put(34,48.5) {\scriptsize{$T/J=0.144$}}
    \put(58,48.5) {\scriptsize{$T/J=0.063$}}
    \put(83,48.5) {\scriptsize{$T/J=0.001$}}

    \put(9,23.0) {\scriptsize{$T/J=0.500$}}
    \put(34,23.0) {\scriptsize{$T/J=0.095$}}
    \put(58,23.0) {\scriptsize{$T/J=0.032$}}
    \put(83,23.0) {\scriptsize{$T/J=0.010$}}
    \put(42,55) {\large{Decreasing $T/J$}}
    \put(4,53){\color{black}\vector(1,0){95}}
    \end{overpic}
\caption{Distribution of the global magnetization direction, $p(\vu*{m})$, from classical Monte Carlo simulations for the two systems labeled by red triangles and red squares in Fig.\ref{fig:fig_phase_diagram}(d). 
Panels (a-d) correspond to the triangular markers in Fig.\ref{fig:fig_phase_diagram}(d), that is to $D/J=1.42$, while panels (e-h) correspond to the square markers in Fig.\ref{fig:fig_phase_diagram}(d) where $D/J=1.73$, shown for decreasing temperatures when read left to right. 
 Here, the magnetization direction is expressed in spherical coordinates $\vu*{m} = (\cos \phi \sin \theta,\sin \phi \sin \theta,\cos \theta)$ and temperature is measured in units of $J$ ($k_{\rm B} = 1$).}
    \label{fig:fig_MChistogram}
\end{figure*}

To illustrate the evolution of the thermal selection via the evolution of the magnetization distribution, we show the temperature evolution of the distribution $p(\vu*{m})$ for two values of the DM interaction: one with $D/J\simeq 1.43$ and another with $D/J\simeq 1.73$. 
As will be made clear later, these two values of $D/J$ have been chosen as they each illustrate distinct temperature-dependent evolution of the magnetization distribution functions.
Figures~\ref{fig:fig_MChistogram} (a)-(d) illustrate the evolution of $p(\vu*{m})$ as a function of temperature for the system with $D/J\simeq 1.43$. 
At temperatures above the critical temperature, $T>T_{\rm c}$, the magnetization distribution shown in Fig.~\ref{fig:fig_MChistogram}(a) is uniformly distributed. 
Just below the critical temperature, $T \lesssim T_{\rm c}$, our MC simulations reveal that the distribution $p(\vu*{m})$, shown in Fig.~\ref{fig:fig_MChistogram}(b), clusters around the points  $\cos(\theta)\in \{0,1\}$ and $\phi\in\{0,\pi/2\}$. 
These clusters indicate that the magnetization aligns along the $\expval{100}$ directions near and not too far below the critical temperature. 
As temperature is decreased, a triangle-shaped pattern develops, where the distribution clusters around $\phi=\pi/4$ and $\cos(\theta)=1/\sqrt{3}$ (see Fig.~\ref{fig:fig_MChistogram}(c) and (d)). 
The center of the triangle, with these $\phi$ and $\cos(\theta)$ values, corresponds to an orientation of the magnetization along $\expval{111}$. 
We note that the variance of the distribution becomes smaller as the temperature is lowered, implying that the selection of the $\expval{111}$ orientation becomes stronger at lower temperatures.

Figures~\ref{fig:fig_MChistogram} (e)-(h) illustrate the evolution of the magnetization distribution $p(\vu*{m})$ as a function of temperature for the system with $D/J\simeq 1.73$. 
We note that for this set of parameters, the system develops three distinct patterns in the distribution of the  magnetization direction as a function of temperature. 
For $T\lesssim T_{\rm c}$, the magnetization is similarly oriented along one of the $\expval{100}$ directions, as found for the case $D/J=1.43$ above [Fig.~\ref{fig:fig_MChistogram}(b)], as shown in Fig.~\ref{fig:fig_MChistogram}(f). 
As the temperature is further decreased, a triangular pattern develops again, indicating a preference to order along the $\expval{111}$ directions (Fig.~\ref{fig:fig_MChistogram}(g)), analogous to Fig.~\ref{fig:fig_MChistogram}(c). 
In contrast, however, as the system is cooled down into the low-temperature region, the magnetization distribution shows a distinct clustering at three lobes, centered around the points $(\cos(\theta),\phi)\in \{(1/\!\!\sqrt{2},0),(0,\pi/4),({1/\!\!\sqrt{2}},\pi/2)\}$. 
These magnetization directions correspond to the $\langle 110 \rangle$ directions. 
No further qualitative changes in the pattern of the distribution occur as the temperature is further decreased and approaches zero.

\begin{figure*}[ht]
\centering
\begin{tikzpicture}
    \draw (0, 0) node[inner sep=0] {\includegraphics[width=\textwidth]{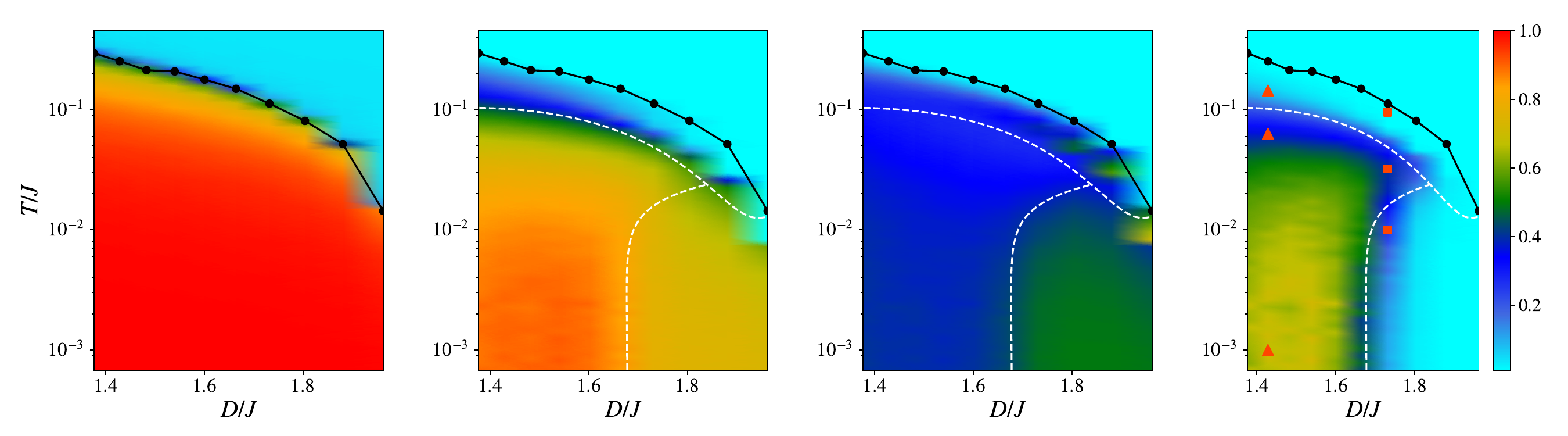}};
    \draw (-8,2.4) node [scale=1]{(a)};
    \draw (-3.6,2.4) node [scale=1]{(b)};
    \draw (0.9,2.4) node [scale=1]{(c)};
    \draw (5.3,2.4) node [scale=1]{(d)};

    \draw (-6.2,2.45) node [scale=1.2]{$\expval{|\vb*{m}|}$};
    \draw (-1.8,2.45) node [scale=1.2]{$\delta M_4$};
    \draw (2.7,2.45) node [scale=1.2]{$M_4$};
    \draw (6.9,2.45) node [scale=1.2]{$M_6$};
    
     \draw (-3.,1.45) node [scale=1,rotate = -7,white]{``$\expval{100}$''};
     \draw (-2.6,-0.5) node [scale=1,white]{``$\expval{111}$''};
     \draw (-1.1,-0.9) node [scale=1,white]{``$\expval{110}$''};

     \draw (1.42,1.44) node [scale=1,rotate = -7,white]{``$\expval{100}$''};
     \draw (1.8,-0.5) node [scale=1,white]{``$\expval{111}$''};
     \draw (3.2,-0.9) node [scale=1,white]{``$\expval{110}$''};

\end{tikzpicture}
        \caption{Thermal evolution of the (a) magnetization per spin $\expval{|\vb*{m}|}$, and the anisotropic cubic parameters (b) $\delta M_4$, (c) $M_{4}$, and (d) $M_6$, as defined in Eqs.~(\ref{eq:M_4}-\ref{eq:M6}). 
        Panels (b) and (c) illustrate the three regions where a distinct easy axis for the magnetization is chosen from ObD. 
        The triangles in panel (d) label the temperatures at which the  magnetization distribution was measured as illustrated in Fig.~\ref{fig:fig_MChistogram}(b-d), and the squares correspond to Fig.~\ref{fig:fig_MChistogram}(f-h). 
        The white dashed lines in panels (b)-(d) are guides to the eye, and the directions are indicated in quotation marks to emphasize the fact that these regions are somewhat loosely identified based on the distinct values of the cubic parameters. 
      }
    \label{fig:fig_phase_diagram}
\end{figure*}
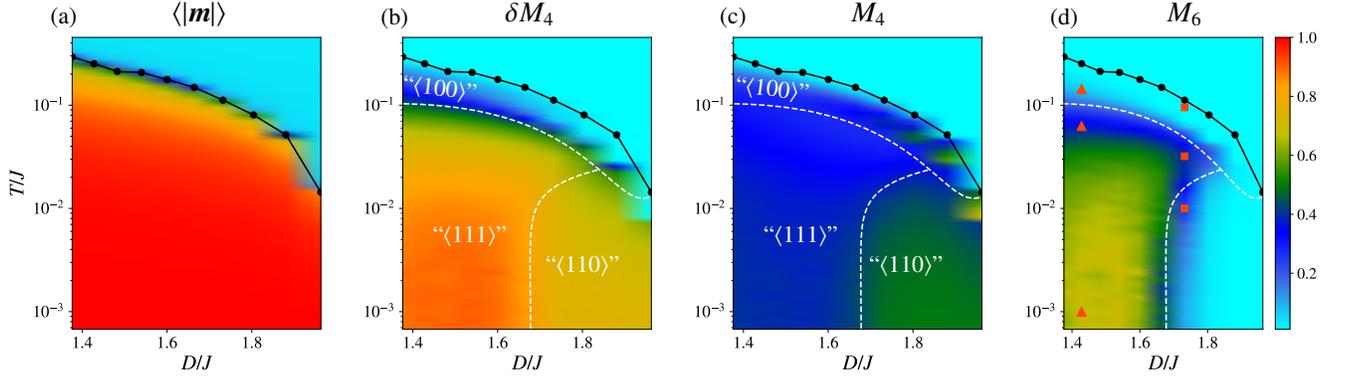

Altogether, the evolution of the magnetization distributions $p(\vu*{m})$ shown in Fig.~\ref{fig:fig_MChistogram} for two representative values of $D/J$ suggests that the colinear ferromagnetic phase of the classical Hamiltonian of Eq.~\eqref{eq:hamiltonian} presents \textit{at least} two distinct ordered phases for a given set of interaction parameters. 
The range in temperature and ratio $D/J$ where these phases are observed, however, is not trivially identified through the sole measurement of the distribution $p(\vu*{m})$. 
To provide an approximate identification of the boundaries between the different magnetization orientations, we measured the temperature evolution of the thermally averaged cubic parameters $\delta M_{4}$, $M_{4}$ and $M_6$ [as defined in Eqs.~(\ref{eq:M_4}-\ref{eq:M6}], to construct a phase diagram as a function of temperature and $D/J$, which we show in Fig.~\ref{fig:fig_phase_diagram}(a-d).
The critical temperature $T_{\rm c}$ is marked by black dots, determined by the onset of a nonzero magnetization $\expval{|\vb*{m}|}$ in Fig.~\ref{fig:fig_phase_diagram}(a). 
We note that the critical temperature collapses to zero in the limit $D/J \to 2^-$, corresponding to the spin liquid phase reported in Refs.~\cite{noculak2023,lozano-gomez2023}. The white dashed lines in Fig.~\ref{fig:fig_phase_diagram}(b-d) are drawn as guides to the eye to separate regions of distinct values of the cubic parameters. 
In Fig.~\ref{fig:fig_phase_diagram}(d), the locations at which the magnetization distributions $p(\vu*{m})$ were calculated are marked with red triangles corresponding to Fig.~\ref{fig:fig_MChistogram}(b-d) and red squares corresponding to Fig.~\ref{fig:fig_MChistogram}(f-h). 
By comparing the regions separated by the white lines with the distributions in Fig.~\ref{fig:fig_MChistogram}, we are able to qualitatively identify that these regions coincide with magnetization along distinct cubic crystalline directions. 
These ordering directions are labelled in Fig.~\ref{fig:fig_phase_diagram}(b,c) in quotation marks, to emphasize that our identification of these regions is approximate. 
We observe that the (extrapolated) zero-temperature boundary separating the $\expval{111}$ and $\expval{110}$ regions is $D/J\sim 1.66$, which is consistent with the CLTE entropy calculation reported in Fig.~\ref{fig:CLTE_entropic_weight}.
Further details regarding the calculation of the cubic $M_4$, $\delta M_4$ and $M_6$ parameters are given in Appendix~\ref{appendix:MC}.

In Fig.~\ref{fig:CLTE_entropic_weight} we found from the CLTE calculations that the entropy difference between the different magnetization direction states selected at low temperature decreases upon approaching the Heisenberg point $D\rightarrow 0^{\pm}$. 
Similarly, we found that our ability to determine what magnetization direction is selected just below $T_c$ becomes very difficult as $D/J \rightarrow 0$ as the $p(\vu*{m})$ histograms such as those in Fig.~\ref{fig:fig_MChistogram} become extremely noisy.
For example, see Fig.~\ref{fig:fig_MChistogram_app} for the case $D/J=-0.84$, which is close to the ferromagnet to all-in-all-out boundary.

\subsection{Quantum selection at \texorpdfstring{$0^+ < T \ll T_{\rm c}$}{0 < T << Tc}}
\label{section:quantum}

Within the classical ferromagnetic regime, the manifold of classical ground states possesses full rotational symmetry, despite the Hamiltonian in Eq.~\eqref{eq:hamiltonian} lacking such a symmetry for an arbitrary spin configuration when the DM interaction is nonzero. 
In a typical scenario, such as this, one might naively expect that quantizing the Hamiltonian in Eq.~\eqref{eq:hamiltonian} would partially lift the accidental degeneracy between classical ground states due to the zero-point fluctuations in Eq.~\eqref{eq:zeropoint}. 
In the present case, we find a somewhat unusual situation, where zero-point fluctuations are absent, owing to the fact that the classical ferromagnetic (product) state is an exact eigenstate of the Hamiltonian. 
To see this, consider a fully polarized product state oriented along a global magnetization unit vector $\vu*{m}$:
\begin{equation}\label{eq:ProductState}
\ket{\vu*{m}} \equiv \bigotimes_{i=1}^{4N} \ket{\uparrow_{\vu*{m}}}_i,   
\end{equation}
where $\ket{\uparrow_{\vu*{m}}}_i$ is the eigenstate of the spin operator {$\vb*{S}_i \cdot \vu*{m}$} corresponding to the maximum eigenvalue.
We find that
\begin{align}
\mathcal{H} \ket{\vu*{m}} &=S^2 E_0 \ket{\vu*{m}} +  iDS \sum_{\langle i,j \rangle} \vb*{d}_{ij}\cdot \left( \vu*{e}_1+i\vu*{e}_2\right) S_i^-  \ket{\vu*{m}} \nonumber \\
&= S^2 E_0 \ket{\vu*{m}}, \label{eq:eigenstate}
\end{align}
where it is straightforward to verify that the sum over nearest-neighbor DM vectors vanishes from Eq.~\eqref{eqn:DMvectors}. 
That is to say, $\ket{\vu*{m}}$ is an \textit{exact} eigenstate of the quantum many-body Hamiltonian in Eq.~\eqref{eq:hamiltonian}. 
Any fully polarized configuration therefore displays no zero-point fluctuations about $\ket{\vu*{m}}$. 
This peculiar scenario motivates our investigation of \emph{thermal} ObD in the ferromagnetic phase of the \emph{quantum model} in the large $S$ limit.

Since the fully polarized product state is an exact eigenstate of the normal-ordered Hamiltonian, we combine Eq.~\eqref{eq:eigenstate} with Eq.~\eqref{eq:LSW_HamiltonianB} to find that $E^{(\hat{\vb*{m}})}_2 = 0$. 
More generally, the classical ground state energy is equivalent to the energy of the ferromagnetic product state in the quantum model, implying that there are no quantum zero-point fluctuations in the fully polarized ferromagnetic phase. 
At nonzero temperature, the ObD selection, now operating in the quantum model, can be exposed by minimizing the free energy difference per unit cell (using Eq.~\eqref{eq:FullFreeEnergy})
\begin{equation}\label{eq:LSWFreeEnergy}
\mathcal{F}^{(\hat{\vb*{m}})} - \mathcal{F}^{(\hat{\vb*{n}})}=  \frac{T}{N} \sum_{\vb*{q}}\sum_{\mu=0}^3 \ln \left( \frac{1-e^{-S\omega^{(\hat{\vb*{m}})}_{\mu,\vb*{q}}/T}}{1-e^{-S\omega^{(\hat{\vb*{n}})}_{\mu,\vb*{q}}/T}}\right)
\end{equation}
with respect to $\hat{\vb*{m}}$, relative to a fixed reference magnetization axis $\hat{\vb*{n}}$. 

\begin{figure}[t!]
    \centering
   \begin{overpic}[width=\columnwidth]{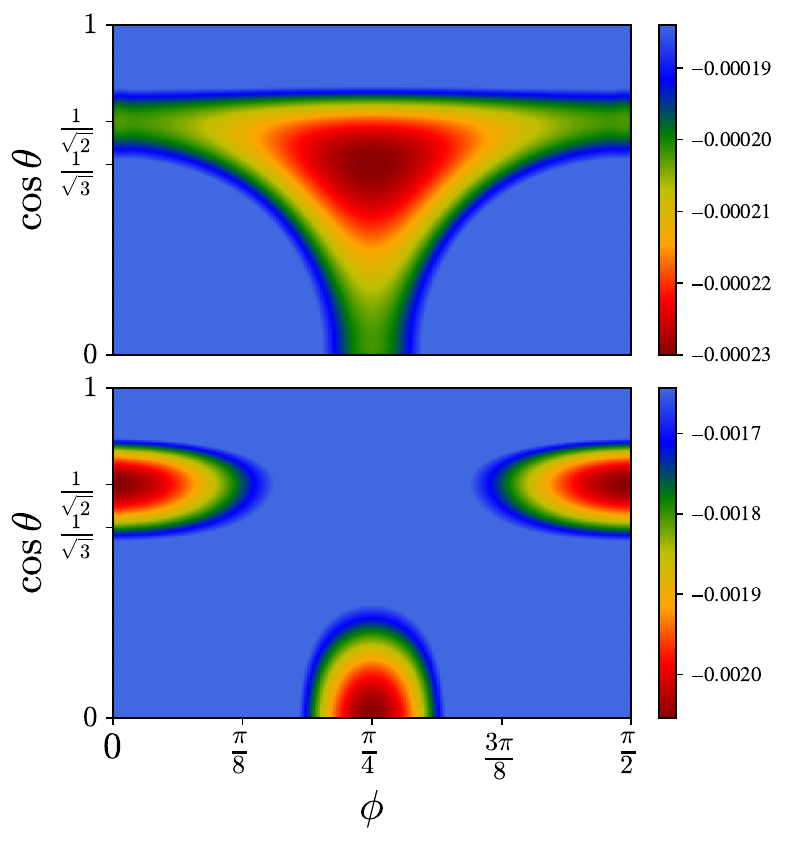}
   \put(14,94) {\textcolor{black}{(a)}}
   \put(14,51) {\textcolor{black}{(b)}}
    \end{overpic}
    \caption{Free energy difference between configurations obtained from linear spin wave theory using Eq.~\eqref{eq:LSWFreeEnergy} with $T/J = 0.1$, $\vu*{n}= \langle 100 \rangle$ and $S=\frac{1}{2}$. 
    Panel (a) depicts the free energy distribution for $D/J = 1.0$, which is minimized by $\vu*{m} = \langle 111 \rangle $. 
    Panel (b) depicts the free energy distribution for $D/J = 1.8$, which is minimized by $\vu*{m} = \langle 110 \rangle $.}
    \label{fig:LSWFreeEnergy}
\end{figure}

Comparing the free energy difference between configurations calculated using Eq.~\eqref{eq:LSWFreeEnergy}, we find that, for a given value of $T/J=0.1$ and depending on the ratio of $D/J$ for $/J<2$, there are two distinct scenarios as to whether the magnetization is directed along either the $\langle 111 \rangle$ or $\langle 110 \rangle$ directions (see Fig.~\ref{fig:LSWFreeEnergy}).
This is similar to what was found with the CLTE calculation for the classical version of the model.
Being limited to a low-temperature description of the magnon excitations well below $T_c$, the (quantum) linear spin wave theory does not capture a selection of the magnetization along the $\langle 100 \rangle$ directions similar to what was found in the Monte Carlo simulations of the classical model for $T \lesssim T_{\rm c}$ (c.f. Fig.~\ref{fig:fig_phase_diagram}(b-d)). 

\begin{figure}[t!]
    \centering
   \begin{overpic}[width=\columnwidth]{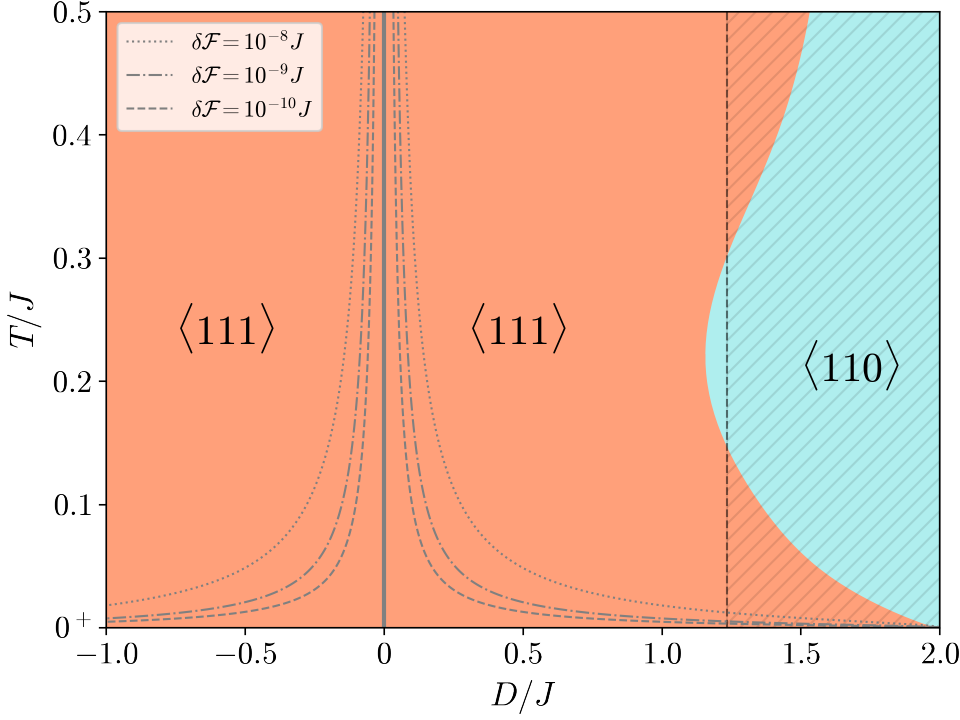}
    \end{overpic}
    \caption{Low temperature phase diagram obtained from linear spin wave theory using Eq.~\eqref{eq:LSWFreeEnergy} and $S=\frac{1}{2}$. The solid gray line at $D/J=0$ is the Heisenberg line, where the model has an exact $\text{O}(3)$ symmetry and no ObD selection.
    Several contours of equal free energy are plotted in the region surrounding the Heisenberg line, where $\delta \mathcal{F}\equiv \mathcal{F}^{\expval{110}}-\mathcal{F}^{\expval{111}}$. 
    The hatched region where $D/J > 1.235$ corresponds to the non-linear spin wave instability as described in Sec.~\ref{sec:NLSWInstability}. The temperature axis is scaled by the ferromagnetic exchange interaction $J$. }
    \label{fig:LSWPhases}
\end{figure}

Upon determining which configurations from Eq.~\eqref{eq:ProductState} minimize the free energy in Eq.~\eqref{eq:LSWFreeEnergy}, we construct a low-temperature phase diagram depicted in Fig.~\ref{fig:LSWPhases} for the case where $S=\frac{1}{2}$.
We find that, for $D/J$ corresponding to the classical ferromagnetic regime in Fig.~\ref{fig:classical_phases}, the $\langle 111\rangle $ directions are selected at $T=0^+$.
On the other hand, we observe that the $\langle 110 \rangle$ directions are the ones selected above a $\langle 111\rangle \rightarrow \langle 110\rangle$ critical temperature that increases as $D/J$ is decreased from  $D/J=2^-$ down to $D/J\sim 1.2$.
Over the same  $1.2\lesssim D/J < 2$ range, the phase diagram exhibits  a region of $\langle 111\rangle$ magnetic order re-entrance (i.e. $\langle 110\rangle \rightarrow \langle 111\rangle $) upon warming with an intermediate region with $\langle 110 \rangle$ order bracketed between a high and low temperature $\langle 111 \rangle$ magnetic order selection. 
 
We note that the thermodynamic mechanism for the low-temperature selection of the magnetization direction (i.e. internal energy versus entropy) in the quantum model is distinct from the classical one. 
In the classical case, one can explicitly identify the ${\vb*{m}}$-dependence of the entropy as the mechanism for the selection at low temperatures.
As there is not an extensive number of quadratic zero modes in the present model 
\footnotetext[103]{In a system where there are an extensive number of quadratic zero modes, there will be a thermal contribution to the internal energy, see for example Ref.~\cite{chalker1992}.}~\cite{Note103,chalker1992}, the equipartition theorem guarantees that the energy per spin is $k_{\rm B} T$, independent of the magnetization direction. 
Conversely, both the energy and entropy in quantum spin wave theory depend on the magnetization direction, contributing to the free energy in Eq.~\eqref{eq:LSWFreeEnergy}, with the low-temperature dependence of each of these two (energy and entropy) contributing terms being discussed in the next subsection. 
More details regarding the thermal selection mechanism in the quantum case can be found in Appendix~\ref{appendix:EnergyEntropy}.

 As expected, we see in Fig.~\ref{fig:LSWPhases} (by tracking the contours of equal free energy) that the free energy difference between magnetization orientations vanishes in two distinct limits, when $D \to 0$ (the Heisenberg limit) or as $T \to 0^+$. 
 We note here that since  the eigenvalue problem of the linear spin wave theory can be exactly solved for the Heisenberg model ($D=0$),  some  analytical inferences about ObD of the quantum model can be made using perturbation theory for $D/J \ll 1$.
 We use this approach in the following subsection to expose analytically what is the relative contribution of the internal energy and entropy, along with their temperature dependence, in selecting the direction of $\hat{\bm m}$ at $T\ll J$ in the limit $D/J \ll 1$.

\subsection{Quantum selection at \texorpdfstring{$T = 0^+$}{T = 0+}}
\label{section:quantumLowT}

We now proceed to perform a low-temperature expansion of the free energy to obtain the temperature dependence of the leading contribution responsible for thermal ObD in the quantum model.
We calculate this contribution perturbatively, expanding about the Heisenberg point ($D=0$). We first note that the linear spin wave Hamiltonian in Eq.~\eqref{eq:LSW_HamiltonianA} can be diagonalized analytically at $\vb*{q}=\vb*{0}$ when $D=0$. 
The matrix
\begin{equation}
    X_{0} \equiv X^{(\hat{\vb*{m}})}(\vb*{q}=\vb*{0})\Big{|}_{D=0} ,
\end{equation}
given in Eq.~\eqref{eq:LSWMatrixElements}, and appearing in the right-hand side of Eq.~\eqref{eq:LSW_HamiltonianA}, has one zero-eigenvalue and three degenerate eigenvalues of $8J$. 
The remaining part of the spin wave Hamiltonian is treated as a perturbation, namely $\delta X_{}^{(\vu*{m})}(\vb*{q}) \equiv X_{}^{(\hat{\vb*{m}})}(\vb*{q})-X_{0}$.

Next, in the framework of non-degenerate perturbation theory, we calculate the energy shift of the zero-energy magnon due to the perturbation $\delta X_{}^{(\vu*{m})}(\vb*{q})$. 
To find the low-temperature contribution of ObD to the free energy, we must find the leading order terms in the magnon dispersion that couple to the magnetization direction. 
This involves calculating the associated energy shift in perturbation theory, and subsequently expanding the dispersion in powers of the wavevector components.
We find that such terms appear at both fourth order in $D/J$, and fourth order in the wavevector components.
The resulting dispersion is expressed as $\varepsilon^{}_2(\vb*{q})+\varepsilon^{(\vu*{m})}_4(\vb*{q})+\mathcal{O}(q^6)$, where $\varepsilon_n(\vb*{q})$ contains all terms of $\mathcal{O}(q^n)$ in the wavevector components. 
The low-temperature contribution of the quartic terms to the free energy is calculated to be
\begin{align}
     \mathcal{F}^{(\hat{\vb*{m}})} \approx T&V \int \frac{\dd[3]q}{(2\pi)^3} \; \ln \left(1-e^{S\varepsilon^{}_2(\vb*{q})/T}\right) \nonumber \\
     &+ VS \int \frac{\dd[3]q}{(2\pi)^3} \; \frac{\varepsilon^{(\vu*{m})}_4(\vb*{q})}{e^{S\varepsilon^{}_2(\vb*{q})/T}-1}.
\end{align}
The leading order term in $\mathcal{F}^{(\hat{\vb*{m}})}$ that depends on the magnetization direction yields
\begin{equation} \label{eq:FreeEnergy72}
    \mathcal{F}^{(\hat{\vb*{m}})} \sim \frac{3S\zeta\left(\frac{7}{2}\right)}{32\pi^{3/2}}\frac{D^4}{J^3}\left(\hat{m}_x^4+\hat{m}_y^4+\hat{m}_z^4 \right) \left(\frac{T}{2JS}\right)^{7/2},
\end{equation}
where $\zeta(z) = \sum_{n=1}^\infty n^{-z}$ is the Riemann zeta function.
This term is minimized when $\vu*{m}=\expval{111}$, which is consistent with the low-temperature selection that was found by direct numerical evaluation of Eq.~\eqref{eq:LSWFreeEnergy}, and reported in 
Fig.~\ref{fig:LSWPhases}.
More details of the perturbative calculation are provided in Appendix~\ref{appendix:PerturbationTheory}. 
We note that Eq.~\eqref{eq:FreeEnergy72} is a sub-leading term in the low-temperature expansion of the free energy, as there is a term proportional to $T^{5/2}$ that is independent of $\vu*{m}$ [see Eq.~\eqref{eq:FreeEnergy52}].
The free energy contribution in Eq.~\eqref{eq:FreeEnergy72} explicitly depends on the magnetization direction, and therefore describes how (in zero magnetic field) the DM interaction lifts the accidental degeneracy of the ferromagnetic ground state to lowest order in $D/J$ and $T/J$. Next, we briefly discuss how ObD manifests itself at low temperature in the presence of an external magnetic field.

The ferromagnetic pyrochlore $\text{Lu}_2\text{V}_2\text{O}_7$ is believed to be well-described by Eq.~\eqref{eq:hamiltonian}.
Interestingly, measurements of the magnetization of $\text{Lu}_2\text{V}_2\text{O}_7$ in Ref.~\cite{onose2010} appear to be anisotropic with respect to the applied field direction, begging the question of whether anisotropic field response may be related to the ObD discussed in the present work and possibly operating in this material.
{Motivated by this question, we generalize the perturbative calculation above to include a weak external magnetic field $\vb*{B}$, which removes the accidental degeneracy of the ground state.
As the model Eq.~\eqref{eq:hamiltonian} does not benefit from a quantum ObD at zero temperature generating a magnetic anisotropy, in the $T=0$ limit the magnetization will immediately align parallel to the external field, as this configuration minimizes the energy. 
This contrasts with the competition between quantum ObD and the Zeeman energy in the Er$_2$Ti$_2$O$_7$ pyrochlore antiferromagnet~\cite{Maryasin2016}.}
Conversely, at small nonzero temperature, the $\vu*{m}=\expval{111}$ configuration is preferred by thermal fluctuations when the field is absent. 
It follows that there must be a rotation of the magnetization direction as a function of the applied field, whereby $\vu*{m}=\expval{111}$ when $\vb*{B} = \vb*{0}$, and $\vu*{m}$ aligns parallel to $\vu*{B}$ when the applied external magnetic field is large in magnitude. 
We generalize the perturbative calculation of the free energy above to include a weak magnetic field in Appendix \ref{appendix:MagnetizationInField}. 
However, we find that such a change of the magnetization direction occurs for extremely weak magnetic fields, where the Zeeman coupling is one million times smaller than the ferromagnetic exchange $\left( \mu_\text{B}^{} g |\vb*{B}|/J \sim 10^{-6} \right)$ even for a sizeable DM interaction $D/J = 0.5$ and at a temperature $T/J = 0.15$ (see Fig.~\ref{fig:MagnetizationInField}(b)).
For reference, the DM interaction for $\text{Lu}_2\text{V}_2\text{O}_7$ is much smaller at $D/J \lesssim 0.13$ \cite{Mena2014,Riedl2016a}.

Finally, it is important to mention that although the fully polarized product state is an exact eigenstate as shown in Eq.~(\ref{eq:eigenstate}), it not necessarily the case that this eigenstate is the true many-body ground state outside of the perturbative regime $|D| \ll J$. 
To address this possibility, we next extend our calculations to the lowest order in non-linear spin wave theory to analyze the presence (or lack of) magnetic order in the many-body ground state.

\subsection{Limit of Stability}\label{sec:NLSWInstability}
\begin{figure*}[t!]
    \centering
    \begin{overpic}[width=\textwidth]{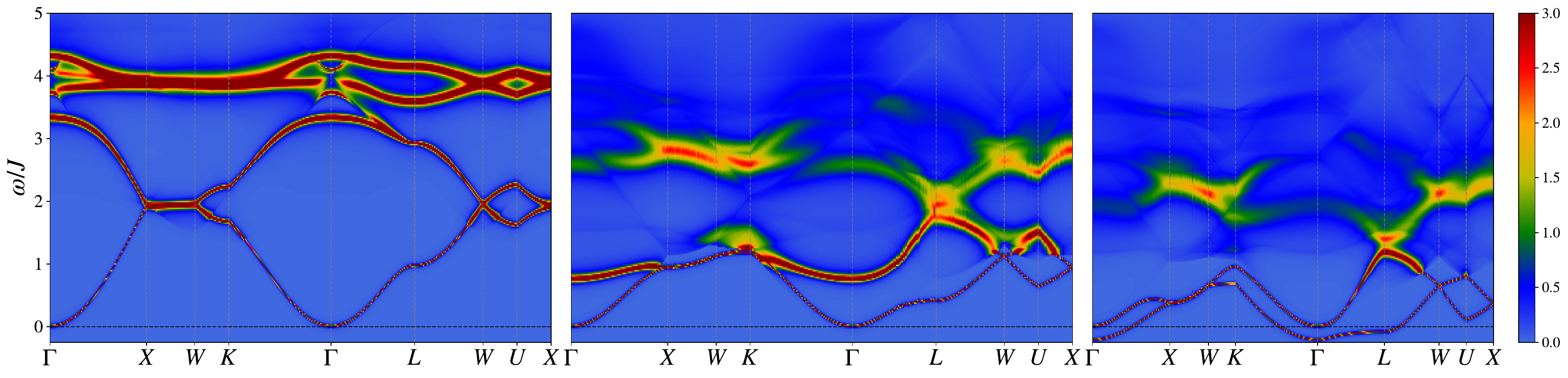}
    \put(19,25) {\textcolor{black}{(a)}}
    \put(52,25) {\textcolor{black}{(b)}}
    \put(82,25) {\textcolor{black}{(c)}}
    \put(23.2,6.1){\color{white}\vector(-3,-4){2}}
    \put(56.4,6.1){\color{white}\vector(-3,-4){2}}
    \put(86.1,6.1){\color{white}\vector(-3,-4){2}}
    \end{overpic}
    \caption{Zero-temperature magnon spectral function $\mathcal{A} (\vb*{k},\omega)$ for $\vu*{m} = \langle 111 \rangle$ and $S=\frac{1}{2}$ to $\mathcal{O}(S^0)$ in non-linear spin wave theory, calculated using Eq.~\eqref{eq:SpectralFunction} for (a) $D/J = 0.25$, (b) $D/J = 1.00$ and (c) $D/J = 1.3$. The white arrows indicate the gapless pseudo-Goldstone mode in the magnon spectrum.}
    \label{fig:NLSW}
\end{figure*}
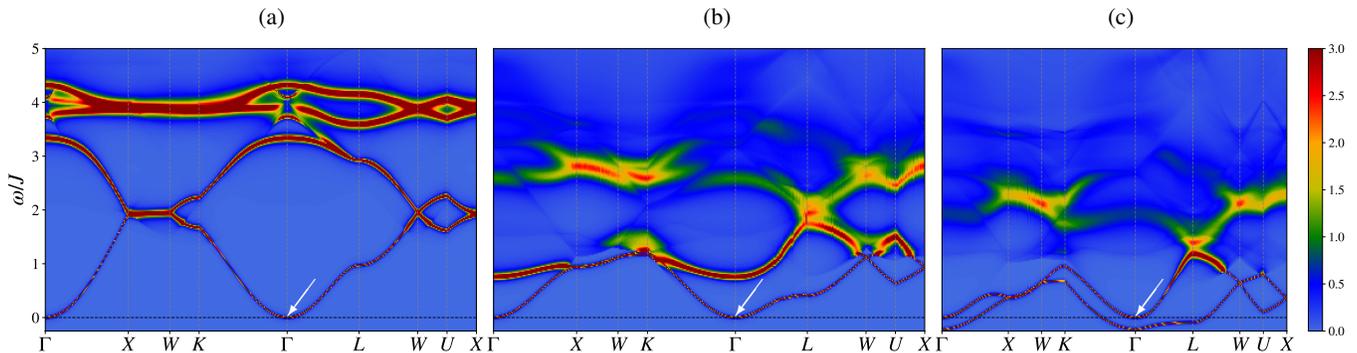

The classical ground states depicted in Fig.~\ref{fig:classical_phases} have thus far been found to be consistent with the magnetic order observed in both the CLTE and classical MC calculations. 
Conversely, while Eq.~\eqref{eq:eigenstate} shows that the ferromagnetic product state is an exact eigenstate, it is not necessarily the true quantum ground state of Eq.~\eqref{eq:hamiltonian} for the whole $-1 < D/J < 2$ range.
For example, recent work investigating the quantum ground states of this model using functional renormalization group methods did identify  a ferromagnetic ordered state for $-1 < D/J \lesssim 1.3$, but found a lack of conventional magnetic order for $\cramped{1.3~\lesssim~D/J~\leq~2}$ \cite{noculak2023,lozano-gomez2023}. 
These last results suggest that while Eq.~\eqref{eq:ProductState} may be the ground state when $-1 < D/J \lesssim 1.3$, there exists the possibility that for spin-$\frac{1}{2}$, an exotic quantum spin liquid of lower energy than the ferromagnetic state may be realized in the range  $1.3~\lesssim~D/J~\leq~2$
\footnote{Using a number of numerical and analytical methods, Ref.~\cite{lozano-gomez2023} found evidence that the $D/J=2$ point for a spin-$\frac{1}{2}$ system may be a quantum spin liquid, described by the combination of a rank-1 and rank-2 emergent gauge field, which corresponds to a triple point in the classical phase diagram where two long-range ordered quadrupolar phases meet with a spin ice state. In the classical $S=\infty$ limit, evidence is compelling
that the point $D/J=2$ is a classical spin liquid~\cite{lozano-gomez2023}}. 
This raises the question of whether linear spin wave theory provides an accurate description of the low-temperature excitations of this model.

Motivated by these results, and to further address the question of stability of the ferromagnetic ordered phase, we go beyond linear spin wave theory to calculate the $\mathcal{O}(S^0)$ correction to the single-magnon spectral function at $T=0$, using Eq.~\eqref{eq:SpectralFunction}. 
Analogous to linear spin wave theory, where a classical instability is characterized by the appearance of soft modes, we investigate whether there are soft modes in the renormalized spectrum in the $-1 < D/J < 2$ range. 

Given that the ferromagnetic configuration is characterized by an $\text{O}(3)$ degeneracy, which persists in the quantum model, at least one gapless mode is guaranteed to be observed in 
the linear spin wave theory at $\vb*{q}=\vb*{0}$ --- the pseudo-Goldstone mode \cite{WEINBERG1972,BURGESS2000}.
In a larger number of frustrated spin models, the presence of quantum fluctuations drives the pseudo-Goldstone mode to become gapped at $\mathcal{O}(S^0)$, as revealed in the self-energy \cite{rau2018}. 
However, the fact that the ferromagnetic configuration is an exact eigenstate of the Hamiltonian implies that this mode remains gapless at all orders of the $1/S$ expansion at $T=0$, a somewhat peculiar property of the current model in Eq.~\eqref{eq:hamiltonian}.
Conversely, we expect the accidental degeneracy of the quantum spin-$\frac{1}{2}$ model to be lifted by thermal fluctuations, leading to the dynamical generation of a pseudo-Goldstone gap at nonzero temperature~\cite{KHATUA2023}. 
Since our self-energy calculations are carried out at $T=0$, the pseudo-Goldstone mode remains gapless. 
We therefore look for instabilities in the form of additional soft modes in the magnon spectral function, $\mathcal{A} (\vb*{k},\omega)$ in Eq.~\eqref{eq:SpectralFunction}. 
Details of this calculation are provided in Appendix~\ref{appendix:SelfEnergy}.

The results of these calculations are depicted in Fig.~\ref{fig:NLSW}, showing the single-magnon spectral function to $\mathcal{O}(S^0)$ in non-linear spin wave theory. 
As $D/J$ becomes large, the optical magnon bands become significantly broadened, displaying a Lorentzian-like form in frequency space. This broadening originates from decay effects, the width of the Lorentzian profile close to the spectral peaks is precisely the decay rate, encoded in the imaginary part of the self-energy, $\Sigma^{(\vu*{m})}(\vb*{k},\omega)$. 
While the self-energy renormalizes the spectral function peaks, we find that there remains a stable and gapless pseudo-Goldstone mode at $\vb*{q}=\vb*{0}$ for all $-1 < D/J < 2$ as expected. This gapless mode is indicated by the white arrows in Fig.~\ref{fig:NLSW}. 

We find that one of the magnon branches becomes soft and shifts to negative frequencies at $\vb*{q}=\vb*{0}$ ($\Gamma$ point) for a large indirect DM interaction ($J>0$, $D>0$), as depicted in Fig.~\ref{fig:NLSW}(b-c).
In particular, this mode goes soft at a critical DM interaction $D/J \approx 1.235$, indicating that the long-range ferromagnetic order becomes unstable at $\mathcal{O}(S^0)$ in the non-linear spin wave theory for $1.235 \lesssim D/J <2$. 
The $D/J$ value for at which this instability occurs was calculated for $\vu*{m} = \langle 100 \rangle$, $\langle 110\rangle$ and $\langle 111 \rangle$ and was found to be independent of the classical magnetization direction $\vu*{m}$. 
For reference, the instability is marked by the vertical dashed line at $D/J \sim 1.235$ on the linear spin wave phase diagram in Fig.~\ref{fig:LSWPhases}. 
Importantly, we note that the instability approximately coincides with the value of the DM interaction where $\langle 110 \rangle$ order is found in the linear spin wave theory at $T>0$ (see Fig.~\ref{fig:LSWPhases}). 
The region of instability $1.235 \lesssim D/J < 2$ is consistent with the pseudo-fermion functional renormalization group calculations for the same model in Ref.~\cite{noculak2023,lozano-gomez2023}, where no long-range magnetic order was detected in the range $1.3 \lesssim D/J < 2$. 
Conversely,  Noculak {\it et al.} \cite{noculak2023} report conventional long-range magnetic order for $D/J \gtrsim 2$, corresponding to the $\Gamma_5$ phase in Fig.~\ref{fig:classical_phases}. 
The breakdown of spin wave theory in the classically ferromagnetic region this model is likely tied to the spin-liquid phase discussed in Ref.~\cite{lozano-gomez2023}.
Our results suggest that the ferromagnetic product state in Eq.~\eqref{eq:ProductState} is the exact ground state of Eq.~\eqref{eq:hamiltonian} for $-1 \lesssim D/J \lesssim 1.235$, and that the phase diagram in Fig.~\eqref{fig:LSWPhases}, along with the establishment of ObD without zero-point fluctuations, is reasonably accurate in this region.

\section{Discussion and Conclusion}

In this work, we studied the classical and quantum order-by-disorder selection of a colinear ferromagnetic phase of the Heisenberg and Dzyaloshinskii-Moriya Hamiltonian on the pyrochlore lattice. 
In the classical model, we found three distinct regions in parameter space \emph{and} as a function of temperature where three distinct magnetization orientations are thermally selected. 
For large and indirect DM interaction $D \lesssim 2J$ with $J>0$, classical Monte Carlo simulations find a cascade of re-orientations of the magnetization direction as a function of temperature, as depicted in Fig.~\ref{fig:fig_MChistogram}, and summarized in the phase diagram of Fig.~\ref{fig:fig_phase_diagram}.
Considering our classical MC results for $T \lesssim T_{\rm c}$, the behavior of the observed ferromagnetic transition is consistent with that of an $\text{O}(3)$ vector model with cubic anisotropy, governed by the Ginzburg-Landau-Wilson (GLW) free energy functional \cite{AHARONY1973}
\begin{equation} \label{eq:GLWfunctional}
\mathcal{F}[ \vb*{m} ] \approx \int \dd[3]{x} \left[ \frac{1}{2} \sum_{\mu=1}^3 \left( \grad{m_\mu} \right)^2 + \frac{r}{2}|\vb*{m}|^2 + \frac{u}{4}|\vb*{m}|^4 + v \sum_{\mu=1}^3 m_\mu^4 \right].
\end{equation}
When $v > 0$, the cubic anisotropy favors the magnetization to be aligned with the $\langle 111 \rangle$ directions. When $v < 0$, the anisotropy favors the magnetization to be aligned along the $\langle 001 \rangle$ directions. 
From a renormalization group perspective, the $\text{O}(n)$ vector model with cubic anisotropy in three dimensions has a critical order parameter dimension $n = n_{\rm c}$, below which the Heisenberg fixed point is stable and above which the cubic fixed point is stable~\cite{CHAIKIN2000}.
Incidentally, $n_{\rm c} \approx 3$ \cite{ferer1981,newman1982,manuelcarmona2000,ADZHEMYAN2019}, and it remains an open debate as to which of the fixed points of Eq.~\eqref{eq:GLWfunctional} is the one that is stable.
In addition, the transition is first order if $u<0$, and may be fluctuation induced first order if $u>0$ and $v<0$, depending on the ratio of the couplings $v/u$ \cite{wallace1973,ketley1973,pelissetto2002}.
In our work, we have not explored the nature of the paramagnetic to ferromagnetic transition (i.e. first or second order and universality class of the latter). 
We leave this topic for future studies.

In contrast, Eq.~\eqref{eq:GLWfunctional} does not predict a  $\langle 110\rangle $ orientation which is observed in our classical MC simulations in the region where $D/J\to 2^-$~\footnote{For the classical model, the instability discussed for the spin-$\frac{1}{2}$ does not occur for any $|D|/|J|<2$}.
This would require, for example, the addition of a (irrelevant, in the renormalization group sense) $6^\text{th}$ order term to the GLW free energy, similar to the terms appearing in Eq.~\eqref{eq:GL-theory}.
Additionally, the critical temperature completely vanishes at the boundary between the colinear ferromagnet and the $\Gamma_5$ phase ($D/J=2$), where a classical spin liquid has recently been reported and characterized by fluctuating rank-1 vector and rank-2 tensor gauge fields~\cite{lozano-gomez2023}. 

In the quantum model, we find that the fully polarized product state in Eq.~\eqref{eq:ProductState} is an exact eigenstate of the Hamiltonian regardless of the magnetization direction.
As a result, the accidental degeneracy of the ferromagnetic phase is not lifted by zero-point fluctuations at $T=0$. 
This accidental degeneracy is however lifted by the low-energy spin wave excitations at nonzero temperature ($T>0$), leading to a selection of a discrete set of magnetization directions as depicted in Fig.~\ref{fig:LSWPhases}. 
This is the first realization of such an ObD mechanism that we are aware of where the selection of an ordered phase in a quantum system is purely thermal --- thus answering the question we set in the Introduction whether one form of ObD can occur without the other.

Again for the quantum model, we find that when $|D| \ll J$, 
the selection of the magnetization direction enters as a sub-leading term in the low-temperature expansion of the free energy.
Using non-linear spin wave theory, we found that the ferromagnetic order does not seem to persist for $D/J \gtrsim 1.235$.  
We note that the instability in non-linear spin wave theory coincides reasonably well with the region of the linear spin wave phase diagram (Fig.~\ref{fig:LSWPhases}) where the magnetization orders in the $\expval{110}$ directions above a certain temperature.
The possible relationship between these two observations is unclear at this point and is left for future work. 
That being said, this result may suggest that the narrow window of $\langle 110\rangle$ order for $T\gtrsim 0.15$ for $1.1 \lesssim D/J \lesssim 1.235$ observed in Fig.~\ref{fig:LSWPhases} may actually not occur in the $S=\frac{1}{2}$ system, but would presumably be present as $S$ is increased above a critical value.

The results from the present work may have implications for real ferromagnetic materials where the DM interaction is the leading perturbation. In particular, the magnetism of the ferromagnetic Mott insulators $\text{Lu}_2\text{V}_2\text{O}_7$ and $\text{Y}_2\text{V}_2\text{O}_7$ is rather well described by Eq.~\eqref{eq:hamiltonian} \cite{Mena2014,Nazipov2016,Nazipov2016a}.  
Recently, it was proposed in Ref.~\cite{KHATUA2023} that the thermal contribution to ObD can be diagnosed by a temperature-dependent energy gap measured via inelastic neutron scattering. 
We expect this would be the strongest experimental signature of ObD in these materials.
However, since $D/J \sim 0.1$ in these materials~\cite{Mena2014,Riedl2016a}, the $T$-dependent ObD gap may be very small and could possibly be overwhelmed by a quantum ($T=0$) ObD gap induced by a small nonzero Kitaev $K$ interaction~\cite{Note80}  with, in addition,
an energetically-induced gap caused by a small pseudo-dipolar interaction $\Gamma$.

Finally, we note that a more general class of spin models may be expected to exhibit ObD without zero-point fluctuations.
Specifically, any ferromagnetic Heisenberg model with DM interactions will satisfy Eq.~\eqref{eq:eigenstate} as long as the sum over DM vectors along the bonds surrounding each site vanishes. 
We then expect that the ferromagnetic product state will remain the exact ground state of such a Hamiltonian as long as the DM interactions are sufficiently small with respect to the Heisenberg exchange $J$. 
Our calculations may then serve as a template to study thermal ObD in these analogous models.

\begin{acknowledgements}
The authors acknowledge useful discussions with Itamar Aharony, Kristian Chung, Yasir Iqbal, Subhankar Khatua, Paul McClarty, Vincent Noculak, Jeffrey Rau, Johannes Reuther, Addison Richards, Rajiv Singh and Mike Zhitomirsky. 
We also thank Vincent Noculak for comments on an early version of the manuscript.
AH acknowledges support from the NSERC of Canada CGS-D Scholarship. 
DLG acknowledges financial support from the DFG through the Hallwachs-R\"ontgen Postdoc Program of the W\"urzburg-Dresden Cluster of Excellence on Complexity and
Topology in Quantum Matter -- \textit{ct.qmat} (EXC 2147, project-id 390858490) and
through SFB 1143 (project-id 247310070).
The work at the University of Waterloo was supported by the NSERC of Canada and the Canada Research Chair (Tier 1, M.J.P.G.) program. 
This research was enabled in part by computing resources provided by the Digital Research Alliance of Canada.
\end{acknowledgements}

\vspace{2mm}

\appendix

\section{Pyrochlore lattice convention} \label{appendix:Pyrochlore}
The pyrochlore lattice is a face-centered cubic (FCC) lattice with four sites, defining a regular tetrahedron, per primitive unit cell. 
The primitive FCC lattice vectors are given by
\begin{equation} \label{eqn:FCCvectors}
\vb*{a}_{1} = \frac{a}{2}\left( \vu*{y}+ \vu*{z} \right)  \qquad \vb*{a}_{2} = \frac{a}{2}\left( \vu*{x}+ \vu*{z} \right) \qquad \vb*{a}_{3} = \frac{a}{2}\left( \vu*{x}+ \vu*{y} \right).
\end{equation}
At each FCC lattice site, there is a spin of type $0$. The remaining three spins within a unit cell are separated by the sublattice vectors $\vb*{r}_{0 \mu} = \vb*{a}_{\mu}/2$. 
The sublattice vectors between a spin of type $\mu$ to a spin of type $\nu$ can then be constructed as $\vb*{r}_{\mu \nu} \equiv \vb*{r}_{0 \nu} - \vb*{r}_{0 \mu}$.

The (indirect) DM vectors in this representation are given by
\begin{align} \label{eqn:DMvectors}
\vb*{d}_{01} &= \vu*{z}-\vu*{y}  \qquad \vb*{d}_{02} = \vu*{x} -\vu*{z} \qquad \vb*{d}_{03} = \vu*{y} - \vu*{z}  \\ \nonumber
\vb*{d}_{21} &= \hat{\vb*{x}}+\hat{\vb*{y}}  \qquad
\vb*{d}_{13} = \vu*{x}+\vu*{z}   \qquad \vb*{d}_{32} =\vu*{y}+\vu*{z},
\end{align}
where the remaining DM vectors can be determined using the antisymmetry property $\vb*{d}_{ji} = -\vb*{d}_{ij}$. 
Note that in this representation, the DM vectors used in the present work are \emph{unnormalized}, with $|\vb*{d}_{ij}| = \sqrt{2}$.

\begin{figure*}[ht!]
    \centering
    \begin{overpic}[width=0.9\textwidth]{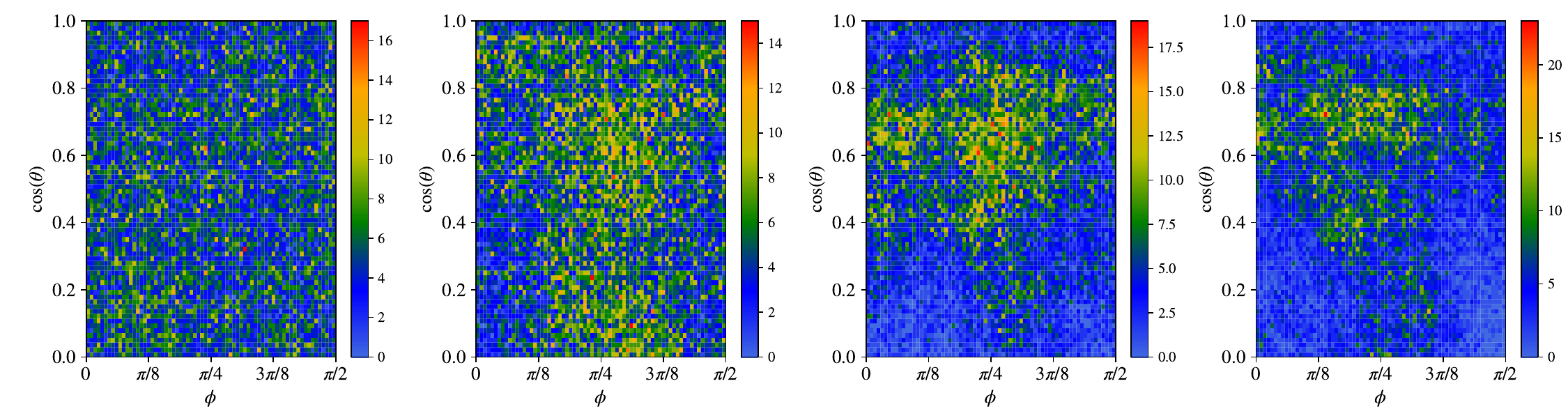}

    \put(5,27) {\textcolor{black}{(a)}}
    \put(29.8,27) {\textcolor{black}{(b)}}
    \put(54.6,27) {\textcolor{black}{(c)}}
    \put(79.4,27) {\textcolor{black}{(d)}}
    
    \put(9,26.0) {\scriptsize{$T/J=0.608$}}
    \put(36,26.0) {\scriptsize{$T/J=0.305$}}
    \put(60,26.0) {\scriptsize{$T/J=0.153$}}
    \put(85,26.0) {\scriptsize{$T/J=0.121$}}
    \put(42,33) {\large{Decreasing $T/J$}}
    \put(4,30){\color{black}\vector(1,0){95}}
    \end{overpic}
\caption{Distribution of the global magnetization direction, $p(\vu*{m})$, from classical Monte Carlo simulations for a system with $D/J=-0.84$ for which $T_c/J \sim 0.64$.
Here, the magnetization direction is expressed in spherical coordinates $\vu*{m} = (\cos \phi \sin \theta,\sin \phi \sin \theta,\cos \theta)$ and temperature is measured in units of $J$ ($k_{\rm B} = 1$).
In panels (b)-(d), which are below $T_c/2$, 
the highest intensity is found near 
$\cos(\theta) = 1/\sqrt{3}, \phi = \pi/4$, suggesting a $\expval{111}$ orientation.
}
    \label{fig:fig_MChistogram_app}
\end{figure*}
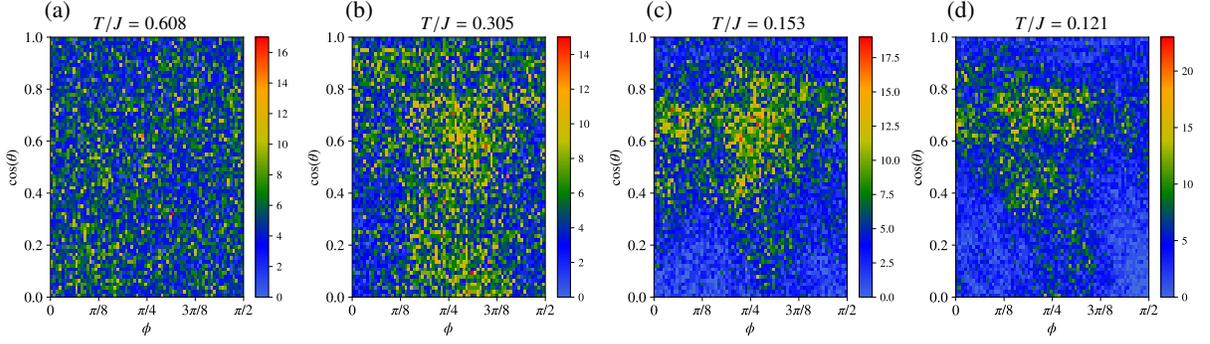

\section{Monte Carlo calculation of \texorpdfstring{$M_4$, $\delta M_4$ and $M_6$}{M4, δM4 and M6}} \label{appendix:MC}

In the main text, we illustrated in Fig.~\ref{fig:fig_MChistogram} the thermal evolution of the distribution $p({\vb*{m}})$ of the global magnetization direction, $\hat{\vb*{m}}$, for two sets of parameters of the Hamiltonian in Eq.~\eqref{eq:hamiltonian}. 
The evolution of these distributions displayed an intricate and rich phase diagram attributed to a thermal ObD selection of the magnetization direction $\hat{\bm m}$ in the classical Hamiltonian in Eq.~\eqref{eq:hamiltonian}. 
To further characterize these phases, we introduced the cubic parameters $M_4$, $\delta M_4$ and $M_6$ in Eqs.~(\ref{eq:M_4}-\ref{eq:M6}) and calculated their averaged value using classical Monte Carlo, as depicted in Fig.~\ref{fig:fig_phase_diagram}. 
In this figure, we mainly focused on the region for which $D/J>1.3$ where the ObD selection is stronger, agreeing with the CLTE analysis in Fig~\ref{fig:CLTE_entropic_weight}. 
In this appendix, we further discuss the thermal evolution of these parameters and the expected values these would have for different magnetization orientations.
Importantly, one should note that for an isotropically distributed magnetization direction, the values of the cubic parameters will be distinct from the case where there is a selected ${\vu*{m}}$ direction, as shown in Table~\ref{table:CubicParams}. 

\begin{table}[th!]
    \centering
    \begin{tabular}{| c | c | c | c |}
        \hline
         & $M_4$ & $\delta M_4$ & $M_6$ \\
        \hline
        Isotropic & $3/5$ & $3/5$ & $9/35$ \\
        $\expval{100}$ & $1$ & $0$ & $0$ \\
        $\expval{110}$ & $1/2$ & $3/4$ & $0$ \\ 
        $\expval{111}$ & $1/3$ & $1$ & $1$ \\
        \hline
    \end{tabular}
    \caption{Expected values of the cubic parameters in Eqs.~(\ref{eq:M_4}-\ref{eq:M6}) for a magnetization along different directions. }
    \label{table:CubicParams}
\end{table}

In Fig.~\ref{fig:fig_phase_diagram}(d), we mark with triangles [squares] the locations at which the distributions $p(\vu*{m})$ shown in Fig.~\ref{fig:fig_MChistogram}(b-d) [Fig.~\ref{fig:fig_MChistogram}(f-h)] were measured below the critical temperature. 
As discussed in the main text, the temperature evolution of the cubic parameters depicted in Fig.~\ref{fig:fig_phase_diagram}(b-d) identifies three distinct regions in both temperature $T$ and DM interaction $D/J$
where the white dashed lines are guides to the eye separating the labeled phases. 
The first, highest, temperature region, which we label ``$\expval{100}$'', is observed just below the critical temperature $T_{\rm c}$ as shown in Fig.~\ref{fig:fig_phase_diagram}. 
This phase is most clearly exposed in the regime $1 \lesssim D/J<2$, and likely extends to lower values of $D/J$ for $D/J>0$~\footnote{The precise determination of the extension of the $\expval{100}$ phase as the temperature drops below $T_c$ is challenged by the strong thermal fluctuations near criticality as well as the weak thermal selection associated to this phase, as well as the $\expval{111}$ (see Fig.~\ref{fig:CLTE_entropic_weight}) as $|D|/J \rightarrow 0^+$.
 As such, the identification of a clearer 
  $\expval{100}/\expval{111}$ phase boundary in this region of the phase diagram would require a significantly more extensive numerical study while also considering larger system sizes.}.  
In this region, 
the $\delta M_4$ and the $M_6$ cubic parameters acquire small values well below the isotropic limit, which are consistent with a $\langle 100\rangle$ magnetization orientation. 
As expected, the deviation of the cubic parameters from the isotropic values increases as $D/J$ increases. 
The second region, which we labeled region ``$\expval{111}$'', is observed at low temperatures in the range $-1<D/J\lesssim 1.66$ excluding the isotropic point $D/J=0$. 
We note that this phase appears to be the only possible orientation  of the magnetization for $D/J<0$ where no sign of a $\expval{100}$ or a $\expval{110}$ region is detected, as illustrated by the magnetization distribution in Fig.~\ref{fig:fig_MChistogram_app} for $D/J=-0.84$ where $T_c/J\sim 0.64$~\footnotetext[126]{In these distributions of the magnetization direction, the highest intensity appears near the point $\cos(\theta) = 1/\sqrt{3}, \phi = \pi/4$, suggesting a $\expval{111}$ orientation.}\cite{Note126}.
In the ``$\expval{111}$'' region, the cubic parameters $\delta M_4$, $M_{4}$, and $M_6$ evolve smoothly, plateauing to the approximate values $9/10$, $1/3$, and $7/10$, respectively, see Fig.~\ref{fig:fig_phase_diagram}(b-d). 
We note that the cubic parameters in this region suggest a $\langle 111 \rangle $ selection where all cubic parameters are non-vanishing. 
The third region, which we label ``$\expval{110}$'', is delimited at low temperatures by a range of interaction parameters $1.66 \lesssim D/J < 2$, and delimited at higher temperatures by regions ``$\expval{100}$'' and ``$\expval{111}$''. 
In the ``$\expval{110}$'' region, the $M_6$ cubic parameter appears to nearly vanish while the $\delta M_4$ and $M_{4}$ parameters plateau to the approximate intermediate values $7/10$ and $1/2$, respectively.
The values of the cubic parameters for this region are consistent with a $\langle 110 \rangle$ magnetization direction. 
Obtaining significantly more precise phase boundaries would require quite large-scale simulation studies involving careful finite-size scaling analysis, an investigation that is outside the scope of the present work.

\section{Perturbative calculation of the free energy} \label{appendix:PerturbationTheory}

\begin{figure}[ht!]
    \centering
   \begin{overpic}[width=\columnwidth]{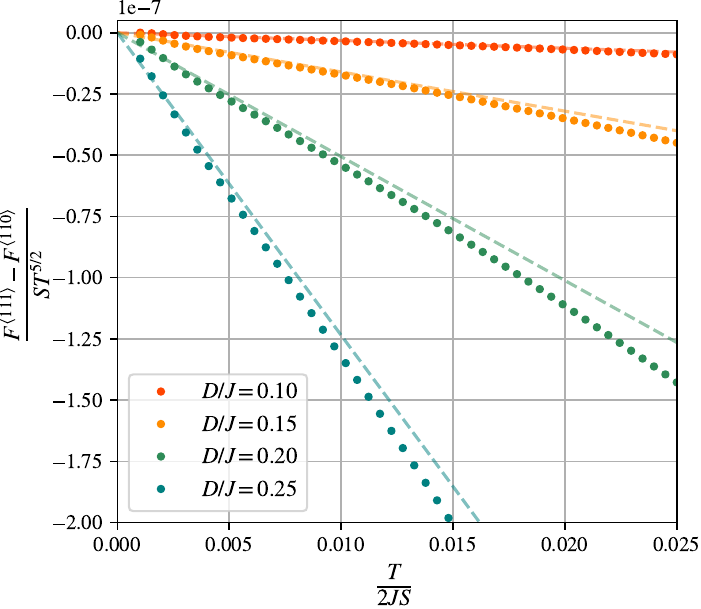}
    \end{overpic}
    \caption{Free energy difference in linear spin wave theory between the $\vu*{m} = \expval{111}$ and $\vu*{m} = \expval{110}$ configurations. 
    The negative slope  $\left( \mathcal{F}^{\langle 111 \rangle}-\mathcal{F}^{\langle 110 \rangle} \right)/(ST^{5/2})$ indicates that the $\vu*{m} = \expval{111}$ directions are selected. 
    The points are obtained by numerically calculating Eq.~\eqref{eq:LSWFreeEnergy}, while the dashed lines are obtained from the perturbative calculation in Eq.~\eqref{eq:FreeEnergy72} [or equivalently, Eq.~\eqref{eq:FreeEnergy-pert-app}].
    We see that both calculations agree in the limit $T\to 0^+$ when $|D| \ll J$.}
    \label{fig:PerturbationComparison}
\end{figure}

Using the matrix elements of the linear spin wave Hamiltonian in Eq.~\eqref{eq:LSWMatrixElements}, we estimate the dispersion of the lowest magnon branch by first diagonalizing the Hamiltonian for the case when $D =0$ and $\vb*{q}=\vb*{0}$, finding
\begin{equation}
    X^{}_{0} \equiv X^{(\hat{\vb*{m}})}(\vb*{q}=\vb*{0})\Big{|}_{D=0} = 0 \dyad{E_0} + 8J \left( 1-\dyad{E_0} \right),
\end{equation}
where $\ket{E_0} = \begin{pmatrix}\frac{1}{2}&\frac{1}{2}&\frac{1}{2}&\frac{1}{2}\end{pmatrix}^\intercal$. We then treat the remaining part of the spin wave Hamiltonian as a perturbation, namely $\delta X^{(\vu*{m})}(\vb*{q}) \equiv X^{(\hat{\vb*{m}})}(\vb*{q})-X^{}_{0}$. It is convenient to define dimensionless expectation values $G_n(\vb*{q},\vu*{m}) \equiv \expval{ \left( J^{-1} \delta X^{(\vu*{m})}(\vb*{q}) \right)^n}{E_0}$. The shift of the lowest energy branch to $4^{\text{th}}$ order in perturbation theory is then given by the expression
\begin{align}
    E^{(\vu*{m})}_{\vb*{q}}/J = \;&G_1 + \frac{1}{8}\left( G_1^2-G_2\right) + \frac{1}{64}\left(2 G_1^3-3 G_2 G_1+G_3\right) \nonumber \\
    &+ \frac{1}{512}\left(5 G_1^4-10 G_2 G_1^2+2G_2^2 +4 G_3 G_1-G_4 \right).
\end{align}
The resulting dispersion is then expanded to $4^\text{th}$ order in the components of the wavevector about $\vb*{q} = \vb*{0}$, where the coefficient of the term $q_x^{n_x}q_y^{n_y}q_z^{n_z}$ is calculated as
\begin{equation}
    \frac{1}{n_x!n_y!n_z!}\frac{\partial^{n_x+n_y+n_z}E^{(\vu*{m})}_{\vb*{q}}}{{\partial q_x^{n_x}} {\partial q_y^{n_y}} {\partial q_z^{n_z}} } \Bigg|_{\vb*{q}=\vb*{0}}.
\end{equation}
This expansion can be written compactly as
\begin{equation}
    E^{(\vu*{m})}_{\vb*{q}} \approx \varepsilon^{}_2(\vb*{q}) + \varepsilon^{(\vu*{m})}_4(\vb*{q}),
\end{equation}
where 
\begin{equation}
   \varepsilon^{}_2(\vb*{q}) \equiv \frac{Ja^2}{8}|\vb*{q}|^2 
\end{equation}
is the part of the dispersion which is quadratic in the wave vector components while $\varepsilon^{(\vu*{m})}_4(\vb*{q})$ contains all of the quartic terms. 
There are no linear or cubic terms due to the inversion symmetry of the pyrochlore lattice, and there is no constant term (corresponding to a spin wave gap) because of the accidental degeneracy of the ground state. 
To write the expression for $\varepsilon^{(\vu*{m})}_4$ more compactly, we define $\eta_{\mu\nu} \equiv 1-\delta_{\mu\nu}$, finding
\begin{align}
   \varepsilon^{(\vu*{m})}_4(\vb*{q}) \equiv &- \frac{a^4}{2048}\sum_{\mu} A^{(\vu*{m})}_\mu q_\mu^4 - \frac{a^4}{4096}\sum_{\mu\nu} \eta_{\mu\nu}B^{(\vu*{m})}_{\mu\nu} q_\mu^2q_\nu^2 \nonumber \\
   &  + \frac{3a^4D^2}{1024J}\left(\hat{m}_x \hat{m}_y q_z+ \hat{m}_z \hat{m}_xq_y + \hat{m}_y \hat{m}_z q_x \right)q_xq_yq_z \nonumber \\ 
   &+ \frac{a^4}{2048}\sum_{\mu\nu} \eta_{\mu\nu}C^{(\vu*{m})}_{\mu\nu} q_\mu^3q_\nu,
\label{eq:QuarticDispersion}
\end{align}
with coefficients
\begin{align}
    A^{(\vu*{m})}_\mu &\equiv \frac{D^4}{2J^3}\sum_{\nu} \eta_{\mu\nu} \hat{m}_\nu^2 \left[\frac{1}{2}\hat{m}_\mu^2 + \sum_{\lambda} \eta_{\mu\lambda}\eta_{\nu\lambda} \hat{m}_\lambda^2 \right] \nonumber \\
    &\qquad+ \frac{D^2}{J}\left(1- \hat{m}_\mu^2\right) -\frac{4J}{3}, \\
    B^{(\vu*{m})}_{\mu\nu} &\equiv \frac{D^4}{J^3}\left(\hat{m}_\mu^4+\hat{m}_\nu^4+\frac{5}{2}\hat{m}_\mu^2\hat{m}_\nu^2\right) \nonumber \\
    &\qquad + \frac{11D^2}{J}\left(\hat{m}_\mu^2+\hat{m}_\nu^2-\frac{2}{11}\right) + 8J, \\
    C^{(\vu*{m})}_{\mu\nu} &\equiv \left(\frac{D^4}{J^3}+\frac{6D^2}{J}\right)\hat{m}_\mu \hat{m}_\nu.
\end{align}
We now proceed to find the contribution of these quartic terms to the free energy, to lowest order in temperature. Assuming $\cramped{T \ll J}$, the dominant contribution to the free energy will come from the spin wave modes in the vicinity of $\vb*{q}=\vb*{0}$. We therefore approximate the free energy as a functional of the quartic part of the dispersion

\begin{equation}
    \mathcal{F}\left[g^{(\vu*{m})}_4(\vb*{q})\right] = TV \int_{\mathbb{R}^3} \frac{\dd[3]q}{(2\pi)^3} \; \ln \left(1-e^{-\frac{JSa^2}{8T}|\vb*{q}|^2}e^{-g^{(\vu*{m})}_4(\vb*{q})}\right),
\end{equation}
where $g^{(\vu*{m})}_4(\vb*{q})\equiv S\varepsilon^{(\vu*{m})}_4(\vb*{q})/T$, and we assume that the integration bounds extend over all ${\vb*{q}}$ space. At low temperature, the contribution of the quartic terms to the free energy becomes sub-dominant, and we can expand the functional to lowest order in $g^{(\vu*{m})}_4(\vb*{q})$ about $g^{(\vu*{m})}_4(\vb*{0}) = 0$ as \cite{Ernzerhof1994}
\begin{align}
    \mathcal{F}\left[g^{(\vu*{m})}_4(\vb*{q})\right] &\approx \mathcal{F}\left[g^{(\vu*{m})}_4(\vb*{0})\right] + \int \dd[3]q \; \frac{\delta \mathcal{F}}{\delta g_4(\vb*{q})}\Big|_{g^{(\vu*{m})}_4(\vb*{q})=g^{(\vu*{m})}_4(\vb*{0})} g^{(\vu*{m})}_4(\vb*{q}) \nonumber \\
    &= TV \int \frac{\dd[3]q}{{(2\pi)^3}} \; \ln \left(1-e^{-\frac{JSa^2}{8T}|\vb*{q}|^2}\right) \nonumber \\
    & \qquad + VS\int \frac{\dd[3]q}{{(2\pi)^3}} \; \frac{\varepsilon^{(\vu*{m})}_4(\vb*{q})}{e^{\frac{JSa^2}{8T}|\vb*{q}|^2}-1}. \label{eq:FreeEnergyExpansion}
\end{align}
The first term in Eq.~\eqref{eq:FreeEnergyExpansion} simply corresponds to the free energy of an ideal Bose gas with a quadratic dispersion. This evaluates to
\begin{equation}\label{eq:FreeEnergy52}
TV \int \frac{\dd[3]q}{{(2\pi)^3}} \; \ln \left(1-e^{-\frac{JSa^2}{8T}|\vb*{q}|^2}\right) = -  \frac{ 4JS \zeta\left(\frac{5}{2}\right)}{\pi^{3/2}} \left(\frac{T}{2JS}\right)^{5/2},
\end{equation}
giving the expected contribution proportional to $T^{5/2}$. 
Note that since the DM interaction does not enter the quadratic part of the dispersion at this order of perturbation theory, this contribution of the free energy does not couple to the magnetization direction $\vu*{m}$. 
The second integral in Eq.~\eqref{eq:FreeEnergyExpansion} evaluates to
\begin{align}
    \frac{3S\zeta\left(\frac{7}{2}\right)}{32\pi^{3/2}}\left[ \left(\hat{m}_x^4+\hat{m}_y^4+\hat{m}_z^4-\frac{7}{3} \right)\frac{D^4}{J^3} - \frac{44D^2}{3J} - 24J \right] \left(\frac{T}{2JS}\right)^{7/2},
    \label{eq:FreeEnergy-pert-app}
\end{align}
leading to the free energy contribution in Eq.~\eqref{eq:FreeEnergy72}, which is minimized for $\vu*{m} = \expval{111}$. A comparison of this perturbative calculation to the numerically integrated free energy in 
Eqs.~\eqref{eq:LSWFreeEnergy} and \eqref{eq:FreeEnergy72}, is shown in Fig.~\ref{fig:PerturbationComparison}, where we see that the two calculations agree in the limit $T \to 0^+$ when $|D| \ll J$. Note that the next term in the free energy expansion is given by
\begin{align}
    \frac{1}{2!}\int \dd[3]k \, \dd[3]q \; &\frac{\delta^2 \mathcal{F}}{\delta g_4(\vb*{k})\delta g_4(\vb*{q})}\Big|_{g_4(\vb*{k})=g_4(\vb*{q})=g_4(\vb*{0})} g_4(\vb*{k})g_4(\vb*{q}),
\end{align}
and leads to a contribution proportional to $T^{9/2}$.

\section{Energy versus entropy as the mechanism for selection} \label{appendix:EnergyEntropy}

Consider two distinct ferromagnetic configurations with ordering directions $\vu*{m}$ and $\vu*{n}$, the free energy difference between configurations can, in the absence of zero-point corrections, be decomposed as 
\begin{equation}
\mathcal{F}^{(\hat{\vb*{m}})} - \mathcal{F}^{(\hat{\vb*{n}})} = \mathcal{E}^{(\hat{\vb*{m}})} - \mathcal{E}^{(\hat{\vb*{n}})} - T \left( \mathcal{S}^{(\hat{\vb*{m}})} - \mathcal{S}^{(\hat{\vb*{n}})} \right),
\end{equation}
where $\mathcal{E}$ and $\mathcal{S}$ are the internal energy and entropy, respectively. In the classical case, the low energy excitations are spin waves, arising from the two quadratic modes of fluctuation about each ordered magnetic moment. 
From the equipartition theorem, the energy is simply
$\mathcal{E}^{(\hat{\vb*{m}})} = 4NT$, where $N$ is the number of FCC unit cells, independent of the ordering direction. 
The low-temperature classical selection is thus dictated solely by the thermodynamic entropy, as depicted in Fig.~\ref{fig:CLTE_entropic_weight}.

\begin{figure}
    \centering
   \begin{overpic}[width=0.9\columnwidth]{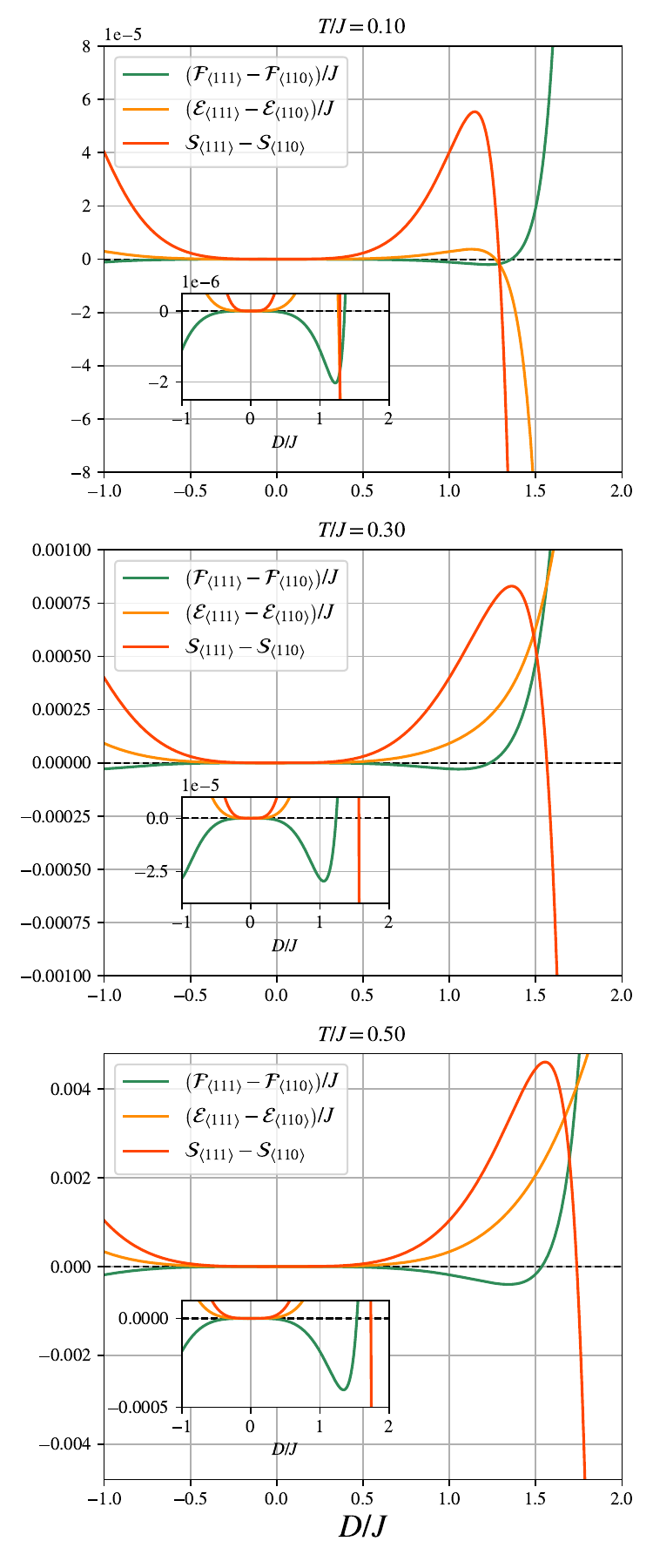}
   \put(7,70.5) {\textcolor{black}{(a)}}
   \put(7,38.5) {\textcolor{black}{(b)}}
   \put(7,6.5) {\textcolor{black}{(c)}}
    \end{overpic}
    \caption{Quantum (linear) spin wave comparison of the thermodynamic free energy, energy, and entropy between the $\expval{111}$ and $\expval{110}$ configurations for (a) $T/J = 0.10$, (b) $T/J = 0.30$, and (c) $T/J = 0.50$. We find that the entropy difference (red curve) changes sign at a different value of $D/J$ when compared with the free energy calculation (green curve), indicating that both energy and entropy play a role in the selection mechanism when $D>J$. The insets depict the same plots with a re-scaled vertical axis to clearly show the profile of the free energy curves.}
    \label{fig:EnergyEntropy}
\end{figure}

In the quantum case, the low-energy excitations are magnons and their contribution to the energy in the linear spin-wave approximation is given by
\begin{equation}\label{eqn:EnsembleEnergy}
\mathcal{E}^{(\hat{\vb*{m}})} =  \sum_{\mu,\vb*{q}} \frac{S\omega^{(\hat{\vb*{m}})}_{\mu,\vb*{q}}} {\exp \left(\frac{S\omega^{(\hat{\vb*{m}})}_{\mu,\vb*{q}}}{T}\right)-1}.
\end{equation}
By computing Eq.~\eqref{eqn:EnsembleEnergy} in conjunction with Eq.~\eqref{eq:LSWFreeEnergy}, we are able to separate the energy and entropy contributions to the free energy. These are depicted in Fig.~\ref{fig:EnergyEntropy}, comparing the thermodynamic quantities for $ \hat{\vb*{m}}$ along $\langle 111\rangle$ and $\hat{\vb*{n}}$ along $\langle 110\rangle$. 
We see from Fig.~\ref{fig:EnergyEntropy} that the free energy changes sign when $D/J > 1$, indicating a change in the ordering direction. Furthermore, we find the entropy difference changes sign at a \emph{different} value of $D/J$, when compared to the free energy calculation. 
Therefore, in the quantum case, the low-temperature selection is no longer determined solely from the entropy contribution when $D>J$, in contrast to the classical calculation. 
That being said, it is worthwhile to note that this $ D \gtrsim J$ region where both the energy and entropy compete to select the magnetic order in linear spin wave theory approximately coincides with the region in Fig.~\ref{fig:LSWPhases} where we find that the ferromagnetic order is unstable in non-linear spin wave theory. 
At this point, however, it is not clear to us if these two observations are related.

To gain some analytical insight into the relative contribution of the energy, ${\cal E}(T)$ and entropy, ${\cal S}(T)$, to the selection of the ${\hat {\bm m}}$ direction at low temperature, we may calculate the low-temperature contributions of both the energy and entropy in the framework of perturbation theory discussed in Appendix 
\ref{appendix:PerturbationTheory}
when $|D| \ll J$, by following the same steps used to derive Eq.~\eqref{eq:FreeEnergyExpansion}. 
The energy and entropy, to the lowest order that couples to the magnetization direction are
\begin{align}
    \mathcal{E}^{(\vu*{m})} &\approx  \frac{ 6 JS \zeta\left(\frac{5}{2}\right)}{\pi^{3/2}} \left(\frac{T}{2JS}\right)^{5/2} \nonumber \\
    &\quad + \frac{15S\zeta\left(\frac{7}{2}\right)}{64\pi^{3/2}}\left( \frac{44D^2}{3J} + 8J \right) \left(\frac{T}{2JS}\right)^{7/2} \nonumber \\
    &\quad + \frac{15S\zeta\left(\frac{7}{2}\right)}{64\pi^{3/2}}\left[ \frac{D^4}{J^3}\left(\frac{7}{3}-\hat{m}_x^4-\hat{m}_y^4-\hat{m}_z^4 \right) \right] \left(\frac{T}{2JS}\right)^{7/2}, \label{eq:Energy72}\\ 
    T\mathcal{S}^{(\vu*{m})} &\approx  \frac{ 10 JS \zeta\left(\frac{5}{2}\right)}{\pi^{3/2}} \left(\frac{T}{2JS}\right)^{5/2} \nonumber \\
    &\quad + \frac{21S\zeta\left(\frac{7}{2}\right)}{64\pi^{3/2}}\left( \frac{44D^2}{3J} + 8J \right) \left(\frac{T}{2JS}\right)^{7/2} \nonumber \\
    &\quad + \frac{21S\zeta\left(\frac{7}{2}\right)}{64\pi^{3/2}}\left[ \frac{D^4}{J^3}\left(\frac{7}{3}-\hat{m}_x^4-\hat{m}_y^4-\hat{m}_z^4 \right) \right] \left(\frac{T}{2JS}\right)^{7/2}. \label{eq:Entropy72}
\end{align}
In this limit, the energy term is \emph{minimized} for $\vu*{m} = \expval{100}$, while the entropy term is \emph{maximized} for $\vu*{m} = \expval{111}$ (note that all of the terms in Eqs.~(\ref{eq:Energy72}-\ref{eq:Entropy72}) are positive as the magnetization direction is normalized $|\vu*{m}|=1$). The net contribution of these terms to the free energy $\mathcal{F}^{(\vu*{m})} = \mathcal{E}^{(\vu*{m})}-T\mathcal{S}^{(\vu*{m})}$ leads to Eq.~\eqref{eq:FreeEnergy72}. 
We see that the coefficient of the entropy term in (third line of Eq.~\eqref{eq:Entropy72}) is always larger in magnitude than the coefficient of the energy term (third line of Eq.~\eqref{eq:Energy72}), implying that \emph{entropy maximization} is the mechanism for the selection
of the magnetization direction in the limit $T\to 0^+$ and $|D|\ll J$.

\section{Magnetization direction in a field} \label{appendix:MagnetizationInField}

\begin{figure*}[t!]
    \centering
    \begin{overpic}[width=\textwidth]{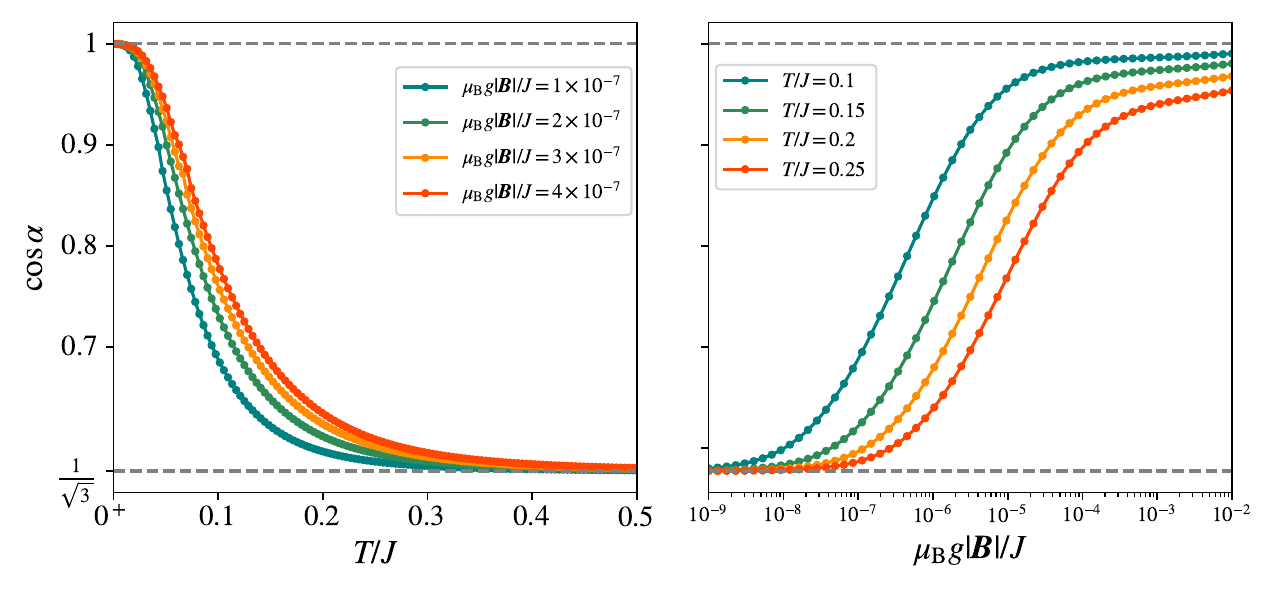}
    \put(28.5,46) {\textcolor{black}{(a)}}
    \put(76,46) {\textcolor{black}{(b)}}
    \end{overpic}
    \caption{Magnetization direction as a function of (a) temperature and (b) magnetic field magnitude obtained by minimizing Eq.~\eqref{eq:FreeEnergyInField} for $D/J = 0.5$ and $\vb*{B}$ oriented along the $\expval{100}$ directions. Here, $\alpha$ is the angle between the external field and the magnetization direction $\vb*{B}\cdot \vu*{m}  = |\vb*{B}|\cos \alpha$, such that $\cos \alpha = 1$ corresponds to $\vu*{m} = \expval{100}$ and $\cos \alpha = 1/\sqrt{3}$ corresponds to $\vu*{m} = \expval{111}$.}
    \label{fig:MagnetizationInField}
\end{figure*}

Here, we generalize the perturbative calculation of the free energy to include an external magnetic field, while still expanding about an arbitrary polarized configuration $\vu*{m}$. 
One must take caution carrying out this expansion as such a configuration is not necessarily metastable and there will be terms that are linear in the bosonic operators appearing in the Holstein-Primakoff expansion. 
Assuming an external magnetic field $\vb*{B}$, we include an additional Zeeman coupling to the Hamiltonian of the form
\begin{equation}
    -\mu^{}_{\text{B}}g \sum_{i}  \vb*{B}\cdot \vb*{S}_i,
\end{equation}
where $\mu^{}_{\text{B}}$ is the Bohr magneton and $g$ is the g-factor of the magnetic ion, which we assume to be isotropic, which would be a reasonable approximation for a 3d transition metal such as V$^{4+}$ in Lu$_2$V$_2$O$_7$~\cite{onose2010,Mena2014,Nazipov2016,Nazipov2016a,AliBiswas2013,Riedl2016a}, for example.
For convenience, we define the projection of the field onto the magnetization direction as $h_0^{(\hat{\vb*{m}})} \equiv \mu^{}_{\text{B}} g \vb*{B}\cdot \vu*{m}/S$ and the two orthogonal components $h_\pm^{(\hat{\vb*{m}})} \equiv \mu^{}_{\text{B}} g \vb*{B}\cdot \vu*{e}_{\pm}/S$, where $\vb*{e}^\pm \equiv \vu*{e}_1 \pm i \vu*{e}_2$. The linear spin wave Hamiltonian is then found to be
\begin{align}
H &= S^2 E_0 - \sqrt{N} S^{3/2} \sum_{\mu} \left[ h_+^{(\hat{\vb*{m}})} a^{\dagger}_{\mu,\vb*{0}} + h_-^{(\hat{\vb*{m}})} a^{}_{\mu,\vb*{0}}   \right] \nonumber \\
& \qquad + S\sum_{\vb*{q},\mu,\nu}  \left( X_{\mu\nu}^{(\hat{\vb*{m}})}(\vb*{q})+ h_0^{(\hat{\vb*{m}})} \delta_{\mu\nu} \right) a^{\dagger}_{\mu,\vb*{q}} a^{}_{\nu,\vb*{q}} \label{eq:LSWwithField0}\\
&= S^2E_{0} - 4 N S^2 h_+^{(\hat{\vb*{m}})} h_-^{(\hat{\vb*{m}})} \expval{\left( X_{}^{(\hat{\vb*{m}})}(\vb*{0})+ h_0^{(\hat{\vb*{m}})} \mathbb{I} \right)^{-1}}{E_0} \nonumber \\
& \qquad + S \sum_{\vb*{q},\mu}  \left(\omega^{(\hat{\vb*{m}})}_{\mu,\vb*{q}}+h_0^{(\hat{\vb*{m}})}\right) b_{\mu,\bm q}^\dagger b^{}_{\mu,\bm q}. \label{eq:LSWwithField}
\end{align}
where the second equality comes from completing the square (involving the $a^\dagger$ and $a$ operators), $X_{\mu\nu}^{(\hat{\vb*{m}})}(\vb*{q})$ is the linear spin wave Hamiltonian from Eq.~\eqref{eq:LSWMatrixElements} and $\ket{E_0} = \begin{pmatrix}
    \frac{1}{2} &\frac{1}{2} &\frac{1}{2} &\frac{1}{2} 
\end{pmatrix}^\intercal$.
We must now re-evaluate the free energy using the Hamiltonian in Eq.~\eqref{eq:LSWwithField}. 
Firstly, to evaluate the expectation value, we decompose the matrices into the sum of Heisenberg and DM couplings, i.e. $X_{}^{(\hat{\vb*{m}})}(\vb*{0})+ h_0^{(\hat{\vb*{m}})} \mathbb{I} = \mathcal{M}_{0}^{(\hat{\vb*{m}})} + \mathcal{M}_{\text{DM}}^{(\hat{\vb*{m}})}$, where 
\begin{equation}
\left[\mathcal{M}_{\text{DM}}^{(\hat{\vb*{m}})}\right]_{\mu\nu} \equiv   2iD (1-\delta_{\mu\nu})  \vb*{d}_{\mu \nu} \cdot \hat{\vb*{m}},
\end{equation}
and
\begin{equation}
\mathcal{M}_{0}^{(\hat{\vb*{m}})} \equiv  X_{}^{(\hat{\vb*{m}})}(\vb*{0})+ h_0^{(\hat{\vb*{m}})} \mathbb{I} - \mathcal{M}_{\text{DM}}^{(\hat{\vb*{m}})}.
\end{equation}
We can then expand the matrix inverse that enters 
in Eq.~\eqref{eq:LSWwithField} as
\begin{equation}
    \left(X_{}^{(\hat{\vb*{m}})}(\vb*{0})+ h_0^{(\hat{\vb*{m}})} \mathbb{I}\right)^{-1} = \left[\mathcal{M}_{0}^{(\hat{\vb*{m}})} \right]^{-1} \sum_{n=0}^\infty \left(-\mathcal{M}_{\text{DM}}^{(\hat{\vb*{m}})}\left[\mathcal{M}_{0}^{(\hat{\vb*{m}})} \right]^{-1}\right)^n.
\end{equation}
The spectral decomposition of the Heisenberg and field contributions is 
\begin{equation}
    \left[\mathcal{M}_{0}^{(\hat{\vb*{m}})} \right]^{-1} = \frac{1}{h_0^{(\hat{\vb*{m}})}} \dyad{E_0}+ \frac{1}{8J+h_0^{(\hat{\vb*{m}})}}\left(1-\dyad{E_0}\right).
\end{equation}
It is straightforward to check that $\mathcal{M}_{\text{DM}}^{(\hat{\vb*{m}})}\left[\mathcal{M}_{0}^{(\hat{\vb*{m}})} \right]^{-1}\ket{E_0}=0$, therefore
\begin{equation}
    \expval{\left( X_{}^{(\hat{\vb*{m}})}(\vb*{0})+ h_0^{(\hat{\vb*{m}})} \mathbb{I} \right)^{-1}}{E_0} = \frac{1}{h_0^{(\hat{\vb*{m}})}}.
\end{equation}
The low-temperature contribution to the free energy per unit cell is then calculated in the same way as described in Appendix~\ref{appendix:PerturbationTheory}, thus obtaining  
\begin{align}
    \mathcal{F}^{(\hat{\vb*{m}})} &\approx S^2\frac{E_0}{N} - 4S^2 \frac{h_+^{(\hat{\vb*{m}})}h_-^{(\hat{\vb*{m}})}}{h_0^{(\hat{\vb*{m}})}} \nonumber \\
    &\quad + \frac{TV}{(2\pi)^3} \int \dd[3]k \; \ln \left(1-e^{-\frac{JSa^2}{8T}|\vb*{k}|^2-\frac{S}{T}h_0^{(\hat{\vb*{m}})}}\right) \nonumber\\
    &\quad + \frac{VS}{(2\pi)^3} \int \dd[3]k \; \frac{\varepsilon_4(\vb*{k})}{e^{\frac{JSa^2}{8T}|\vb*{k}|^2+\frac{S}{T}h_0^{(\hat{\vb*{m}})}}-1}. \label{eq:FreeEnergyFieldIntegrals}
\end{align}
Evaluating the integrals in Eq.~\eqref{eq:FreeEnergyFieldIntegrals}, we obtain an expression for the free energy to order $\left(\frac{D}{J}\right)^{4} \left(\frac{T}{J}\right)^{7/2}$
\begin{widetext}
\begin{align}
    \mathcal{F}^{(\hat{\vb*{m}})} &\approx  - 2\mu^{}_{\rm B}g S\left(\frac{|\vb*{B}|^2+(\vb*{B}\cdot \vu*{m})^2}{\vb*{B}\cdot \vu*{m}}\right) -  \frac{ 4JS}{\pi^{3/2}} \left(\frac{T}{2JS}\right)^{5/2}\text{Li}^{}_{\frac{5}{2}}\left(e^{-\frac{\mu^{}_{\rm B}g}{T} \vb*{B}\cdot \vu*{m}}\right) \nonumber \\
    &\quad + \frac{3S}{32\pi^{3/2}}\left[ \frac{D^4}{J^3}\left(\hat{m}_x^4+\hat{m}_y^4+\hat{m}_z^4-\frac{7}{3} \right)- \frac{44D^2}{3J} - 24J \right] \left(\frac{T}{2JS}\right)^{7/2}\text{Li}^{}_{\frac{7}{2}}\left(e^{-\frac{\mu^{}_{\rm B}g}{T} \vb*{B}\cdot \vu*{m}}\right) .\label{eq:FreeEnergyInField}
\end{align}
\end{widetext}
where $\text{Li}_{\alpha}(z) \equiv \sum_{n=1}^\infty n^{-\alpha} z^n $ is the polylogarithm. 
We may then minimize Eq.~\eqref{eq:FreeEnergyInField} with respect to the $\vu*{m}$ to find the magnetization direction at finite temperature. 
The results of this minimization procedure are depicted in Fig.~\ref{fig:MagnetizationInField} for an external magnetic field oriented along the $[100]$ direction. 
At finite temperature, we observe a crossover between magnetization oriented along the $\expval{111}$ directions, favored by thermal fluctuations, to the $\expval{100}$ directions as the magnetic field is increased (see Fig.~\ref{fig:MagnetizationInField}(b)). 
This change in the magnetization direction is induced by a very weak magnetic field. 
For illustration purposes, taking for example $D/J = 0.5$ and $T/J = 0.15$, the magnetization changes direction for a magnitude $\left( \mu_\text{B}^{} g |\vb*{B}|/J \sim 10^{-6} \right)$. 
Using a ferromagnetic exchange $J \sim 8 \text{ meV}$ and g-factor $g \sim 2$ (as in $\text{Lu}_2\text{V}_2\text{O}_7$ \cite{Mena2014,Riedl2016a}), this would correspond to a magnetic field of magnitude $|\vb*{B}| \sim 70 \text{ }\mu\text{T}$ at a temperature $T \sim 14 \text{ K}$.
This reflects the weakness of thermal order by disorder of the quantum model against the energetic perturbation (here the magnetic Zeeman energy), at least for this $D/J=0.5$ value, which is already quite large considering a system such as Lu$_2$V$_2$O$_7$~\cite{onose2010,Mena2014,Nazipov2016,Nazipov2016a,AliBiswas2013,Riedl2016a}.

\section{Calculation of the magnon self-energy} \label{appendix:SelfEnergy}
In this appendix, we explicitly derive the expression for the magnon self-energy to $\mathcal{O}(S^0)$ using the imaginary-time Green's function formalism. 
The three and four-magnon interaction vertices in Eqs.~(\ref{eq:ThreeMagnonHamiltonian}-\ref{eq:FourMagnonHamiltonian}) are given by
\begin{align}
 Y_{\mu\nu\lambda}^{(\vu*{m})}(\vb*{k} ,\vb*{q}) &= i\sqrt{2} D \vb*{d}_{\mu \nu} \cdot \vb*{e}^- \left(\delta_{\nu \lambda} \cos(\vb*{k} \cdot \vb*{r}_{\mu \nu}) - \delta_{\mu \lambda}\cos(\vb*{q} \cdot \vb*{r}_{\mu \nu}) \right), \\
V_{\mu\nu\lambda \rho}^{(\hat{\vb*{m}})}(\vb*{k} ,\vb*{q},\vb*{Q}) &= -2J(1-\delta_{\mu \nu})\Big[  \delta_{\mu \rho}\delta_{\nu \lambda}\cos(\vb*{Q} \cdot \vb*{r}_{\mu \nu}) \nonumber \\
&\qquad \qquad + \delta_{\mu \lambda}\delta_{\nu \rho}\cos((\vb*{k}-\vb*{q}+\vb*{Q}) \cdot \vb*{r}_{\mu \nu}) \Big] \nonumber \\
&\quad + (J-i D \vb*{d}_{\mu \nu} \cdot \vu*{m} )(1-\delta_{\mu \rho}) \Big[ \delta_{\mu \nu}\delta_{\mu \lambda}\cos(\vb*{k} \cdot \vb*{r}_{\mu \rho}) \nonumber \\
&\qquad \qquad + \delta_{\lambda \rho}\delta_{\lambda \nu}\cos((\vb*{k}+\vb*{Q}) \cdot \vb*{r}_{\mu \rho}) \Big] \nonumber \\
&\quad + (J-i D \vb*{d}_{\nu \lambda} \cdot \vu*{m} )(1-\delta_{\nu \lambda}) \Big[ \delta_{\mu \nu}\delta_{\mu \rho}\cos(\vb*{q} \cdot \vb*{r}_{\nu \lambda}) \nonumber \nonumber \\
&\qquad \qquad + \delta_{\mu \lambda}\delta_{\mu \rho}\cos((\vb*{q}-\vb*{Q}) \cdot \vb*{r}_{\nu \lambda}) \Big] ,
\end{align}
where $\vb*{e}^- \equiv \vu*{e}_1 - i \vu*{e}_2$.

We calculate the self-energy to $\mathcal{O}(S^0)$ by evaluating the diagrams depicted in Fig.~\ref{fig:FeynmanBubble} in the imaginary time formalism at finite temperature $T = 1/\beta$, and subsequently extrapolate to zero temperature ($\beta \to \infty$). 
First, we express the interactions in terms of the operators $a_{\vb*{q},\mu} = \sum_\nu U^{(\vu*{m})}_{\mu,\nu}(\vb*{q})b_{\vb*{q},\nu}$ that diagonalize the linear spin wave theory. 
Equations ~(\ref{eq:ThreeMagnonHamiltonian}-\ref{eq:FourMagnonHamiltonian}) then become
\begin{equation}
\mathcal{H}^{(\hat{\vb*{m}})}_{3} = \frac{1}{2! \sqrt{N}} \sum_{\vb*{k} \vb*{q}} \sum_{\mu \nu \lambda} \left[ T_{\mu\nu\lambda}^{(\vu*{m})}(\vb*{k} ,\vb*{q}) b^\dagger_{\vb*{k},\mu} b^\dagger_{\vb*{q},\nu} b^{}_{\vb*{k}+\vb*{q},\lambda} + \text{H.c.}\right] 
\end{equation}
and
\begin{equation}
\mathcal{H}^{(\hat{\vb*{m}})}_{4} = \frac{1}{N(2!)^2} \sum_{\vb*{k} \vb*{q} \vb*{Q}} \sum_{\mu \nu \lambda \rho} W_{\mu\nu\lambda \rho}^{(\vu*{m})}(\vb*{k} ,\vb*{q},\vb*{Q}) b^\dagger_{\vb*{k}+\vb*{Q},\mu} b^\dagger_{\vb*{q}-\vb*{Q},\nu} b^{}_{\vb*{q},\lambda} b^{}_{\vb*{k},\rho} ,
\end{equation}
with the interaction vertices expressed in the energy basis (i.e. involving the $b$ and $b^\dagger$ operators) defined as
\begin{equation}
T_{\mu\nu\lambda}^{(\hat{\vb*{m}})}(\vb*{k} ,\vb*{q}) =  \sum_{\alpha \beta \gamma} \overline{U}^{(\hat{\vb*{m}})}_{\alpha \mu } (\bm k) \overline{U}^{(\hat{\vb*{m}})}_{\beta \nu} (\bm q)U^{(\hat{\vb*{m}})}_{\gamma \lambda} (\vb*{k}+\vb*{q}) Y_{\alpha \beta \gamma}^{(\vu*{m})}(\vb*{k} ,\vb*{q}),
\end{equation}
and
\begin{align}
W_{\mu\nu\lambda \rho}^{(\hat{\vb*{m}})}(\vb*{k} ,\vb*{q},\vb*{Q}) &= \sum_{\alpha \beta \gamma \sigma}\Bigg[ \overline{U}^{(\hat{\vb*{m}})}_{\alpha \mu } (\vb*{k}+\vb*{Q}) \overline{U}^{(\hat{\vb*{m}})}_{\beta \nu} (\vb*{q}-\vb*{Q}) \nonumber \\
&\qquad \times U^{(\hat{\vb*{m}})}_{\gamma \lambda} (\vb*{q}) U^{(\hat{\vb*{m}})}_{\sigma \rho} (\vb*{k}) V_{\alpha\beta\gamma \sigma}^{(\vu*{m})}(\vb*{k} ,\vb*{q},\vb*{Q}) \Bigg].
\end{align}
The bare imaginary-time Green's function in the energy basis is given by
\begin{equation}
\mathcal{G}^{(\vu*{m})}_{\lambda \rho}(\vb*{k},i\omega) = \frac{\delta_{\lambda \rho}}{-i\omega + S \omega_{\lambda,\vb*{k}}^{(\vu*{m})}}.
\end{equation}
We decompose the imaginary-time self-energy into the sum of each of the diagram contributions in Fig.~\ref{fig:FeynmanBubble} as $\Sigma^{(\vu*{m})} = \Sigma^a + \Sigma^b + \Sigma^c$.
In order of appearance in Fig.~\ref{fig:FeynmanBubble} , the diagrams are evaluated as
\begin{align}
\Sigma^a_{\mu \nu}(\vb*{k},i\omega) &= -\frac{S}{2N\beta}\sum_{\vb*{q},n}\sum_{\lambda \rho} \Bigg[ T_{\lambda\rho\mu}^{(\vu*{m})}(\vb*{q} ,\vb*{k}-\vb*{q}) \overline{T}_{\lambda\rho\nu}^{(\vu*{m})}(\vb*{q} ,\vb*{k}-\vb*{q}) \nonumber\\
&\qquad \times \mathcal{G}^{(\vu*{m})}_{\lambda \lambda}(\vb*{q},i\Omega_n) \mathcal{G}^{(\vu*{m})}_{\rho \rho}(\vb*{k}-\vb*{q},i\omega-i\Omega_n) \Bigg] \nonumber\\
&\quad - \frac{S}{N\beta}\sum_{\vb*{q},n}\sum_{\lambda \rho} \Bigg[ T_{\nu\rho\lambda}^{(\vu*{m})}(\vb*{k} ,\vb*{q}-\vb*{k}) \overline{T}_{\mu\rho\lambda}^{(\vu*{m})}(\vb*{k} ,\vb*{q}-\vb*{k}) \nonumber\\
&\qquad \times \mathcal{G}^{(\vu*{m})}_{\lambda \lambda}(\vb*{q},i\Omega_n) \mathcal{G}^{(\vu*{m})}_{\rho \rho}(\vb*{q}-\vb*{k},i\Omega_n-i\omega) \Bigg], 
\label{eq:sigma-F8} 
\\
\Sigma^b_{\mu \nu}(\vb*{k},i\omega) &= -\frac{S}{N\beta}\sum_{\vb*{q},n}\sum_{\lambda \rho}  \mathcal{G}^{(\vu*{m})}_{\lambda \lambda}(\vb*{0},i0) \mathcal{G}^{(\vu*{m})}_{\rho \rho}(\vb*{q},i\Omega_n)  \nonumber \\
&\quad \times \left[ T_{\nu\lambda\mu}^{(\vu*{m})}(\vb*{k} ,\vb*{0}) \overline{T}_{\rho\lambda\rho}^{(\vu*{m})}(\vb*{q} ,\vb*{0})+\overline{T}_{\mu\lambda\nu}^{(\vu*{m})}(\vb*{k} ,\vb*{0}) T_{\rho\lambda\rho}^{(\vu*{m})}(\vb*{q} ,\vb*{0})\right] , 
\label{eq:sigma-F9}
\\
\Sigma^c_{\mu \nu}(\vb*{k},i\omega) &= \frac{S}{N\beta} \sum_{\vb*{q},n} \sum_{\lambda} W_{\nu \lambda \lambda \mu}^{(\vu*{m})}(\vb*{k} ,\vb*{q},\vb*{0}) \mathcal{G}^{(\vu*{m})}_{\lambda \lambda}(\vb*{q},i\Omega_{n}) 
\label{eq:sigma-F10},
\end{align}
where $\Omega_n = 2\pi n/\beta$ are the bosonic Matsubara frequencies. We can then evaluate each of the frequency summations and subsequently take the $\beta \to \infty$ limit as
\begin{equation}
\frac{1}{\beta}\sum_{n}\mathcal{G}^{(\vu*{m})}_{\lambda \lambda}(\vb*{q},i\Omega_{n}) = n_{\rm B}\left(S \omega_{\lambda,\vb*{q}}^{(\vu*{m})}\right) \xrightarrow{\beta \to \infty} 0, \\
\end{equation}
which enters in Eqs.~\eqref{eq:sigma-F9}  and \eqref{eq:sigma-F10}. 
We also have
\begin{align}
&\frac{1}{\beta}\sum_{n}\mathcal{G}^{(\vu*{m})}_{\lambda \lambda}(\vb*{q},i\Omega_n) \mathcal{G}^{(\vu*{m})}_{\rho \rho}(\vb*{q}-\vb*{k},i\Omega_n-i\omega) \nonumber \\
& \qquad= \frac{n_{\rm B}\left(S \omega_{\lambda,\vb*{q}}^{(\vu*{m})}\right)-n_{\rm B}\left(S \omega_{\rho,\vb*{q}-\vb*{k}}^{(\vu*{m})}\right)}{i\omega-S \omega_{\rho,\vb*{q}}^{(\vu*{m})} + S \omega_{\lambda,\vb*{q}-\vb*{k}}^{(\vu*{m})}}\xrightarrow{\beta \to \infty} 0, 
\end{align}
which enters in Eq.~\eqref{eq:sigma-F8}, along with
\begin{align}
&\frac{1}{\beta}\sum_{n}\mathcal{G}^{(\vu*{m})}_{\lambda \lambda}(\vb*{q},i\Omega_n) \mathcal{G}^{(\vu*{m})}_{\rho \rho}(\vb*{k}-\vb*{q},i\omega-i\Omega_n) \nonumber \\
& \qquad= \frac{n_{\rm B}\left(S \omega_{\lambda,\vb*{q}}^{(\vu*{m})}\right)-n_{\rm B}\left(- S \omega_{\rho,\vb*{k}-\vb*{q}}^{(\vu*{m})}\right)}{-i\omega +S \omega_{\rho,\vb*{q}}^{(\vu*{m})}+ S \omega_{\lambda,\vb*{k}-\vb*{q}}^{(\vu*{m})}} \nonumber \\
&\qquad \xrightarrow{\beta \to \infty} \frac{1}{-i\omega +S \omega_{\rho,\vb*{q}}^{(\vu*{m})}+ S \omega_{\lambda,\vb*{k}-\vb*{q}}^{(\vu*{m})}},
\end{align}
where $n_{\rm B}\left(z \right) \equiv \left(e^{\beta z}-1 \right)^{-1}$ is the Bose distribution function.
It thus finally follows that $\Sigma^b = \Sigma^c = 0$, and 
\begin{equation}
\Sigma_{\mu\nu}^{a}(\vb*{k},i\omega) = \frac{S}{2N} \sum_{\vb*{q}} \sum_{\lambda \rho}  \frac{ T_{\lambda \rho \mu}^{(\vu*{m})}(\vb*{q} ,\vb*{k}-\vb*{q})  \overline{T}_{\lambda \rho \nu}^{(\vu*{m})}(\vb*{q} ,\vb*{k}-\vb*{q})}{ i\omega - S \omega^{(\hat{\vb*{m}})}_{\lambda,\vb*{k}-\vb*{q}} - S\omega^{(\hat{\vb*{m}})}_{\rho,\vb*{q}}  }.
\end{equation}
The retarded self-energy is then obtained by analytic continuation which, making the substitution $i\omega \to \omega + i0^+$ leads to Eq.~\eqref{eqn:SelfEnergy}. The instability discussed in Sec.~\ref{sec:NLSWInstability} then corresponds to a pole at $\omega < 0$ of the single-magnon Green's function (Eq.~\eqref{eq:GreensFunction}, with $S=\frac{1}{2}$) to $\mathcal{O}(S^0)$.

\bibliography{refs}

\begin{thebibliography}{127}%
\makeatletter
\providecommand \@ifxundefined [1]{%
 \@ifx{#1\undefined}
}%
\providecommand \@ifnum [1]{%
 \ifnum #1\expandafter \@firstoftwo
 \else \expandafter \@secondoftwo
 \fi
}%
\providecommand \@ifx [1]{%
 \ifx #1\expandafter \@firstoftwo
 \else \expandafter \@secondoftwo
 \fi
}%
\providecommand \natexlab [1]{#1}%
\providecommand \enquote  [1]{``#1''}%
\providecommand \bibnamefont  [1]{#1}%
\providecommand \bibfnamefont [1]{#1}%
\providecommand \citenamefont [1]{#1}%
\providecommand \href@noop [0]{\@secondoftwo}%
\providecommand \href [0]{\begingroup \@sanitize@url \@href}%
\providecommand \@href[1]{\@@startlink{#1}\@@href}%
\providecommand \@@href[1]{\endgroup#1\@@endlink}%
\providecommand \@sanitize@url [0]{\catcode `\\12\catcode `\$12\catcode `\&12\catcode `\#12\catcode `\^12\catcode `\_12\catcode `\%12\relax}%
\providecommand \@@startlink[1]{}%
\providecommand \@@endlink[0]{}%
\providecommand \url  [0]{\begingroup\@sanitize@url \@url }%
\providecommand \@url [1]{\endgroup\@href {#1}{\urlprefix }}%
\providecommand \urlprefix  [0]{URL }%
\providecommand \Eprint [0]{\href }%
\providecommand \doibase [0]{https://doi.org/}%
\providecommand \selectlanguage [0]{\@gobble}%
\providecommand \bibinfo  [0]{\@secondoftwo}%
\providecommand \bibfield  [0]{\@secondoftwo}%
\providecommand \translation [1]{[#1]}%
\providecommand \BibitemOpen [0]{}%
\providecommand \bibitemStop [0]{}%
\providecommand \bibitemNoStop [0]{.\EOS\space}%
\providecommand \EOS [0]{\spacefactor3000\relax}%
\providecommand \BibitemShut  [1]{\csname bibitem#1\endcsname}%
\let\auto@bib@innerbib\@empty
%</preamble>
\bibitem [{\citenamefont {Villain}(1979)}]{villain1979}%
  \BibitemOpen
  \bibfield  {author} {\bibinfo {author} {\bibfnamefont {J.}~\bibnamefont {Villain}},\ }\bibfield  {title} {\bibinfo {title} {Insulating spin glasses},\ }\href {https://doi.org/10.1007/BF01325811} {\bibfield  {journal} {\bibinfo  {journal} {Z. Phys. B}\ }\textbf {\bibinfo {volume} {33}},\ \bibinfo {pages} {31} (\bibinfo {year} {1979})}\BibitemShut {NoStop}%
\bibitem [{\citenamefont {Moessner}\ and\ \citenamefont {Chalker}(1998)}]{moessner1998}%
  \BibitemOpen
  \bibfield  {author} {\bibinfo {author} {\bibfnamefont {R.}~\bibnamefont {Moessner}}\ and\ \bibinfo {author} {\bibfnamefont {J.~T.}\ \bibnamefont {Chalker}},\ }\bibfield  {title} {\bibinfo {title} {Low-temperature properties of classical geometrically frustrated antiferromagnets},\ }\href {https://doi.org/10.1103/PhysRevB.58.12049} {\bibfield  {journal} {\bibinfo  {journal} {Phys. Rev. B}\ }\textbf {\bibinfo {volume} {58}},\ \bibinfo {pages} {12049} (\bibinfo {year} {1998})}\BibitemShut {NoStop}%
\bibitem [{\citenamefont {Canals}\ and\ \citenamefont {Lacroix}(1998)}]{Canals1998}%
  \BibitemOpen
  \bibfield  {author} {\bibinfo {author} {\bibfnamefont {B.}~\bibnamefont {Canals}}\ and\ \bibinfo {author} {\bibfnamefont {C.}~\bibnamefont {Lacroix}},\ }\bibfield  {title} {\bibinfo {title} {Pyrochlore antiferromagnet: A three-dimensional quantum spin liquid},\ }\href {https://doi.org/10.1103/PhysRevLett.80.2933} {\bibfield  {journal} {\bibinfo  {journal} {Phys. Rev. Lett.}\ }\textbf {\bibinfo {volume} {80}},\ \bibinfo {pages} {2933} (\bibinfo {year} {1998})}\BibitemShut {NoStop}%
\bibitem [{\citenamefont {Kitaev}(2006)}]{kitaev2006a}%
  \BibitemOpen
  \bibfield  {author} {\bibinfo {author} {\bibfnamefont {A.}~\bibnamefont {Kitaev}},\ }\bibfield  {title} {\bibinfo {title} {Anyons in an exactly solved model and beyond},\ }\href {https://doi.org/10.1016/j.aop.2005.10.005} {\bibfield  {journal} {\bibinfo  {journal} {Ann. Phys.}\ }\textbf {\bibinfo {volume} {321}},\ \bibinfo {pages} {2} (\bibinfo {year} {2006})}\BibitemShut {NoStop}%
\bibitem [{\citenamefont {Balents}(2010)}]{Balents2010}%
  \BibitemOpen
  \bibfield  {author} {\bibinfo {author} {\bibfnamefont {L.}~\bibnamefont {Balents}},\ }\bibfield  {title} {\bibinfo {title} {Spin liquids in frustrated magnets},\ }\href {https://doi.org/10.1038/nature08917} {\bibfield  {journal} {\bibinfo  {journal} {Nature}\ }\textbf {\bibinfo {volume} {464}},\ \bibinfo {pages} {199} (\bibinfo {year} {2010})}\BibitemShut {NoStop}%
\bibitem [{\citenamefont {Gingras}\ and\ \citenamefont {McClarty}(2014)}]{gingras2014}%
  \BibitemOpen
  \bibfield  {author} {\bibinfo {author} {\bibfnamefont {M.~J.~P.}\ \bibnamefont {Gingras}}\ and\ \bibinfo {author} {\bibfnamefont {P.~A.}\ \bibnamefont {McClarty}},\ }\bibfield  {title} {\bibinfo {title} {Quantum spin ice: A search for gapless quantum spin liquids in pyrochlore magnets},\ }\href {https://doi.org/10.1088/0034-4885/77/5/056501} {\bibfield  {journal} {\bibinfo  {journal} {Rep. Prog. Phys.}\ }\textbf {\bibinfo {volume} {77}},\ \bibinfo {pages} {056501} (\bibinfo {year} {2014})}\BibitemShut {NoStop}%
\bibitem [{\citenamefont {Imai}\ and\ \citenamefont {Lee}(2016)}]{Imai2016}%
  \BibitemOpen
  \bibfield  {author} {\bibinfo {author} {\bibfnamefont {T.}~\bibnamefont {Imai}}\ and\ \bibinfo {author} {\bibfnamefont {Y.~S.}\ \bibnamefont {Lee}},\ }\bibfield  {title} {\bibinfo {title} {{Do quantum spin liquids exist?}},\ }\href {https://doi.org/10.1063/PT.3.3266} {\bibfield  {journal} {\bibinfo  {journal} {Phys. Today}\ }\textbf {\bibinfo {volume} {69}},\ \bibinfo {pages} {30} (\bibinfo {year} {2016})}\BibitemShut {NoStop}%
\bibitem [{\citenamefont {Savary}\ and\ \citenamefont {Balents}(2017)}]{Savary2017}%
  \BibitemOpen
  \bibfield  {author} {\bibinfo {author} {\bibfnamefont {L.}~\bibnamefont {Savary}}\ and\ \bibinfo {author} {\bibfnamefont {L.}~\bibnamefont {Balents}},\ }\bibfield  {title} {\bibinfo {title} {Quantum spin liquids: A review},\ }\href {https://doi.org/10.1088/0034-4885/80/1/016502} {\bibfield  {journal} {\bibinfo  {journal} {Reports on Progress in Physics}\ }\textbf {\bibinfo {volume} {80}},\ \bibinfo {pages} {016502} (\bibinfo {year} {2017})}\BibitemShut {NoStop}%
\bibitem [{\citenamefont {Knolle}\ and\ \citenamefont {Moessner}(2019)}]{Knolle2019}%
  \BibitemOpen
  \bibfield  {author} {\bibinfo {author} {\bibfnamefont {J.}~\bibnamefont {Knolle}}\ and\ \bibinfo {author} {\bibfnamefont {R.}~\bibnamefont {Moessner}},\ }\bibfield  {title} {\bibinfo {title} {A field guide to spin liquids},\ }\href {https://doi.org/10.1146/annurev-conmatphys-031218-013401} {\bibfield  {journal} {\bibinfo  {journal} {Annu. Rev. Condens. Matter Phys.}\ }\textbf {\bibinfo {volume} {10}},\ \bibinfo {pages} {451} (\bibinfo {year} {2019})}\BibitemShut {NoStop}%
\bibitem [{Note9()}]{Note9}%
  \BibitemOpen
  \bibinfo {note} {Since frozen random disorder can also lift the accidental classical degeneracy of the parent disorder-free system~\cite {villain1979,henley1989}, the term order-by-disorder may be confusing in this context and should perhaps be referred to as ``order-by-fluctuations''.}\BibitemShut {Stop}%
\bibitem [{\citenamefont {Henley}(1989)}]{henley1989}%
  \BibitemOpen
  \bibfield  {author} {\bibinfo {author} {\bibfnamefont {C.~L.}\ \bibnamefont {Henley}},\ }\bibfield  {title} {\bibinfo {title} {Ordering due to disorder in a frustrated vector antiferromagnet},\ }\href {https://doi.org/10.1103/PhysRevLett.62.2056} {\bibfield  {journal} {\bibinfo  {journal} {Phys. Rev. Lett.}\ }\textbf {\bibinfo {volume} {62}},\ \bibinfo {pages} {2056} (\bibinfo {year} {1989})}\BibitemShut {NoStop}%
\bibitem [{\citenamefont {Villain}\ \emph {et~al.}(1980)\citenamefont {Villain}, \citenamefont {Bidaux}, \citenamefont {Carton},\ and\ \citenamefont {Conte}}]{villain1980}%
  \BibitemOpen
  \bibfield  {author} {\bibinfo {author} {\bibfnamefont {J.}~\bibnamefont {Villain}}, \bibinfo {author} {\bibfnamefont {R.}~\bibnamefont {Bidaux}}, \bibinfo {author} {\bibfnamefont {J.-P.}\ \bibnamefont {Carton}},\ and\ \bibinfo {author} {\bibfnamefont {R.}~\bibnamefont {Conte}},\ }\bibfield  {title} {\bibinfo {title} {Order as an effect of disorder},\ }\href {https://doi.org/10.1051/jphys:0198000410110126300} {\bibfield  {journal} {\bibinfo  {journal} {Journal de Physique}\ }\textbf {\bibinfo {volume} {41}},\ \bibinfo {pages} {1263} (\bibinfo {year} {1980})}\BibitemShut {NoStop}%
\bibitem [{\citenamefont {Belorizky}\ \emph {et~al.}(1980)\citenamefont {Belorizky}, \citenamefont {Casalegno},\ and\ \citenamefont {Niez}}]{belorizky1980}%
  \BibitemOpen
  \bibfield  {author} {\bibinfo {author} {\bibfnamefont {E.}~\bibnamefont {Belorizky}}, \bibinfo {author} {\bibfnamefont {R.}~\bibnamefont {Casalegno}},\ and\ \bibinfo {author} {\bibfnamefont {J.~J.}\ \bibnamefont {Niez}},\ }\bibfield  {title} {\bibinfo {title} {Calculation of the {Spin} {Wave} {Energy} {Gap} at $\vb*{k} = 0$ for a {Simple} {Cubic} {Ferromagnet} with {Anisotropic} {Exchange} {Interactions}},\ }\href {https://doi.org/10.1002/pssb.2221020135} {\bibfield  {journal} {\bibinfo  {journal} {Phys. Status Solidi B}\ }\textbf {\bibinfo {volume} {102}},\ \bibinfo {pages} {365} (\bibinfo {year} {1980})}\BibitemShut {NoStop}%
\bibitem [{\citenamefont {Shender}(1982)}]{Shender1982}%
  \BibitemOpen
  \bibfield  {author} {\bibinfo {author} {\bibfnamefont {E.~F.}\ \bibnamefont {Shender}},\ }\bibfield  {title} {\bibinfo {title} {Antiferromagnetic garnets with fluctuationally interacting sublattices},\ }\href {http://www.jetp.ras.ru/cgi-bin/dn/e_056_01_0178.pdf} {\bibfield  {journal} {\bibinfo  {journal} {Sov. Phys. JETP}\ }\textbf {\bibinfo {volume} {56}},\ \bibinfo {pages} {178} (\bibinfo {year} {1982})}\BibitemShut {NoStop}%
\bibitem [{\citenamefont {Prakash}\ and\ \citenamefont {Henley}(1990)}]{prakash1990}%
  \BibitemOpen
  \bibfield  {author} {\bibinfo {author} {\bibfnamefont {S.}~\bibnamefont {Prakash}}\ and\ \bibinfo {author} {\bibfnamefont {C.~L.}\ \bibnamefont {Henley}},\ }\bibfield  {title} {\bibinfo {title} {Ordering due to disorder in dipolar magnets on two-dimensional lattices},\ }\href {https://doi.org/10.1103/PhysRevB.42.6574} {\bibfield  {journal} {\bibinfo  {journal} {Phys. Rev. B}\ }\textbf {\bibinfo {volume} {42}},\ \bibinfo {pages} {6574} (\bibinfo {year} {1990})}\BibitemShut {NoStop}%
\bibitem [{\citenamefont {Henley}(1987)}]{Henley1987}%
  \BibitemOpen
  \bibfield  {author} {\bibinfo {author} {\bibfnamefont {C.~L.}\ \bibnamefont {Henley}},\ }\bibfield  {title} {\bibinfo {title} {{{Ordering} by disorder: {Ground‐state} selection in fcc vector antiferromagnets}},\ }\href {https://doi.org/10.1063/1.338570} {\bibfield  {journal} {\bibinfo  {journal} {J. Appl. Phys.}\ }\textbf {\bibinfo {volume} {61}},\ \bibinfo {pages} {3962} (\bibinfo {year} {1987})}\BibitemShut {NoStop}%
\bibitem [{\citenamefont {Tessman}(1954)}]{tessman1954}%
  \BibitemOpen
  \bibfield  {author} {\bibinfo {author} {\bibfnamefont {J.~R.}\ \bibnamefont {Tessman}},\ }\bibfield  {title} {\bibinfo {title} {Magnetic {Anisotropy} at $0^\circ${K}},\ }\href {https://doi.org/10.1103/PhysRev.96.1192} {\bibfield  {journal} {\bibinfo  {journal} {Phys. Rev.}\ }\textbf {\bibinfo {volume} {96}},\ \bibinfo {pages} {1192} (\bibinfo {year} {1954})}\BibitemShut {NoStop}%
\bibitem [{\citenamefont {Kittel}(1991)}]{kittel1991}%
  \BibitemOpen
  \bibfield  {author} {\bibinfo {author} {\bibfnamefont {C.}~\bibnamefont {Kittel}},\ }\href@noop {} {\emph {\bibinfo {title} {Quantum {Theory} of {Solids}}}},\ \bibinfo {edition} {2nd}\ ed.\ (\bibinfo  {publisher} {Wiley},\ \bibinfo {address} {New York, NY},\ \bibinfo {year} {1991})\BibitemShut {NoStop}%
\bibitem [{Note19()}]{Note19}%
  \BibitemOpen
  \bibinfo {note} {From that perspective, and as the present work illustrates through the specific model considered, quantum order-by-disorder ($T=0$) is not \protect \textit {per se} a ``real'' physical phenomenon. Rather, it is a statement reflecting a mathematical procedure that perturbatively corrects the approximation of a classical (direct product) ground state for a quantum spin model. Conversely, thermal order-by-disorder ($T>0$) being driven by ``real'' thermal fluctuations, can be viewed as a genuine physical phenomenon.}\BibitemShut {Stop}%
\bibitem [{\citenamefont {Reimers}\ and\ \citenamefont {Berlinsky}(1993)}]{reimers1993}%
  \BibitemOpen
  \bibfield  {author} {\bibinfo {author} {\bibfnamefont {J.~N.}\ \bibnamefont {Reimers}}\ and\ \bibinfo {author} {\bibfnamefont {A.~J.}\ \bibnamefont {Berlinsky}},\ }\bibfield  {title} {\bibinfo {title} {Order by disorder in the classical {Heisenberg} {kagomé} {Antiferromagnet}},\ }\href {https://doi.org/10.1103/PhysRevB.48.9539} {\bibfield  {journal} {\bibinfo  {journal} {Phys. Rev. B}\ }\textbf {\bibinfo {volume} {48}},\ \bibinfo {pages} {9539} (\bibinfo {year} {1993})}\BibitemShut {NoStop}%
\bibitem [{\citenamefont {Bramwell}\ \emph {et~al.}(1994)\citenamefont {Bramwell}, \citenamefont {Gingras},\ and\ \citenamefont {Reimers}}]{bramwell1994}%
  \BibitemOpen
  \bibfield  {author} {\bibinfo {author} {\bibfnamefont {S.~T.}\ \bibnamefont {Bramwell}}, \bibinfo {author} {\bibfnamefont {M.~J.~P.}\ \bibnamefont {Gingras}},\ and\ \bibinfo {author} {\bibfnamefont {J.~N.}\ \bibnamefont {Reimers}},\ }\bibfield  {title} {\bibinfo {title} {Order by disorder in an anisotropic pyrochlore lattice antiferromagnet},\ }\href {https://doi.org/10.1063/1.355676} {\bibfield  {journal} {\bibinfo  {journal} {J. Appl. Phys.}\ }\textbf {\bibinfo {volume} {75}},\ \bibinfo {pages} {5523} (\bibinfo {year} {1994})}\BibitemShut {NoStop}%
\bibitem [{\citenamefont {Elhajal}\ \emph {et~al.}(2005)\citenamefont {Elhajal}, \citenamefont {Canals}, \citenamefont {Sunyer},\ and\ \citenamefont {Lacroix}}]{Elhajal2005}%
  \BibitemOpen
  \bibfield  {author} {\bibinfo {author} {\bibfnamefont {M.}~\bibnamefont {Elhajal}}, \bibinfo {author} {\bibfnamefont {B.}~\bibnamefont {Canals}}, \bibinfo {author} {\bibfnamefont {R.}~\bibnamefont {Sunyer}},\ and\ \bibinfo {author} {\bibfnamefont {C.}~\bibnamefont {Lacroix}},\ }\bibfield  {title} {\bibinfo {title} {Ordering in the pyrochlore antiferromagnet due to {Dzyaloshinsky}-{Moriya} interactions},\ }\href {https://doi.org/10.1103/PhysRevB.71.094420} {\bibfield  {journal} {\bibinfo  {journal} {Phys. Rev. B}\ }\textbf {\bibinfo {volume} {71}},\ \bibinfo {pages} {094420} (\bibinfo {year} {2005})}\BibitemShut {NoStop}%
\bibitem [{\citenamefont {Bergman}\ \emph {et~al.}(2007)\citenamefont {Bergman}, \citenamefont {Alicea}, \citenamefont {Gull}, \citenamefont {Trebst},\ and\ \citenamefont {Balents}}]{bergman2007}%
  \BibitemOpen
  \bibfield  {author} {\bibinfo {author} {\bibfnamefont {D.}~\bibnamefont {Bergman}}, \bibinfo {author} {\bibfnamefont {J.}~\bibnamefont {Alicea}}, \bibinfo {author} {\bibfnamefont {E.}~\bibnamefont {Gull}}, \bibinfo {author} {\bibfnamefont {S.}~\bibnamefont {Trebst}},\ and\ \bibinfo {author} {\bibfnamefont {L.}~\bibnamefont {Balents}},\ }\bibfield  {title} {\bibinfo {title} {Order-by-disorder and spiral spin-liquid in frustrated diamond-lattice antiferromagnets},\ }\href {https://doi.org/10.1038/nphys622} {\bibfield  {journal} {\bibinfo  {journal} {Nat. Phys.}\ }\textbf {\bibinfo {volume} {3}},\ \bibinfo {pages} {487} (\bibinfo {year} {2007})}\BibitemShut {NoStop}%
\bibitem [{\citenamefont {Chern}(2010)}]{chern2010}%
  \BibitemOpen
  \bibfield  {author} {\bibinfo {author} {\bibfnamefont {G.-W.}\ \bibnamefont {Chern}},\ }\bibfield  {title} {\bibinfo {title} {Pyrochlore antiferromagnet with antisymmetric exchange interactions: critical behavior and order from disorder},\ }\href {http://arxiv.org/abs/1008.3038} {\bibfield  {journal} {\bibinfo  {journal} {arXiv:1008.3038 [cond-mat.str-el]}\ } (\bibinfo {year} {2010})}\BibitemShut {NoStop}%
\bibitem [{\citenamefont {Gvozdikova}\ \emph {et~al.}(2011)\citenamefont {Gvozdikova}, \citenamefont {Melchy},\ and\ \citenamefont {Zhitomirsky}}]{Gvozdikova2011}%
  \BibitemOpen
  \bibfield  {author} {\bibinfo {author} {\bibfnamefont {M.~V.}\ \bibnamefont {Gvozdikova}}, \bibinfo {author} {\bibfnamefont {P.-E.}\ \bibnamefont {Melchy}},\ and\ \bibinfo {author} {\bibfnamefont {M.~E.}\ \bibnamefont {Zhitomirsky}},\ }\bibfield  {title} {\bibinfo {title} {Magnetic phase diagrams of classical triangular and kagome antiferromagnets},\ }\href {https://doi.org/10.1088/0953-8984/23/16/164209} {\bibfield  {journal} {\bibinfo  {journal} {J. Phys. Condens. Matter}\ }\textbf {\bibinfo {volume} {23}},\ \bibinfo {pages} {164209} (\bibinfo {year} {2011})}\BibitemShut {NoStop}%
\bibitem [{\citenamefont {Oitmaa}\ \emph {et~al.}(2013)\citenamefont {Oitmaa}, \citenamefont {Singh}, \citenamefont {Javanparast}, \citenamefont {Day}, \citenamefont {Bagheri},\ and\ \citenamefont {Gingras}}]{oitmaa2013}%
  \BibitemOpen
  \bibfield  {author} {\bibinfo {author} {\bibfnamefont {J.}~\bibnamefont {Oitmaa}}, \bibinfo {author} {\bibfnamefont {R.~R.~P.}\ \bibnamefont {Singh}}, \bibinfo {author} {\bibfnamefont {B.}~\bibnamefont {Javanparast}}, \bibinfo {author} {\bibfnamefont {A.~G.~R.}\ \bibnamefont {Day}}, \bibinfo {author} {\bibfnamefont {B.~V.}\ \bibnamefont {Bagheri}},\ and\ \bibinfo {author} {\bibfnamefont {M.~J.~P.}\ \bibnamefont {Gingras}},\ }\bibfield  {title} {\bibinfo {title} {Phase transition and thermal order-by-disorder in the pyrochlore antiferromagnet {Er}$_{\textrm{2}}${Ti}$_{\textrm{2}}${O}$_{\textrm{7}}$: {A} high-temperature series expansion study},\ }\href {https://doi.org/10.1103/PhysRevB.88.220404} {\bibfield  {journal} {\bibinfo  {journal} {Phys. Rev. B}\ }\textbf {\bibinfo {volume} {88}},\ \bibinfo {pages} {220404} (\bibinfo {year} {2013})}\BibitemShut {NoStop}%
\bibitem [{\citenamefont {Zhitomirsky}\ \emph {et~al.}(2014)\citenamefont {Zhitomirsky}, \citenamefont {Holdsworth},\ and\ \citenamefont {Moessner}}]{zhitomirsky2014}%
  \BibitemOpen
  \bibfield  {author} {\bibinfo {author} {\bibfnamefont {M.~E.}\ \bibnamefont {Zhitomirsky}}, \bibinfo {author} {\bibfnamefont {P.~C.~W.}\ \bibnamefont {Holdsworth}},\ and\ \bibinfo {author} {\bibfnamefont {R.}~\bibnamefont {Moessner}},\ }\bibfield  {title} {\bibinfo {title} {Nature of finite-temperature transition in anisotropic pyrochlore {Er}$_{\textrm{2}}${Ti}$_{\textrm{2}}${O}$_{\textrm{7}}$},\ }\href {https://doi.org/10.1103/PhysRevB.89.140403} {\bibfield  {journal} {\bibinfo  {journal} {Phys. Rev. B}\ }\textbf {\bibinfo {volume} {89}},\ \bibinfo {pages} {140403} (\bibinfo {year} {2014})}\BibitemShut {NoStop}%
\bibitem [{\citenamefont {McClarty}\ \emph {et~al.}(2014)\citenamefont {McClarty}, \citenamefont {Stasiak},\ and\ \citenamefont {Gingras}}]{mcclarty2014}%
  \BibitemOpen
  \bibfield  {author} {\bibinfo {author} {\bibfnamefont {P.~A.}\ \bibnamefont {McClarty}}, \bibinfo {author} {\bibfnamefont {P.}~\bibnamefont {Stasiak}},\ and\ \bibinfo {author} {\bibfnamefont {M.~J.~P.}\ \bibnamefont {Gingras}},\ }\bibfield  {title} {\bibinfo {title} {Order-by-disorder in the ${XY}$ pyrochlore antiferromagnet},\ }\href {https://doi.org/10.1103/PhysRevB.89.024425} {\bibfield  {journal} {\bibinfo  {journal} {Phys. Rev. B}\ }\textbf {\bibinfo {volume} {89}},\ \bibinfo {pages} {024425} (\bibinfo {year} {2014})}\BibitemShut {NoStop}%
\bibitem [{\citenamefont {Francini}\ \emph {et~al.}(2025)\citenamefont {Francini}, \citenamefont {Janssen},\ and\ \citenamefont {Lozano-G\'omez}}]{Francini2024nematicR2}%
  \BibitemOpen
  \bibfield  {author} {\bibinfo {author} {\bibfnamefont {N.}~\bibnamefont {Francini}}, \bibinfo {author} {\bibfnamefont {L.}~\bibnamefont {Janssen}},\ and\ \bibinfo {author} {\bibfnamefont {D.}~\bibnamefont {Lozano-G\'omez}},\ }\bibfield  {title} {\bibinfo {title} {Higher-rank spin liquids and spin nematics from competing orders in pyrochlore magnets},\ }\href {https://doi.org/10.1103/PhysRevB.111.085140} {\bibfield  {journal} {\bibinfo  {journal} {Phys. Rev. B}\ }\textbf {\bibinfo {volume} {111}},\ \bibinfo {pages} {085140} (\bibinfo {year} {2025})}\BibitemShut {NoStop}%
\bibitem [{\citenamefont {Chandra}\ and\ \citenamefont {Doucot}(1988)}]{chandra1988}%
  \BibitemOpen
  \bibfield  {author} {\bibinfo {author} {\bibfnamefont {P.}~\bibnamefont {Chandra}}\ and\ \bibinfo {author} {\bibfnamefont {B.}~\bibnamefont {Doucot}},\ }\bibfield  {title} {\bibinfo {title} {Possible spin-liquid state at large ${S}$ for the frustrated square {Heisenberg} lattice},\ }\href {https://doi.org/10.1103/PhysRevB.38.9335} {\bibfield  {journal} {\bibinfo  {journal} {Phys. Rev. B}\ }\textbf {\bibinfo {volume} {38}},\ \bibinfo {pages} {9335} (\bibinfo {year} {1988})}\BibitemShut {NoStop}%
\bibitem [{\citenamefont {Kubo}\ and\ \citenamefont {Kishi}(1991)}]{kubo1991}%
  \BibitemOpen
  \bibfield  {author} {\bibinfo {author} {\bibfnamefont {K.}~\bibnamefont {Kubo}}\ and\ \bibinfo {author} {\bibfnamefont {T.}~\bibnamefont {Kishi}},\ }\bibfield  {title} {\bibinfo {title} {Ordering {Due} to {Quantum} {Fluctuations} in the {Frustrated} {Heisenberg} {Model}},\ }\href {https://doi.org/10.1143/JPSJ.60.567} {\bibfield  {journal} {\bibinfo  {journal} {J. Phys. Soc. Jpn.}\ }\textbf {\bibinfo {volume} {60}},\ \bibinfo {pages} {567} (\bibinfo {year} {1991})}\BibitemShut {NoStop}%
\bibitem [{\citenamefont {Sachdev}(1992)}]{sachdev1991}%
  \BibitemOpen
  \bibfield  {author} {\bibinfo {author} {\bibfnamefont {S.}~\bibnamefont {Sachdev}},\ }\bibfield  {title} {\bibinfo {title} {Kagom\'e and triangular-lattice {Heisenberg} antiferromagnets: {Ordering} from quantum fluctuations and quantum-disordered ground states with unconfined bosonic spinons},\ }\href {https://doi.org/10.1103/PhysRevB.45.12377} {\bibfield  {journal} {\bibinfo  {journal} {Phys. Rev. B}\ }\textbf {\bibinfo {volume} {45}},\ \bibinfo {pages} {12377} (\bibinfo {year} {1992})}\BibitemShut {NoStop}%
\bibitem [{\citenamefont {Chubukov}\ and\ \citenamefont {Golosov}(1991)}]{Chubukov_1991}%
  \BibitemOpen
  \bibfield  {author} {\bibinfo {author} {\bibfnamefont {A.~V.}\ \bibnamefont {Chubukov}}\ and\ \bibinfo {author} {\bibfnamefont {D.~I.}\ \bibnamefont {Golosov}},\ }\bibfield  {title} {\bibinfo {title} {Quantum theory of an antiferromagnet on a triangular lattice in a magnetic field},\ }\href {https://doi.org/10.1088/0953-8984/3/1/005} {\bibfield  {journal} {\bibinfo  {journal} {J. Phys. Condens. Matter}\ }\textbf {\bibinfo {volume} {3}},\ \bibinfo {pages} {69} (\bibinfo {year} {1991})}\BibitemShut {NoStop}%
\bibitem [{\citenamefont {Chubukov}(1992)}]{chubukov1992}%
  \BibitemOpen
  \bibfield  {author} {\bibinfo {author} {\bibfnamefont {A.}~\bibnamefont {Chubukov}},\ }\bibfield  {title} {\bibinfo {title} {Order from {Disorder} in a {Kagomé} {Antiferromagnet}},\ }\href {https://doi.org/10.1103/PhysRevLett.69.832} {\bibfield  {journal} {\bibinfo  {journal} {Phys. Rev. Lett.}\ }\textbf {\bibinfo {volume} {69}},\ \bibinfo {pages} {832} (\bibinfo {year} {1992})}\BibitemShut {NoStop}%
\bibitem [{\citenamefont {Henley}(1994)}]{henley1994}%
  \BibitemOpen
  \bibfield  {author} {\bibinfo {author} {\bibfnamefont {C.~L.}\ \bibnamefont {Henley}},\ }\bibfield  {title} {\bibinfo {title} {Selection by {Quantum} {Fluctuations} of {Dipolar} {Order} in a {Diamond} {Lattice}},\ }\href {https://doi.org/10.1103/PhysRevLett.73.2788} {\bibfield  {journal} {\bibinfo  {journal} {Phys. Rev. Lett.}\ }\textbf {\bibinfo {volume} {73}},\ \bibinfo {pages} {2788} (\bibinfo {year} {1994})}\BibitemShut {NoStop}%
\bibitem [{\citenamefont {Lecheminant}\ \emph {et~al.}(1995)\citenamefont {Lecheminant}, \citenamefont {Bernu}, \citenamefont {Lhuillier},\ and\ \citenamefont {Pierre}}]{Lecheminant1995}%
  \BibitemOpen
  \bibfield  {author} {\bibinfo {author} {\bibfnamefont {P.}~\bibnamefont {Lecheminant}}, \bibinfo {author} {\bibfnamefont {B.}~\bibnamefont {Bernu}}, \bibinfo {author} {\bibfnamefont {C.}~\bibnamefont {Lhuillier}},\ and\ \bibinfo {author} {\bibfnamefont {L.}~\bibnamefont {Pierre}},\ }\bibfield  {title} {\bibinfo {title} {{$J_1-J_2$} quantum {Heisenberg} antiferromagnet on the triangular lattice: {A} group-symmetry analysis of order by disorder},\ }\href {https://doi.org/10.1103/PhysRevB.52.6647} {\bibfield  {journal} {\bibinfo  {journal} {Phys. Rev. B}\ }\textbf {\bibinfo {volume} {52}},\ \bibinfo {pages} {6647} (\bibinfo {year} {1995})}\BibitemShut {NoStop}%
\bibitem [{\citenamefont {Sobral}\ and\ \citenamefont {Lacroix}(1997)}]{sobral1997}%
  \BibitemOpen
  \bibfield  {author} {\bibinfo {author} {\bibfnamefont {R.~R.}\ \bibnamefont {Sobral}}\ and\ \bibinfo {author} {\bibfnamefont {C.}~\bibnamefont {Lacroix}},\ }\bibfield  {title} {\bibinfo {title} {Order by disorder in the pyrochlore antiferromagnets},\ }\href {https://doi.org/10.1016/S0038-1098(97)00212-3} {\bibfield  {journal} {\bibinfo  {journal} {Solid State Communications}\ }\textbf {\bibinfo {volume} {103}},\ \bibinfo {pages} {407} (\bibinfo {year} {1997})}\BibitemShut {NoStop}%
\bibitem [{\citenamefont {Champion}\ \emph {et~al.}(2003)\citenamefont {Champion}, \citenamefont {Harris}, \citenamefont {Holdsworth}, \citenamefont {Wills}, \citenamefont {Balakrishnan}, \citenamefont {Bramwell}, \citenamefont {Čižmár}, \citenamefont {Fennell}, \citenamefont {Gardner}, \citenamefont {Lago}, \citenamefont {McMorrow}, \citenamefont {Orendáč}, \citenamefont {Orendáčová}, \citenamefont {Paul}, \citenamefont {Smith}, \citenamefont {Telling},\ and\ \citenamefont {Wildes}}]{champion2003}%
  \BibitemOpen
  \bibfield  {author} {\bibinfo {author} {\bibfnamefont {J.~D.~M.}\ \bibnamefont {Champion}}, \bibinfo {author} {\bibfnamefont {M.~J.}\ \bibnamefont {Harris}}, \bibinfo {author} {\bibfnamefont {P.~C.~W.}\ \bibnamefont {Holdsworth}}, \bibinfo {author} {\bibfnamefont {A.~S.}\ \bibnamefont {Wills}}, \bibinfo {author} {\bibfnamefont {G.}~\bibnamefont {Balakrishnan}}, \bibinfo {author} {\bibfnamefont {S.~T.}\ \bibnamefont {Bramwell}}, \bibinfo {author} {\bibfnamefont {E.}~\bibnamefont {Čižmár}}, \bibinfo {author} {\bibfnamefont {T.}~\bibnamefont {Fennell}}, \bibinfo {author} {\bibfnamefont {J.~S.}\ \bibnamefont {Gardner}}, \bibinfo {author} {\bibfnamefont {J.}~\bibnamefont {Lago}}, \bibinfo {author} {\bibfnamefont {D.~F.}\ \bibnamefont {McMorrow}}, \bibinfo {author} {\bibfnamefont {M.}~\bibnamefont {Orendáč}}, \bibinfo {author} {\bibfnamefont {A.}~\bibnamefont {Orendáčová}}, \bibinfo {author} {\bibfnamefont {D.~M.}\ \bibnamefont {Paul}}, \bibinfo {author} {\bibfnamefont {R.~I.}\ \bibnamefont {Smith}},
  \bibinfo {author} {\bibfnamefont {M.~T.~F.}\ \bibnamefont {Telling}},\ and\ \bibinfo {author} {\bibfnamefont {A.}~\bibnamefont {Wildes}},\ }\bibfield  {title} {\bibinfo {title} {Er$_{\textrm{2}}${Ti}$_{\textrm{2}}${O}$_{\textrm{7}}$: {Evidence} of quantum order by disorder in a frustrated antiferromagnet},\ }\href {https://doi.org/10.1103/PhysRevB.68.020401} {\bibfield  {journal} {\bibinfo  {journal} {Phys. Rev. B}\ }\textbf {\bibinfo {volume} {68}},\ \bibinfo {pages} {020401} (\bibinfo {year} {2003})}\BibitemShut {NoStop}%
\bibitem [{\citenamefont {Baskaran}\ \emph {et~al.}(2008)\citenamefont {Baskaran}, \citenamefont {Sen},\ and\ \citenamefont {Shankar}}]{baskaran2008}%
  \BibitemOpen
  \bibfield  {author} {\bibinfo {author} {\bibfnamefont {G.}~\bibnamefont {Baskaran}}, \bibinfo {author} {\bibfnamefont {D.}~\bibnamefont {Sen}},\ and\ \bibinfo {author} {\bibfnamefont {R.}~\bibnamefont {Shankar}},\ }\bibfield  {title} {\bibinfo {title} {Spin-${S}$ {Kitaev} model: {Classical} ground states, order from disorder, and exact correlation functions},\ }\href {https://doi.org/10.1103/PhysRevB.78.115116} {\bibfield  {journal} {\bibinfo  {journal} {Phys. Rev. B}\ }\textbf {\bibinfo {volume} {78}},\ \bibinfo {pages} {115116} (\bibinfo {year} {2008})}\BibitemShut {NoStop}%
\bibitem [{\citenamefont {Bernier}\ \emph {et~al.}(2008)\citenamefont {Bernier}, \citenamefont {Lawler},\ and\ \citenamefont {Kim}}]{bernier2008}%
  \BibitemOpen
  \bibfield  {author} {\bibinfo {author} {\bibfnamefont {J.-S.}\ \bibnamefont {Bernier}}, \bibinfo {author} {\bibfnamefont {M.~J.}\ \bibnamefont {Lawler}},\ and\ \bibinfo {author} {\bibfnamefont {Y.~B.}\ \bibnamefont {Kim}},\ }\bibfield  {title} {\bibinfo {title} {Quantum {Order} by {Disorder} in {Frustrated} {Diamond} {Lattice} {Antiferromagnets}},\ }\href {https://doi.org/10.1103/PhysRevLett.101.047201} {\bibfield  {journal} {\bibinfo  {journal} {Phys. Rev. Lett.}\ }\textbf {\bibinfo {volume} {101}},\ \bibinfo {pages} {047201} (\bibinfo {year} {2008})}\BibitemShut {NoStop}%
\bibitem [{\citenamefont {Mulder}\ \emph {et~al.}(2010)\citenamefont {Mulder}, \citenamefont {Ganesh}, \citenamefont {Capriotti},\ and\ \citenamefont {Paramekanti}}]{mulder2010}%
  \BibitemOpen
  \bibfield  {author} {\bibinfo {author} {\bibfnamefont {A.}~\bibnamefont {Mulder}}, \bibinfo {author} {\bibfnamefont {R.}~\bibnamefont {Ganesh}}, \bibinfo {author} {\bibfnamefont {L.}~\bibnamefont {Capriotti}},\ and\ \bibinfo {author} {\bibfnamefont {A.}~\bibnamefont {Paramekanti}},\ }\bibfield  {title} {\bibinfo {title} {Spiral order by disorder and lattice nematic order in a frustrated {Heisenberg} antiferromagnet on the honeycomb lattice},\ }\href {https://doi.org/10.1103/PhysRevB.81.214419} {\bibfield  {journal} {\bibinfo  {journal} {Phys. Rev. B}\ }\textbf {\bibinfo {volume} {81}},\ \bibinfo {pages} {214419} (\bibinfo {year} {2010})}\BibitemShut {NoStop}%
\bibitem [{\citenamefont {Savary}\ \emph {et~al.}(2012)\citenamefont {Savary}, \citenamefont {Ross}, \citenamefont {Gaulin}, \citenamefont {Ruff},\ and\ \citenamefont {Balents}}]{savary2012}%
  \BibitemOpen
  \bibfield  {author} {\bibinfo {author} {\bibfnamefont {L.}~\bibnamefont {Savary}}, \bibinfo {author} {\bibfnamefont {K.~A.}\ \bibnamefont {Ross}}, \bibinfo {author} {\bibfnamefont {B.~D.}\ \bibnamefont {Gaulin}}, \bibinfo {author} {\bibfnamefont {J.~P.~C.}\ \bibnamefont {Ruff}},\ and\ \bibinfo {author} {\bibfnamefont {L.}~\bibnamefont {Balents}},\ }\bibfield  {title} {\bibinfo {title} {Order by {Quantum} {Disorder} in {Er}$_{\textrm{2}}${Ti}$_{\textrm{2}}${O}$_{\textrm{7}}$},\ }\href {https://doi.org/10.1103/PhysRevLett.109.167201} {\bibfield  {journal} {\bibinfo  {journal} {Phys. Rev. Lett.}\ }\textbf {\bibinfo {volume} {109}},\ \bibinfo {pages} {167201} (\bibinfo {year} {2012})}\BibitemShut {NoStop}%
\bibitem [{\citenamefont {Zhitomirsky}\ \emph {et~al.}(2012)\citenamefont {Zhitomirsky}, \citenamefont {Gvozdikova}, \citenamefont {Holdsworth},\ and\ \citenamefont {Moessner}}]{zhitomirsky2012}%
  \BibitemOpen
  \bibfield  {author} {\bibinfo {author} {\bibfnamefont {M.~E.}\ \bibnamefont {Zhitomirsky}}, \bibinfo {author} {\bibfnamefont {M.~V.}\ \bibnamefont {Gvozdikova}}, \bibinfo {author} {\bibfnamefont {P.~C.~W.}\ \bibnamefont {Holdsworth}},\ and\ \bibinfo {author} {\bibfnamefont {R.}~\bibnamefont {Moessner}},\ }\bibfield  {title} {\bibinfo {title} {Quantum {Order} by {Disorder} and {Accidental} {Soft} {Mode} in {Er}$_{\textrm{2}}${Ti}$_{\textrm{2}}${O}$_{\textrm{7}}$},\ }\href {https://doi.org/10.1103/PhysRevLett.109.077204} {\bibfield  {journal} {\bibinfo  {journal} {Phys. Rev. Lett.}\ }\textbf {\bibinfo {volume} {109}},\ \bibinfo {pages} {077204} (\bibinfo {year} {2012})}\BibitemShut {NoStop}%
\bibitem [{\citenamefont {Chernyshev}\ and\ \citenamefont {Zhitomirsky}(2014)}]{Chernyshev2014}%
  \BibitemOpen
  \bibfield  {author} {\bibinfo {author} {\bibfnamefont {A.~L.}\ \bibnamefont {Chernyshev}}\ and\ \bibinfo {author} {\bibfnamefont {M.~E.}\ \bibnamefont {Zhitomirsky}},\ }\bibfield  {title} {\bibinfo {title} {{Quantum} {Selection} of {Order} in an {$XXZ$} {Antiferromagnet} on a {Kagome} {Lattice}},\ }\href {https://doi.org/10.1103/PhysRevLett.113.237202} {\bibfield  {journal} {\bibinfo  {journal} {Phys. Rev. Lett.}\ }\textbf {\bibinfo {volume} {113}},\ \bibinfo {pages} {237202} (\bibinfo {year} {2014})}\BibitemShut {NoStop}%
\bibitem [{\citenamefont {Rousochatzakis}\ \emph {et~al.}(2015)\citenamefont {Rousochatzakis}, \citenamefont {Reuther}, \citenamefont {Thomale}, \citenamefont {Rachel},\ and\ \citenamefont {Perkins}}]{Rousochatzakis2015}%
  \BibitemOpen
  \bibfield  {author} {\bibinfo {author} {\bibfnamefont {I.}~\bibnamefont {Rousochatzakis}}, \bibinfo {author} {\bibfnamefont {J.}~\bibnamefont {Reuther}}, \bibinfo {author} {\bibfnamefont {R.}~\bibnamefont {Thomale}}, \bibinfo {author} {\bibfnamefont {S.}~\bibnamefont {Rachel}},\ and\ \bibinfo {author} {\bibfnamefont {N.~B.}\ \bibnamefont {Perkins}},\ }\bibfield  {title} {\bibinfo {title} {{Phase} {Diagram} and {Quantum} {Order} by {Disorde}r in the {Kitaev} ${K}_{1}\ensuremath{-}{K}_{2}$ {Honeycomb} {Magnet}},\ }\href {https://doi.org/10.1103/PhysRevX.5.041035} {\bibfield  {journal} {\bibinfo  {journal} {Phys. Rev. X}\ }\textbf {\bibinfo {volume} {5}},\ \bibinfo {pages} {041035} (\bibinfo {year} {2015})}\BibitemShut {NoStop}%
\bibitem [{\citenamefont {Rau}\ \emph {et~al.}(2018)\citenamefont {Rau}, \citenamefont {McClarty},\ and\ \citenamefont {Moessner}}]{rau2018}%
  \BibitemOpen
  \bibfield  {author} {\bibinfo {author} {\bibfnamefont {J.~G.}\ \bibnamefont {Rau}}, \bibinfo {author} {\bibfnamefont {P.~A.}\ \bibnamefont {McClarty}},\ and\ \bibinfo {author} {\bibfnamefont {R.}~\bibnamefont {Moessner}},\ }\bibfield  {title} {\bibinfo {title} {Pseudo-{Goldstone} {Gaps} and {Order}-by-{Quantum} {Disorder} in {Frustrated} {Magnets}},\ }\href {https://doi.org/10.1103/PhysRevLett.121.237201} {\bibfield  {journal} {\bibinfo  {journal} {Phys. Rev. Lett.}\ }\textbf {\bibinfo {volume} {121}},\ \bibinfo {pages} {237201} (\bibinfo {year} {2018})}\BibitemShut {NoStop}%
\bibitem [{\citenamefont {Placke}\ \emph {et~al.}(2020)\citenamefont {Placke}, \citenamefont {Moessner},\ and\ \citenamefont {Benton}}]{Placke2020}%
  \BibitemOpen
  \bibfield  {author} {\bibinfo {author} {\bibfnamefont {B.}~\bibnamefont {Placke}}, \bibinfo {author} {\bibfnamefont {R.}~\bibnamefont {Moessner}},\ and\ \bibinfo {author} {\bibfnamefont {O.}~\bibnamefont {Benton}},\ }\bibfield  {title} {\bibinfo {title} {Hierarchy of energy scales and field-tunable order by disorder in dipolar-octupolar pyrochlores},\ }\href {https://doi.org/10.1103/PhysRevB.102.245102} {\bibfield  {journal} {\bibinfo  {journal} {Phys. Rev. B}\ }\textbf {\bibinfo {volume} {102}},\ \bibinfo {pages} {245102} (\bibinfo {year} {2020})}\BibitemShut {NoStop}%
\bibitem [{\citenamefont {Chen}\ and\ \citenamefont {Wang}(2020)}]{Chen2020}%
  \BibitemOpen
  \bibfield  {author} {\bibinfo {author} {\bibfnamefont {G.}~\bibnamefont {Chen}}\ and\ \bibinfo {author} {\bibfnamefont {X.}~\bibnamefont {Wang}},\ }\bibfield  {title} {\bibinfo {title} {Electron quasi-itinerancy intertwined with quantum order by disorder in pyrochlore iridate magnetism},\ }\href {https://doi.org/10.1103/PhysRevResearch.2.043273} {\bibfield  {journal} {\bibinfo  {journal} {Phys. Rev. Res.}\ }\textbf {\bibinfo {volume} {2}},\ \bibinfo {pages} {043273} (\bibinfo {year} {2020})}\BibitemShut {NoStop}%
\bibitem [{\citenamefont {Liu}\ \emph {et~al.}(2020)\citenamefont {Liu}, \citenamefont {Huang},\ and\ \citenamefont {Chen}}]{Liu2020}%
  \BibitemOpen
  \bibfield  {author} {\bibinfo {author} {\bibfnamefont {C.}~\bibnamefont {Liu}}, \bibinfo {author} {\bibfnamefont {C.-J.}\ \bibnamefont {Huang}},\ and\ \bibinfo {author} {\bibfnamefont {G.}~\bibnamefont {Chen}},\ }\bibfield  {title} {\bibinfo {title} {Intrinsic quantum {I}sing model on a triangular lattice magnet $\mathrm{Tm}\mathrm{Mg}\mathrm{Ga}{\mathrm{o}}_{4}$},\ }\href {https://doi.org/10.1103/PhysRevResearch.2.043013} {\bibfield  {journal} {\bibinfo  {journal} {Phys. Rev. Res.}\ }\textbf {\bibinfo {volume} {2}},\ \bibinfo {pages} {043013} (\bibinfo {year} {2020})}\BibitemShut {NoStop}%
\bibitem [{\citenamefont {Khatua}\ \emph {et~al.}(2021)\citenamefont {Khatua}, \citenamefont {Srinivasan},\ and\ \citenamefont {Ganesh}}]{Khatua2021}%
  \BibitemOpen
  \bibfield  {author} {\bibinfo {author} {\bibfnamefont {S.}~\bibnamefont {Khatua}}, \bibinfo {author} {\bibfnamefont {S.}~\bibnamefont {Srinivasan}},\ and\ \bibinfo {author} {\bibfnamefont {R.}~\bibnamefont {Ganesh}},\ }\bibfield  {title} {\bibinfo {title} {State selection in frustrated magnets},\ }\href {https://doi.org/10.1103/PhysRevB.103.174412} {\bibfield  {journal} {\bibinfo  {journal} {Phys. Rev. B}\ }\textbf {\bibinfo {volume} {103}},\ \bibinfo {pages} {174412} (\bibinfo {year} {2021})}\BibitemShut {NoStop}%
\bibitem [{\citenamefont {Khatua}\ \emph {et~al.}(2023)\citenamefont {Khatua}, \citenamefont {Gingras},\ and\ \citenamefont {Rau}}]{KHATUA2023}%
  \BibitemOpen
  \bibfield  {author} {\bibinfo {author} {\bibfnamefont {S.}~\bibnamefont {Khatua}}, \bibinfo {author} {\bibfnamefont {M.~J.~P.}\ \bibnamefont {Gingras}},\ and\ \bibinfo {author} {\bibfnamefont {J.~G.}\ \bibnamefont {Rau}},\ }\bibfield  {title} {\bibinfo {title} {Pseudo-{Goldstone} {Modes} and {Dynamical} {Gap} {Generation} from {Order} by {Thermal} {Disorder}},\ }\href {https://doi.org/10.1103/PhysRevLett.130.266702} {\bibfield  {journal} {\bibinfo  {journal} {Phys. Rev. Lett.}\ }\textbf {\bibinfo {volume} {130}},\ \bibinfo {pages} {266702} (\bibinfo {year} {2023})}\BibitemShut {NoStop}%
\bibitem [{\citenamefont {Khatua}\ \emph {et~al.}(2024)\citenamefont {Khatua}, \citenamefont {Howson}, \citenamefont {Gingras},\ and\ \citenamefont {Rau}}]{khatua2024}%
  \BibitemOpen
  \bibfield  {author} {\bibinfo {author} {\bibfnamefont {S.}~\bibnamefont {Khatua}}, \bibinfo {author} {\bibfnamefont {G.~C.}\ \bibnamefont {Howson}}, \bibinfo {author} {\bibfnamefont {M.~J.~P.}\ \bibnamefont {Gingras}},\ and\ \bibinfo {author} {\bibfnamefont {J.~G.}\ \bibnamefont {Rau}},\ }\bibfield  {title} {\bibinfo {title} {Ground state properties of the {{Heisenberg-compass}} model on the square lattice},\ }\href {https://journals.aps.org/prb/abstract/10.1103/PhysRevB.110.104426} {\bibfield  {journal} {\bibinfo  {journal} {Phys. Rev. B}\ }\textbf {\bibinfo {volume} {110}},\ \bibinfo {pages} {104426} (\bibinfo {year} {2024})}\BibitemShut {NoStop}%
\bibitem [{\citenamefont {Lee}\ \emph {et~al.}(2014)\citenamefont {Lee}, \citenamefont {Lee}, \citenamefont {Paramekanti},\ and\ \citenamefont {Kim}}]{Lee2014}%
  \BibitemOpen
  \bibfield  {author} {\bibinfo {author} {\bibfnamefont {S.}~\bibnamefont {Lee}}, \bibinfo {author} {\bibfnamefont {E.~K.-H.}\ \bibnamefont {Lee}}, \bibinfo {author} {\bibfnamefont {A.}~\bibnamefont {Paramekanti}},\ and\ \bibinfo {author} {\bibfnamefont {Y.~B.}\ \bibnamefont {Kim}},\ }\bibfield  {title} {\bibinfo {title} {Order-by-disorder and magnetic field response in the {H}eisenberg-{K}itaev model on a hyperhoneycomb lattice},\ }\href {https://doi.org/10.1103/PhysRevB.89.014424} {\bibfield  {journal} {\bibinfo  {journal} {Phys. Rev. B}\ }\textbf {\bibinfo {volume} {89}},\ \bibinfo {pages} {014424} (\bibinfo {year} {2014})}\BibitemShut {NoStop}%
\bibitem [{\citenamefont {Danu}\ \emph {et~al.}(2016)\citenamefont {Danu}, \citenamefont {Nambiar},\ and\ \citenamefont {Ganesh}}]{danu2016}%
  \BibitemOpen
  \bibfield  {author} {\bibinfo {author} {\bibfnamefont {B.}~\bibnamefont {Danu}}, \bibinfo {author} {\bibfnamefont {G.}~\bibnamefont {Nambiar}},\ and\ \bibinfo {author} {\bibfnamefont {R.}~\bibnamefont {Ganesh}},\ }\bibfield  {title} {\bibinfo {title} {Extended degeneracy and order by disorder in the square lattice ${J}_1-{J}_2-{J}_3$ model},\ }\href {https://doi.org/10.1103/PhysRevB.94.094438} {\bibfield  {journal} {\bibinfo  {journal} {Phys. Rev. B}\ }\textbf {\bibinfo {volume} {94}},\ \bibinfo {pages} {094438} (\bibinfo {year} {2016})}\BibitemShut {NoStop}%
\bibitem [{\citenamefont {Schick}\ \emph {et~al.}(2020)\citenamefont {Schick}, \citenamefont {Ziman},\ and\ \citenamefont {Zhitomirsky}}]{schick2020}%
  \BibitemOpen
  \bibfield  {author} {\bibinfo {author} {\bibfnamefont {R.}~\bibnamefont {Schick}}, \bibinfo {author} {\bibfnamefont {T.}~\bibnamefont {Ziman}},\ and\ \bibinfo {author} {\bibfnamefont {M.~E.}\ \bibnamefont {Zhitomirsky}},\ }\bibfield  {title} {\bibinfo {title} {Quantum versus thermal fluctuations in the fcc antiferromagnet: {Alternative} routes to order by disorder},\ }\href {https://doi.org/10.1103/PhysRevB.102.220405} {\bibfield  {journal} {\bibinfo  {journal} {Phys. Rev. B}\ }\textbf {\bibinfo {volume} {102}},\ \bibinfo {pages} {220405} (\bibinfo {year} {2020})}\BibitemShut {NoStop}%
\bibitem [{\citenamefont {Noculak}\ \emph {et~al.}(2023)\citenamefont {Noculak}, \citenamefont {Lozano-Gómez}, \citenamefont {Oitmaa}, \citenamefont {Singh}, \citenamefont {Iqbal}, \citenamefont {Gingras},\ and\ \citenamefont {Reuther}}]{noculak2023}%
  \BibitemOpen
  \bibfield  {author} {\bibinfo {author} {\bibfnamefont {V.}~\bibnamefont {Noculak}}, \bibinfo {author} {\bibfnamefont {D.}~\bibnamefont {Lozano-Gómez}}, \bibinfo {author} {\bibfnamefont {J.}~\bibnamefont {Oitmaa}}, \bibinfo {author} {\bibfnamefont {R.~R.~P.}\ \bibnamefont {Singh}}, \bibinfo {author} {\bibfnamefont {Y.}~\bibnamefont {Iqbal}}, \bibinfo {author} {\bibfnamefont {M.~J.~P.}\ \bibnamefont {Gingras}},\ and\ \bibinfo {author} {\bibfnamefont {J.}~\bibnamefont {Reuther}},\ }\bibfield  {title} {\bibinfo {title} {Classical and quantum phases of the pyrochlore ${S}=\frac{1}{2}$ magnet with {Heisenberg} and {Dzyaloshinskii}-{Moriya} interactions},\ }\href {https://doi.org/10.1103/PhysRevB.107.214414} {\bibfield  {journal} {\bibinfo  {journal} {Phys. Rev. B}\ }\textbf {\bibinfo {volume} {107}},\ \bibinfo {pages} {214414} (\bibinfo {year} {2023})}\BibitemShut {NoStop}%
\bibitem [{\citenamefont {Fyodorov}\ and\ \citenamefont {Shender}(1991)}]{Fyodorov1991}%
  \BibitemOpen
  \bibfield  {author} {\bibinfo {author} {\bibfnamefont {Y.~V.}\ \bibnamefont {Fyodorov}}\ and\ \bibinfo {author} {\bibfnamefont {E.~F.}\ \bibnamefont {Shender}},\ }\bibfield  {title} {\bibinfo {title} {Random-field effects in antiferromagnets with classically degenerate ground states},\ }\href {https://doi.org/10.1088/0953-8984/3/46/013} {\bibfield  {journal} {\bibinfo  {journal} {J. Phys. Condens. Matter}\ }\textbf {\bibinfo {volume} {3}},\ \bibinfo {pages} {9123} (\bibinfo {year} {1991})}\BibitemShut {NoStop}%
\bibitem [{\citenamefont {Savary}\ \emph {et~al.}(2011)\citenamefont {Savary}, \citenamefont {Gull}, \citenamefont {Trebst}, \citenamefont {Alicea}, \citenamefont {Bergman},\ and\ \citenamefont {Balents}}]{Savary2011}%
  \BibitemOpen
  \bibfield  {author} {\bibinfo {author} {\bibfnamefont {L.}~\bibnamefont {Savary}}, \bibinfo {author} {\bibfnamefont {E.}~\bibnamefont {Gull}}, \bibinfo {author} {\bibfnamefont {S.}~\bibnamefont {Trebst}}, \bibinfo {author} {\bibfnamefont {J.}~\bibnamefont {Alicea}}, \bibinfo {author} {\bibfnamefont {D.}~\bibnamefont {Bergman}},\ and\ \bibinfo {author} {\bibfnamefont {L.}~\bibnamefont {Balents}},\ }\bibfield  {title} {\bibinfo {title} {Impurity effects in highly frustrated diamond-lattice antiferromagnets},\ }\href {https://doi.org/10.1103/PhysRevB.84.064438} {\bibfield  {journal} {\bibinfo  {journal} {Phys. Rev. B}\ }\textbf {\bibinfo {volume} {84}},\ \bibinfo {pages} {064438} (\bibinfo {year} {2011})}\BibitemShut {NoStop}%
\bibitem [{\citenamefont {Maryasin}\ and\ \citenamefont {Zhitomirsky}(2013)}]{Maryasin2013}%
  \BibitemOpen
  \bibfield  {author} {\bibinfo {author} {\bibfnamefont {V.~S.}\ \bibnamefont {Maryasin}}\ and\ \bibinfo {author} {\bibfnamefont {M.~E.}\ \bibnamefont {Zhitomirsky}},\ }\bibfield  {title} {\bibinfo {title} {Triangular {Antiferromagnet} with {Nonmagnetic} {Impurities}},\ }\href {https://doi.org/10.1103/PhysRevLett.111.247201} {\bibfield  {journal} {\bibinfo  {journal} {Phys. Rev. Lett.}\ }\textbf {\bibinfo {volume} {111}},\ \bibinfo {pages} {247201} (\bibinfo {year} {2013})}\BibitemShut {NoStop}%
\bibitem [{\citenamefont {Andreanov}\ and\ \citenamefont {McClarty}(2015)}]{Andreanov}%
  \BibitemOpen
  \bibfield  {author} {\bibinfo {author} {\bibfnamefont {A.}~\bibnamefont {Andreanov}}\ and\ \bibinfo {author} {\bibfnamefont {P.~A.}\ \bibnamefont {McClarty}},\ }\bibfield  {title} {\bibinfo {title} {Order induced by dilution in pyrochlore {XY} antiferromagnets},\ }\href {https://doi.org/10.1103/PhysRevB.91.064401} {\bibfield  {journal} {\bibinfo  {journal} {Phys. Rev. B}\ }\textbf {\bibinfo {volume} {91}},\ \bibinfo {pages} {064401} (\bibinfo {year} {2015})}\BibitemShut {NoStop}%
\bibitem [{\citenamefont {Maryasin}\ and\ \citenamefont {Zhitomirsky}(2015)}]{Maryasin2015}%
  \BibitemOpen
  \bibfield  {author} {\bibinfo {author} {\bibfnamefont {V.~S.}\ \bibnamefont {Maryasin}}\ and\ \bibinfo {author} {\bibfnamefont {M.~E.}\ \bibnamefont {Zhitomirsky}},\ }\bibfield  {title} {\bibinfo {title} {Collective impurity effects in the {H}eisenberg triangular antiferromagnet},\ }\href {https://doi.org/10.1088/1742-6596/592/1/012112} {\bibfield  {journal} {\bibinfo  {journal} {J. Phys. Conf. Ser.}\ }\textbf {\bibinfo {volume} {592}},\ \bibinfo {pages} {012112} (\bibinfo {year} {2015})}\BibitemShut {NoStop}%
\bibitem [{\citenamefont {Smirnov}\ \emph {et~al.}(2017)\citenamefont {Smirnov}, \citenamefont {Soldatov}, \citenamefont {Petrenko}, \citenamefont {Takata}, \citenamefont {Kida}, \citenamefont {Hagiwara}, \citenamefont {Shapiro},\ and\ \citenamefont {Zhitomirsky}}]{Smirnov2017}%
  \BibitemOpen
  \bibfield  {author} {\bibinfo {author} {\bibfnamefont {A.~I.}\ \bibnamefont {Smirnov}}, \bibinfo {author} {\bibfnamefont {T.~A.}\ \bibnamefont {Soldatov}}, \bibinfo {author} {\bibfnamefont {O.~A.}\ \bibnamefont {Petrenko}}, \bibinfo {author} {\bibfnamefont {A.}~\bibnamefont {Takata}}, \bibinfo {author} {\bibfnamefont {T.}~\bibnamefont {Kida}}, \bibinfo {author} {\bibfnamefont {M.}~\bibnamefont {Hagiwara}}, \bibinfo {author} {\bibfnamefont {A.~Y.}\ \bibnamefont {Shapiro}},\ and\ \bibinfo {author} {\bibfnamefont {M.~E.}\ \bibnamefont {Zhitomirsky}},\ }\bibfield  {title} {\bibinfo {title} {{Order} by {Quenched} {Disorder} in the {Model} {Triangular} {Antiferromagnet} {RbFe}({MoO$_4$})$_2$},\ }\href {https://doi.org/10.1103/PhysRevLett.119.047204} {\bibfield  {journal} {\bibinfo  {journal} {Phys. Rev. Lett.}\ }\textbf {\bibinfo {volume} {119}},\ \bibinfo {pages} {047204} (\bibinfo {year} {2017})}\BibitemShut {NoStop}%
\bibitem [{\citenamefont {Andrade}\ \emph {et~al.}(2018)\citenamefont {Andrade}, \citenamefont {Hoyos}, \citenamefont {Rachel},\ and\ \citenamefont {Vojta}}]{AndradePRL2018}%
  \BibitemOpen
  \bibfield  {author} {\bibinfo {author} {\bibfnamefont {E.~C.}\ \bibnamefont {Andrade}}, \bibinfo {author} {\bibfnamefont {J.~A.}\ \bibnamefont {Hoyos}}, \bibinfo {author} {\bibfnamefont {S.}~\bibnamefont {Rachel}},\ and\ \bibinfo {author} {\bibfnamefont {M.}~\bibnamefont {Vojta}},\ }\bibfield  {title} {\bibinfo {title} {{Cluster-Glass} {Phase} in {Pyrochlore} ${XY}$ {Antiferromagnets} with {Quenched} {Disorder}},\ }\href {https://doi.org/10.1103/PhysRevLett.120.097204} {\bibfield  {journal} {\bibinfo  {journal} {Phys. Rev. Lett.}\ }\textbf {\bibinfo {volume} {120}},\ \bibinfo {pages} {097204} (\bibinfo {year} {2018})}\BibitemShut {NoStop}%
\bibitem [{\citenamefont {C\^onsoli}\ and\ \citenamefont {Vojta}(2024)}]{consoli2024}%
  \BibitemOpen
  \bibfield  {author} {\bibinfo {author} {\bibfnamefont {P.~M.}\ \bibnamefont {C\^onsoli}}\ and\ \bibinfo {author} {\bibfnamefont {M.}~\bibnamefont {Vojta}},\ }\bibfield  {title} {\bibinfo {title} {Disorder effects in spiral spin liquids: Long-range spin textures, {F}riedel-like oscillations, and spiral spin glasses},\ }\href {https://doi.org/10.1103/PhysRevB.109.064423} {\bibfield  {journal} {\bibinfo  {journal} {Phys. Rev. B}\ }\textbf {\bibinfo {volume} {109}},\ \bibinfo {pages} {064423} (\bibinfo {year} {2024})}\BibitemShut {NoStop}%
\bibitem [{\citenamefont {Ross}\ \emph {et~al.}(2014)\citenamefont {Ross}, \citenamefont {Qiu}, \citenamefont {Copley}, \citenamefont {Dabkowska},\ and\ \citenamefont {Gaulin}}]{ross2014}%
  \BibitemOpen
  \bibfield  {author} {\bibinfo {author} {\bibfnamefont {K.~A.}\ \bibnamefont {Ross}}, \bibinfo {author} {\bibfnamefont {Y.}~\bibnamefont {Qiu}}, \bibinfo {author} {\bibfnamefont {J.~R.~D.}\ \bibnamefont {Copley}}, \bibinfo {author} {\bibfnamefont {H.~A.}\ \bibnamefont {Dabkowska}},\ and\ \bibinfo {author} {\bibfnamefont {B.~D.}\ \bibnamefont {Gaulin}},\ }\bibfield  {title} {\bibinfo {title} {Order by {Disorder} {Spin} {Wave} {Gap} in the ${XY}$ {Pyrochlore} {Magnet} {Er}$_{\textrm{2}}${Ti}$_{\textrm{2}}${O}$_{\textrm{7}}$},\ }\href {https://doi.org/10.1103/PhysRevLett.112.057201} {\bibfield  {journal} {\bibinfo  {journal} {Phys. Rev. Lett.}\ }\textbf {\bibinfo {volume} {112}},\ \bibinfo {pages} {057201} (\bibinfo {year} {2014})}\BibitemShut {NoStop}%
\bibitem [{Note64()}]{Note64}%
  \BibitemOpen
  \bibinfo {note} {Although selection, at least in part, due to virtual crystal field fluctuations has not been ruled out \cite {mcclarty2009a,Petit2014,Rau2016}}\BibitemShut {NoStop}%
\bibitem [{\citenamefont {McClarty}\ \emph {et~al.}(2009)\citenamefont {McClarty}, \citenamefont {Curnoe},\ and\ \citenamefont {Gingras}}]{mcclarty2009a}%
  \BibitemOpen
  \bibfield  {author} {\bibinfo {author} {\bibfnamefont {P.~A.}\ \bibnamefont {McClarty}}, \bibinfo {author} {\bibfnamefont {S.~H.}\ \bibnamefont {Curnoe}},\ and\ \bibinfo {author} {\bibfnamefont {M.~J.~P.}\ \bibnamefont {Gingras}},\ }\bibfield  {title} {\bibinfo {title} {Energetic selection of ordered states in a model of the {{Er$_{\textrm{2}}${Ti}$_{\textrm{2}}${O}$_{\textrm{7}}$}} frustrated pyrochlore {{XY}} antiferromagnet},\ }\href {https://doi.org/10.1088/1742-6596/145/1/012032} {\bibfield  {journal} {\bibinfo  {journal} {J. Phys. Conf. Ser.}\ }\textbf {\bibinfo {volume} {145}},\ \bibinfo {pages} {012032} (\bibinfo {year} {2009})}\BibitemShut {NoStop}%
\bibitem [{\citenamefont {Petit}\ \emph {et~al.}(2014)\citenamefont {Petit}, \citenamefont {Robert}, \citenamefont {Guitteny}, \citenamefont {Bonville}, \citenamefont {Decorse}, \citenamefont {Ollivier}, \citenamefont {Mutka}, \citenamefont {Gingras},\ and\ \citenamefont {Mirebeau}}]{Petit2014}%
  \BibitemOpen
  \bibfield  {author} {\bibinfo {author} {\bibfnamefont {S.}~\bibnamefont {Petit}}, \bibinfo {author} {\bibfnamefont {J.}~\bibnamefont {Robert}}, \bibinfo {author} {\bibfnamefont {S.}~\bibnamefont {Guitteny}}, \bibinfo {author} {\bibfnamefont {P.}~\bibnamefont {Bonville}}, \bibinfo {author} {\bibfnamefont {C.}~\bibnamefont {Decorse}}, \bibinfo {author} {\bibfnamefont {J.}~\bibnamefont {Ollivier}}, \bibinfo {author} {\bibfnamefont {H.}~\bibnamefont {Mutka}}, \bibinfo {author} {\bibfnamefont {M.~J.~P.}\ \bibnamefont {Gingras}},\ and\ \bibinfo {author} {\bibfnamefont {I.}~\bibnamefont {Mirebeau}},\ }\bibfield  {title} {\bibinfo {title} {{Order} by disorder or energetic selection of the ground state in the {$XY$} pyrochlore antiferromagnet {Er}$_2${Ti}$_2${O}$_7$: {An} inelastic neutron scattering study},\ }\href {https://doi.org/10.1103/PhysRevB.90.060410} {\bibfield  {journal} {\bibinfo  {journal} {Phys. Rev. B}\ }\textbf {\bibinfo {volume} {90}},\ \bibinfo {pages} {060410} (\bibinfo {year} {2014})}\BibitemShut
  {NoStop}%
\bibitem [{\citenamefont {Rau}\ \emph {et~al.}(2016{\natexlab{a}})\citenamefont {Rau}, \citenamefont {Petit},\ and\ \citenamefont {Gingras}}]{Rau2016}%
  \BibitemOpen
  \bibfield  {author} {\bibinfo {author} {\bibfnamefont {J.~G.}\ \bibnamefont {Rau}}, \bibinfo {author} {\bibfnamefont {S.}~\bibnamefont {Petit}},\ and\ \bibinfo {author} {\bibfnamefont {M.~J.~P.}\ \bibnamefont {Gingras}},\ }\bibfield  {title} {\bibinfo {title} {Order by virtual crystal field fluctuations in pyrochlore {{$XY$}} antiferromagnets},\ }\href {https://doi.org/10.1103/PhysRevB.93.184408} {\bibfield  {journal} {\bibinfo  {journal} {Phys. Rev. B}\ }\textbf {\bibinfo {volume} {93}},\ \bibinfo {pages} {184408} (\bibinfo {year} {2016}{\natexlab{a}})}\BibitemShut {NoStop}%
\bibitem [{\citenamefont {Elliot}\ \emph {et~al.}(2021)\citenamefont {Elliot}, \citenamefont {McClarty}, \citenamefont {Prabhakaran}, \citenamefont {Johnson}, \citenamefont {Walker}, \citenamefont {Manuel},\ and\ \citenamefont {Coldea}}]{elliot2021}%
  \BibitemOpen
  \bibfield  {author} {\bibinfo {author} {\bibfnamefont {M.}~\bibnamefont {Elliot}}, \bibinfo {author} {\bibfnamefont {P.~A.}\ \bibnamefont {McClarty}}, \bibinfo {author} {\bibfnamefont {D.}~\bibnamefont {Prabhakaran}}, \bibinfo {author} {\bibfnamefont {R.~D.}\ \bibnamefont {Johnson}}, \bibinfo {author} {\bibfnamefont {H.~C.}\ \bibnamefont {Walker}}, \bibinfo {author} {\bibfnamefont {P.}~\bibnamefont {Manuel}},\ and\ \bibinfo {author} {\bibfnamefont {R.}~\bibnamefont {Coldea}},\ }\bibfield  {title} {\bibinfo {title} {{Order-by-Disorder} from {Bond-Dependent} {Exchange} and {Intensity} {Signature} of {Nodal} {Quasiparticles} in a {Honeycomb Cobaltate}},\ }\href {https://doi.org/10.1038/s41467-021-23851-0} {\bibfield  {journal} {\bibinfo  {journal} {Nature Communications}\ }\textbf {\bibinfo {volume} {12}},\ \bibinfo {pages} {3936} (\bibinfo {year} {2021})}\BibitemShut {NoStop}%
\bibitem [{\citenamefont {Kim}\ \emph {et~al.}(1999)\citenamefont {Kim}, \citenamefont {Aharony}, \citenamefont {Birgeneau}, \citenamefont {Chou}, \citenamefont {Entin-Wohlman}, \citenamefont {Erwin}, \citenamefont {Greven}, \citenamefont {Harris}, \citenamefont {Kastner}, \citenamefont {Korenblit}, \citenamefont {Lee},\ and\ \citenamefont {Shirane}}]{kim1999}%
  \BibitemOpen
  \bibfield  {author} {\bibinfo {author} {\bibfnamefont {Y.~J.}\ \bibnamefont {Kim}}, \bibinfo {author} {\bibfnamefont {A.}~\bibnamefont {Aharony}}, \bibinfo {author} {\bibfnamefont {R.~J.}\ \bibnamefont {Birgeneau}}, \bibinfo {author} {\bibfnamefont {F.~C.}\ \bibnamefont {Chou}}, \bibinfo {author} {\bibfnamefont {O.}~\bibnamefont {Entin-Wohlman}}, \bibinfo {author} {\bibfnamefont {R.~W.}\ \bibnamefont {Erwin}}, \bibinfo {author} {\bibfnamefont {M.}~\bibnamefont {Greven}}, \bibinfo {author} {\bibfnamefont {A.~B.}\ \bibnamefont {Harris}}, \bibinfo {author} {\bibfnamefont {M.~A.}\ \bibnamefont {Kastner}}, \bibinfo {author} {\bibfnamefont {I.~Y.}\ \bibnamefont {Korenblit}}, \bibinfo {author} {\bibfnamefont {Y.~S.}\ \bibnamefont {Lee}},\ and\ \bibinfo {author} {\bibfnamefont {G.}~\bibnamefont {Shirane}},\ }\bibfield  {title} {\bibinfo {title} {Ordering due to {Quantum} {Fluctuations} in {Sr}$_{\textrm{2}}${Cu}$_{\textrm{3}}${O}$_{\textrm{4}}${Cl}$_{\textrm{2}}$},\ }\href
  {https://doi.org/10.1103/PhysRevLett.83.852} {\bibfield  {journal} {\bibinfo  {journal} {Phys. Rev. Lett.}\ }\textbf {\bibinfo {volume} {83}},\ \bibinfo {pages} {852} (\bibinfo {year} {1999})}\BibitemShut {NoStop}%
\bibitem [{\citenamefont {Brueckel}\ \emph {et~al.}(1988)\citenamefont {Brueckel}, \citenamefont {Dorner}, \citenamefont {Gukasov}, \citenamefont {Plakhty}, \citenamefont {Prandl}, \citenamefont {Shender},\ and\ \citenamefont {Smirnow}}]{brueckel1988}%
  \BibitemOpen
  \bibfield  {author} {\bibinfo {author} {\bibfnamefont {T.}~\bibnamefont {Brueckel}}, \bibinfo {author} {\bibfnamefont {B.}~\bibnamefont {Dorner}}, \bibinfo {author} {\bibfnamefont {A.~G.}\ \bibnamefont {Gukasov}}, \bibinfo {author} {\bibfnamefont {V.~P.}\ \bibnamefont {Plakhty}}, \bibinfo {author} {\bibfnamefont {W.}~\bibnamefont {Prandl}}, \bibinfo {author} {\bibfnamefont {E.~F.}\ \bibnamefont {Shender}},\ and\ \bibinfo {author} {\bibfnamefont {O.~P.}\ \bibnamefont {Smirnow}},\ }\bibfield  {title} {\bibinfo {title} {Dynamical interaction of antiferromagnetic subsystems: a neutron scattering study of the spinwave spectrum of the garnet {Fe}$_{\textrm{2}}${Ca}$_{\textrm{3}}$({GeO}$_{\textrm{4}}$)$_{\textrm{3}}$},\ }\href {https://doi.org/10.1007/BF01314529} {\bibfield  {journal} {\bibinfo  {journal} {Z. Phys. B}\ }\textbf {\bibinfo {volume} {72}},\ \bibinfo {pages} {477} (\bibinfo {year} {1988})}\BibitemShut {NoStop}%
\bibitem [{\citenamefont {Sarkis}\ \emph {et~al.}(2020)\citenamefont {Sarkis}, \citenamefont {Rau}, \citenamefont {Sanjeewa}, \citenamefont {Powell}, \citenamefont {Kolis}, \citenamefont {Marbey}, \citenamefont {Hill}, \citenamefont {Rodriguez-Rivera}, \citenamefont {Nair}, \citenamefont {Yahne}, \citenamefont {Säubert}, \citenamefont {Gingras},\ and\ \citenamefont {Ross}}]{sarkis2020}%
  \BibitemOpen
  \bibfield  {author} {\bibinfo {author} {\bibfnamefont {C.~L.}\ \bibnamefont {Sarkis}}, \bibinfo {author} {\bibfnamefont {J.~G.}\ \bibnamefont {Rau}}, \bibinfo {author} {\bibfnamefont {L.~D.}\ \bibnamefont {Sanjeewa}}, \bibinfo {author} {\bibfnamefont {M.}~\bibnamefont {Powell}}, \bibinfo {author} {\bibfnamefont {J.}~\bibnamefont {Kolis}}, \bibinfo {author} {\bibfnamefont {J.}~\bibnamefont {Marbey}}, \bibinfo {author} {\bibfnamefont {S.}~\bibnamefont {Hill}}, \bibinfo {author} {\bibfnamefont {J.~A.}\ \bibnamefont {Rodriguez-Rivera}}, \bibinfo {author} {\bibfnamefont {H.~S.}\ \bibnamefont {Nair}}, \bibinfo {author} {\bibfnamefont {D.~R.}\ \bibnamefont {Yahne}}, \bibinfo {author} {\bibfnamefont {S.}~\bibnamefont {Säubert}}, \bibinfo {author} {\bibfnamefont {M.~J.~P.}\ \bibnamefont {Gingras}},\ and\ \bibinfo {author} {\bibfnamefont {K.~A.}\ \bibnamefont {Ross}},\ }\bibfield  {title} {\bibinfo {title} {Unravelling competing microscopic interactions at a phase boundary: {A} single-crystal study of the metastable
  antiferromagnetic pyrochlore {Yb}$_{\textrm{2}}${Ge}$_{\textrm{2}}${O}$_{\textrm{7}}$},\ }\href {https://doi.org/10.1103/PhysRevB.102.134418} {\bibfield  {journal} {\bibinfo  {journal} {Phys. Rev. B}\ }\textbf {\bibinfo {volume} {102}},\ \bibinfo {pages} {134418} (\bibinfo {year} {2020})}\BibitemShut {NoStop}%
\bibitem [{\citenamefont {Inami}\ \emph {et~al.}(1996)\citenamefont {Inami}, \citenamefont {Ajiro},\ and\ \citenamefont {Goto}}]{Inami1996}%
  \BibitemOpen
  \bibfield  {author} {\bibinfo {author} {\bibfnamefont {T.}~\bibnamefont {Inami}}, \bibinfo {author} {\bibfnamefont {Y.}~\bibnamefont {Ajiro}},\ and\ \bibinfo {author} {\bibfnamefont {T.}~\bibnamefont {Goto}},\ }\bibfield  {title} {\bibinfo {title} {{Magnetization} {Process} of the {Triangular} {Lattice} {Antiferromagnets}, $\text{RbFe(MoO}_4\text{)}_2$ and $\text{CsFe(SO}_4\text{)}_2$},\ }\href {https://doi.org/10.1143/JPSJ.65.2374} {\bibfield  {journal} {\bibinfo  {journal} {J. Phys. Soc. Jpn}\ }\textbf {\bibinfo {volume} {65}},\ \bibinfo {pages} {2374} (\bibinfo {year} {1996})}\BibitemShut {NoStop}%
\bibitem [{\citenamefont {Smirnov}\ \emph {et~al.}(2007)\citenamefont {Smirnov}, \citenamefont {Yashiro}, \citenamefont {Kimura}, \citenamefont {Hagiwara}, \citenamefont {Narumi}, \citenamefont {Kindo}, \citenamefont {Kikkawa}, \citenamefont {Katsumata}, \citenamefont {Shapiro},\ and\ \citenamefont {Demianets}}]{Smirnov2007}%
  \BibitemOpen
  \bibfield  {author} {\bibinfo {author} {\bibfnamefont {A.~I.}\ \bibnamefont {Smirnov}}, \bibinfo {author} {\bibfnamefont {H.}~\bibnamefont {Yashiro}}, \bibinfo {author} {\bibfnamefont {S.}~\bibnamefont {Kimura}}, \bibinfo {author} {\bibfnamefont {M.}~\bibnamefont {Hagiwara}}, \bibinfo {author} {\bibfnamefont {Y.}~\bibnamefont {Narumi}}, \bibinfo {author} {\bibfnamefont {K.}~\bibnamefont {Kindo}}, \bibinfo {author} {\bibfnamefont {A.}~\bibnamefont {Kikkawa}}, \bibinfo {author} {\bibfnamefont {K.}~\bibnamefont {Katsumata}}, \bibinfo {author} {\bibfnamefont {A.~Y.}\ \bibnamefont {Shapiro}},\ and\ \bibinfo {author} {\bibfnamefont {L.~N.}\ \bibnamefont {Demianets}},\ }\bibfield  {title} {\bibinfo {title} {Triangular lattice antiferromagnet $\mathrm{Rb}\mathrm{Fe}{(\mathrm{Mo}{\mathrm{O}}_{4})}_{2}$ in high magnetic fields},\ }\href {https://doi.org/10.1103/PhysRevB.75.134412} {\bibfield  {journal} {\bibinfo  {journal} {Phys. Rev. B}\ }\textbf {\bibinfo {volume} {75}},\ \bibinfo {pages} {134412} (\bibinfo {year}
  {2007})}\BibitemShut {NoStop}%
\bibitem [{\citenamefont {Susuki}\ \emph {et~al.}(2013)\citenamefont {Susuki}, \citenamefont {Kurita}, \citenamefont {Tanaka}, \citenamefont {Nojiri}, \citenamefont {Matsuo}, \citenamefont {Kindo},\ and\ \citenamefont {Tanaka}}]{Susuki2013}%
  \BibitemOpen
  \bibfield  {author} {\bibinfo {author} {\bibfnamefont {T.}~\bibnamefont {Susuki}}, \bibinfo {author} {\bibfnamefont {N.}~\bibnamefont {Kurita}}, \bibinfo {author} {\bibfnamefont {T.}~\bibnamefont {Tanaka}}, \bibinfo {author} {\bibfnamefont {H.}~\bibnamefont {Nojiri}}, \bibinfo {author} {\bibfnamefont {A.}~\bibnamefont {Matsuo}}, \bibinfo {author} {\bibfnamefont {K.}~\bibnamefont {Kindo}},\ and\ \bibinfo {author} {\bibfnamefont {H.}~\bibnamefont {Tanaka}},\ }\bibfield  {title} {\bibinfo {title} {{Magnetization} {Process} and {Collective} {Excitations} in the {$S=1/2$} {Triangular-Lattice} {Heisenberg} {Antiferromagnet} $\text{Ba}_3\text{CoSb}_2\text{O}_9$},\ }\href {https://doi.org/10.1103/PhysRevLett.110.267201} {\bibfield  {journal} {\bibinfo  {journal} {Phys. Rev. Lett.}\ }\textbf {\bibinfo {volume} {110}},\ \bibinfo {pages} {267201} (\bibinfo {year} {2013})}\BibitemShut {NoStop}%
\bibitem [{\citenamefont {Ranjith}\ \emph {et~al.}(2019)\citenamefont {Ranjith}, \citenamefont {Luther}, \citenamefont {Reimann}, \citenamefont {Schmidt}, \citenamefont {Schlender}, \citenamefont {Sichelschmidt}, \citenamefont {Yasuoka}, \citenamefont {Strydom}, \citenamefont {Skourski}, \citenamefont {Wosnitza}, \citenamefont {K\"uhne}, \citenamefont {Doert},\ and\ \citenamefont {Baenitz}}]{Ranjith2019}%
  \BibitemOpen
  \bibfield  {author} {\bibinfo {author} {\bibfnamefont {K.~M.}\ \bibnamefont {Ranjith}}, \bibinfo {author} {\bibfnamefont {S.}~\bibnamefont {Luther}}, \bibinfo {author} {\bibfnamefont {T.}~\bibnamefont {Reimann}}, \bibinfo {author} {\bibfnamefont {B.}~\bibnamefont {Schmidt}}, \bibinfo {author} {\bibfnamefont {P.}~\bibnamefont {Schlender}}, \bibinfo {author} {\bibfnamefont {J.}~\bibnamefont {Sichelschmidt}}, \bibinfo {author} {\bibfnamefont {H.}~\bibnamefont {Yasuoka}}, \bibinfo {author} {\bibfnamefont {A.~M.}\ \bibnamefont {Strydom}}, \bibinfo {author} {\bibfnamefont {Y.}~\bibnamefont {Skourski}}, \bibinfo {author} {\bibfnamefont {J.}~\bibnamefont {Wosnitza}}, \bibinfo {author} {\bibfnamefont {H.}~\bibnamefont {K\"uhne}}, \bibinfo {author} {\bibfnamefont {T.}~\bibnamefont {Doert}},\ and\ \bibinfo {author} {\bibfnamefont {M.}~\bibnamefont {Baenitz}},\ }\bibfield  {title} {\bibinfo {title} {Anisotropic field-induced ordering in the triangular-lattice quantum spin liquid $\text{NaYbSe}_2$},\ }\href
  {https://doi.org/10.1103/PhysRevB.100.224417} {\bibfield  {journal} {\bibinfo  {journal} {Phys. Rev. B}\ }\textbf {\bibinfo {volume} {100}},\ \bibinfo {pages} {224417} (\bibinfo {year} {2019})}\BibitemShut {NoStop}%
\bibitem [{\citenamefont {Lenz}\ \emph {et~al.}(2024)\citenamefont {Lenz}, \citenamefont {Fabrizio},\ and\ \citenamefont {Casula}}]{lenz2024}%
  \BibitemOpen
  \bibfield  {author} {\bibinfo {author} {\bibfnamefont {B.}~\bibnamefont {Lenz}}, \bibinfo {author} {\bibfnamefont {M.}~\bibnamefont {Fabrizio}},\ and\ \bibinfo {author} {\bibfnamefont {M.}~\bibnamefont {Casula}},\ }\bibfield  {title} {\bibinfo {title} {Order from disorder phenomena in {BaCoS$_2$}},\ }\href {https://doi.org/10.1038/s42005-023-01514-4} {\bibfield  {journal} {\bibinfo  {journal} {Communications Physics}\ }\textbf {\bibinfo {volume} {7}},\ \bibinfo {pages} {1} (\bibinfo {year} {2024})}\BibitemShut {NoStop}%
\bibitem [{\citenamefont {Lozano-Gómez}\ \emph {et~al.}(2024)\citenamefont {Lozano-Gómez}, \citenamefont {Noculak}, \citenamefont {Oitmaa}, \citenamefont {Singh}, \citenamefont {Iqbal}, \citenamefont {Reuther},\ and\ \citenamefont {Gingras}}]{lozano-gomez2023}%
  \BibitemOpen
  \bibfield  {author} {\bibinfo {author} {\bibfnamefont {D.}~\bibnamefont {Lozano-Gómez}}, \bibinfo {author} {\bibfnamefont {V.}~\bibnamefont {Noculak}}, \bibinfo {author} {\bibfnamefont {J.}~\bibnamefont {Oitmaa}}, \bibinfo {author} {\bibfnamefont {R.~R.~P.}\ \bibnamefont {Singh}}, \bibinfo {author} {\bibfnamefont {Y.}~\bibnamefont {Iqbal}}, \bibinfo {author} {\bibfnamefont {J.}~\bibnamefont {Reuther}},\ and\ \bibinfo {author} {\bibfnamefont {M.~J.~P.}\ \bibnamefont {Gingras}},\ }\bibfield  {title} {\bibinfo {title} {Competing gauge fields and entropically driven spin liquid to spin liquid transition in non-{K}ramers pyrochlores},\ }\href {https://doi.org/10.1073/pnas.2403487121} {\bibfield  {journal} {\bibinfo  {journal} {Proc. Natl. Acad. Sci. U.S.A.}\ }\textbf {\bibinfo {volume} {121}},\ \bibinfo {pages} {e2403487121} (\bibinfo {year} {2024})}\BibitemShut {NoStop}%
\bibitem [{Note1()}]{Note1}%
  \BibitemOpen
  \bibinfo {note} {In this work we do not consider the possibility of antiferromagnetic $J<0$~\cite {noculak2023}.}\BibitemShut {Stop}%
\bibitem [{\citenamefont {Moriya}(1960)}]{Moriya1960}%
  \BibitemOpen
  \bibfield  {author} {\bibinfo {author} {\bibfnamefont {T.}~\bibnamefont {Moriya}},\ }\bibfield  {title} {\bibinfo {title} {Anisotropic {Superexchange} {Interaction} and {Weak} {Ferromagnetism}},\ }\href {https://doi.org/10.1103/PhysRev.120.91} {\bibfield  {journal} {\bibinfo  {journal} {Phys. Rev.}\ }\textbf {\bibinfo {volume} {120}},\ \bibinfo {pages} {91} (\bibinfo {year} {1960})}\BibitemShut {NoStop}%
\bibitem [{\citenamefont {Yan}\ \emph {et~al.}(2017)\citenamefont {Yan}, \citenamefont {Benton}, \citenamefont {Jaubert},\ and\ \citenamefont {Shannon}}]{Yan2017}%
  \BibitemOpen
  \bibfield  {author} {\bibinfo {author} {\bibfnamefont {H.}~\bibnamefont {Yan}}, \bibinfo {author} {\bibfnamefont {O.}~\bibnamefont {Benton}}, \bibinfo {author} {\bibfnamefont {L.}~\bibnamefont {Jaubert}},\ and\ \bibinfo {author} {\bibfnamefont {N.}~\bibnamefont {Shannon}},\ }\bibfield  {title} {\bibinfo {title} {Theory of multiple-phase competition in pyrochlore magnets with anisotropic exchange with application to {Yb}$_{\textrm{2}}${Ti}$_{\textrm{2}}${O}$_{\textrm{7}}$, {Er}$_{\textrm{2}}${Ti}$_{\textrm{2}}${O}$_{\textrm{7}}$, and {Er}$_{\textrm{2}}${Sn}$_{\textrm{2}}${O}$_{\textrm{7}}$},\ }\href {https://doi.org/10.1103/PhysRevB.95.094422} {\bibfield  {journal} {\bibinfo  {journal} {Phys. Rev. B}\ }\textbf {\bibinfo {volume} {95}},\ \bibinfo {pages} {094422} (\bibinfo {year} {2017})}\BibitemShut {NoStop}%
\bibitem [{Note2()}]{Note2}%
  \BibitemOpen
  \bibinfo {note} {Although at $D/J=-1$ there is a degeneracy between the internal energy of the all-in-all-out and colinear ferromagnetic phases, we note that, according to our classical Monte Carlo simulation, the colinear ferromagnetic phase is selected at $D/J=-1$ at $T_c$ down to the $T\to 0^+$ limit.}\BibitemShut {Stop}%
\bibitem [{Note3()}]{Note3}%
  \BibitemOpen
  \bibinfo {note} {There is an exception to this statement when $D=0$. In this case, Eq.~\protect \eqref {eq:hamiltonian} is simply the ferromagnetic Heisenberg model which has an exact $O(3)$ symmetry, and therefore the ground state degeneracy is no longer accidental}\BibitemShut {NoStop}%
\bibitem [{\citenamefont {Coffey}\ \emph {et~al.}(1991)\citenamefont {Coffey}, \citenamefont {Rice},\ and\ \citenamefont {Zhang}}]{Coffey1991a}%
  \BibitemOpen
  \bibfield  {author} {\bibinfo {author} {\bibfnamefont {D.}~\bibnamefont {Coffey}}, \bibinfo {author} {\bibfnamefont {T.~M.}\ \bibnamefont {Rice}},\ and\ \bibinfo {author} {\bibfnamefont {F.~C.}\ \bibnamefont {Zhang}},\ }\bibfield  {title} {\bibinfo {title} {Dzyaloshinskii-{M}oriya interaction in the cuprates},\ }\href {https://doi.org/10.1103/PhysRevB.44.10112} {\bibfield  {journal} {\bibinfo  {journal} {Phys. Rev. B}\ }\textbf {\bibinfo {volume} {44}},\ \bibinfo {pages} {10112} (\bibinfo {year} {1991})}\BibitemShut {NoStop}%
\bibitem [{\citenamefont {Riedl}\ \emph {et~al.}(2016)\citenamefont {Riedl}, \citenamefont {Guterding}, \citenamefont {Jeschke}, \citenamefont {Gingras},\ and\ \citenamefont {Valent{\'i}}}]{Riedl2016a}%
  \BibitemOpen
  \bibfield  {author} {\bibinfo {author} {\bibfnamefont {K.}~\bibnamefont {Riedl}}, \bibinfo {author} {\bibfnamefont {D.}~\bibnamefont {Guterding}}, \bibinfo {author} {\bibfnamefont {H.~O.}\ \bibnamefont {Jeschke}}, \bibinfo {author} {\bibfnamefont {M.~J.~P.}\ \bibnamefont {Gingras}},\ and\ \bibinfo {author} {\bibfnamefont {R.}~\bibnamefont {Valent{\'i}}},\ }\bibfield  {title} {\bibinfo {title} {{Ab} {Initio} {Determination} of {Spin} {{Hamiltonians}} with {Anisotropic} {Exchange} {Interactions}: {{The}} {Case} of the {Pyrochlore} {Ferromagnet} {Lu}$_{\textrm{2}}${V}$_{\textrm{2}}${O}$_{\textrm{7}}$},\ }\href {https://doi.org/10.1103/PhysRevB.94.014410} {\bibfield  {journal} {\bibinfo  {journal} {Phys. Rev. B}\ }\textbf {\bibinfo {volume} {94}},\ \bibinfo {pages} {014410} (\bibinfo {year} {2016})}\BibitemShut {NoStop}%
\bibitem [{\citenamefont {Ross}\ \emph {et~al.}(2011)\citenamefont {Ross}, \citenamefont {Savary}, \citenamefont {Gaulin},\ and\ \citenamefont {Balents}}]{Ross2011}%
  \BibitemOpen
  \bibfield  {author} {\bibinfo {author} {\bibfnamefont {K.~A.}\ \bibnamefont {Ross}}, \bibinfo {author} {\bibfnamefont {L.}~\bibnamefont {Savary}}, \bibinfo {author} {\bibfnamefont {B.~D.}\ \bibnamefont {Gaulin}},\ and\ \bibinfo {author} {\bibfnamefont {L.}~\bibnamefont {Balents}},\ }\bibfield  {title} {\bibinfo {title} {Quantum excitations in quantum spin ice},\ }\href {https://doi.org/10.1103/PhysRevX.1.021002} {\bibfield  {journal} {\bibinfo  {journal} {Phys. Rev. X}\ }\textbf {\bibinfo {volume} {1}},\ \bibinfo {pages} {021002} (\bibinfo {year} {2011})}\BibitemShut {NoStop}%
\bibitem [{Note80()}]{Note80}%
  \BibitemOpen
  \bibinfo {note} {Here, we refer to the Kitaev interaction for the pyrochlore lattice in the same way that it is used in Ref.~\cite {Rau2018B}. The colinear ferromagnet is still the classical ground state configuration when a small but non-zero Kitaev interaction is present. However, the ferromagnetic product state will no longer be an exact eigenstate of the quantum Hamiltonian. This, in turn, will generate zero-point fluctuations proportional to $K$, and there will be quantum ObD present in addition to thermal ObD, as is expected in a typical quantum ObD scenario.}\BibitemShut {Stop}%
\bibitem [{\citenamefont {Rau}\ and\ \citenamefont {Gingras}(2018)}]{Rau2018B}%
  \BibitemOpen
  \bibfield  {author} {\bibinfo {author} {\bibfnamefont {J.~G.}\ \bibnamefont {Rau}}\ and\ \bibinfo {author} {\bibfnamefont {M.~J.~P.}\ \bibnamefont {Gingras}},\ }\bibfield  {title} {\bibinfo {title} {{Frustration} and anisotropic exchange in ytterbium magnets with edge-shared octahedra},\ }\href {https://doi.org/10.1103/PhysRevB.98.054408} {\bibfield  {journal} {\bibinfo  {journal} {Phys. Rev. B}\ }\textbf {\bibinfo {volume} {98}},\ \bibinfo {pages} {054408} (\bibinfo {year} {2018})}\BibitemShut {NoStop}%
\bibitem [{\citenamefont {Rau}\ \emph {et~al.}(2016{\natexlab{b}})\citenamefont {Rau}, \citenamefont {Wu}, \citenamefont {May}, \citenamefont {Poudel}, \citenamefont {Winn}, \citenamefont {Garlea}, \citenamefont {Huq}, \citenamefont {Whitfield}, \citenamefont {Taylor}, \citenamefont {Lumsden}, \citenamefont {Gingras},\ and\ \citenamefont {Christianson}}]{rau2016b}%
  \BibitemOpen
  \bibfield  {author} {\bibinfo {author} {\bibfnamefont {J.~G.}\ \bibnamefont {Rau}}, \bibinfo {author} {\bibfnamefont {L.~S.}\ \bibnamefont {Wu}}, \bibinfo {author} {\bibfnamefont {A.~F.}\ \bibnamefont {May}}, \bibinfo {author} {\bibfnamefont {L.}~\bibnamefont {Poudel}}, \bibinfo {author} {\bibfnamefont {B.}~\bibnamefont {Winn}}, \bibinfo {author} {\bibfnamefont {V.~O.}\ \bibnamefont {Garlea}}, \bibinfo {author} {\bibfnamefont {A.}~\bibnamefont {Huq}}, \bibinfo {author} {\bibfnamefont {P.}~\bibnamefont {Whitfield}}, \bibinfo {author} {\bibfnamefont {A.~E.}\ \bibnamefont {Taylor}}, \bibinfo {author} {\bibfnamefont {M.~D.}\ \bibnamefont {Lumsden}}, \bibinfo {author} {\bibfnamefont {M.~J.~P.}\ \bibnamefont {Gingras}},\ and\ \bibinfo {author} {\bibfnamefont {A.~D.}\ \bibnamefont {Christianson}},\ }\bibfield  {title} {\bibinfo {title} {Anisotropic exchange within decoupled tetrahedra in the quantum breathing pyrochlore {Ba}$_3${Yb}$_2${Zn}$_5${O}$_{11}$},\ }\href {https://doi.org/10.1103/PhysRevLett.116.257204}
  {\bibfield  {journal} {\bibinfo  {journal} {Phys. Rev. Lett.}\ }\textbf {\bibinfo {volume} {116}},\ \bibinfo {pages} {257204} (\bibinfo {year} {2016}{\natexlab{b}})}\BibitemShut {NoStop}%
\bibitem [{\citenamefont {Rau}\ and\ \citenamefont {Gingras}(2019)}]{Rau2019Review}%
  \BibitemOpen
  \bibfield  {author} {\bibinfo {author} {\bibfnamefont {J.~G.}\ \bibnamefont {Rau}}\ and\ \bibinfo {author} {\bibfnamefont {M.~J.}\ \bibnamefont {Gingras}},\ }\bibfield  {title} {\bibinfo {title} {Frustrated {{Quantum Rare-Earth Pyrochlores}}},\ }\href {https://doi.org/10.1146/annurev-conmatphys-022317-110520} {\bibfield  {journal} {\bibinfo  {journal} {Annual Review of Condensed Matter Physics}\ }\textbf {\bibinfo {volume} {10}},\ \bibinfo {pages} {357} (\bibinfo {year} {2019})}\BibitemShut {NoStop}%
\bibitem [{\citenamefont {Shamoto}\ \emph {et~al.}(2002)\citenamefont {Shamoto}, \citenamefont {Nakano}, \citenamefont {Nozue},\ and\ \citenamefont {Kajitani}}]{Shamoto2002}%
  \BibitemOpen
  \bibfield  {author} {\bibinfo {author} {\bibfnamefont {S.-i.}\ \bibnamefont {Shamoto}}, \bibinfo {author} {\bibfnamefont {T.}~\bibnamefont {Nakano}}, \bibinfo {author} {\bibfnamefont {Y.}~\bibnamefont {Nozue}},\ and\ \bibinfo {author} {\bibfnamefont {T.}~\bibnamefont {Kajitani}},\ }\bibfield  {title} {\bibinfo {title} {{Substitution} {Effects} on {Ferromagnetic} {Mott} {Insulator} {Lu}$_{\textrm{2}}${V}$_{\textrm{2}}${O}$_{\textrm{7}}$},\ }\href {https://doi.org/10.1016/S0022-3697(02)00071-9} {\bibfield  {journal} {\bibinfo  {journal} {J. Phys. Chem. Solids}\ }\textbf {\bibinfo {volume} {63}},\ \bibinfo {pages} {1047} (\bibinfo {year} {2002})}\BibitemShut {NoStop}%
\bibitem [{\citenamefont {Ali~Biswas}\ and\ \citenamefont {Jana}(2013)}]{AliBiswas2013}%
  \BibitemOpen
  \bibfield  {author} {\bibinfo {author} {\bibfnamefont {A.}~\bibnamefont {Ali~Biswas}}\ and\ \bibinfo {author} {\bibfnamefont {Y.}~\bibnamefont {Jana}},\ }\bibfield  {title} {\bibinfo {title} {{Crystal-Field}, {Exchange} {Interactions} and {Magnetism} in {Pyrochlore} {Ferromagnet} {R}$_{\textrm{2}}${V}$_{\textrm{2}}${O}$_{\textrm{7}}$ ({R}$^{3+}$={Y}, {Lu})},\ }\href {https://doi.org/10.1016/j.jmmm.2012.10.012} {\bibfield  {journal} {\bibinfo  {journal} {J. Magn. Magn. Mater.}\ }\textbf {\bibinfo {volume} {329}},\ \bibinfo {pages} {118} (\bibinfo {year} {2013})}\BibitemShut {NoStop}%
\bibitem [{\citenamefont {Nazipov}\ \emph {et~al.}(2016{\natexlab{a}})\citenamefont {Nazipov}, \citenamefont {Nikiforov},\ and\ \citenamefont {Chernyshev}}]{Nazipov2016}%
  \BibitemOpen
  \bibfield  {author} {\bibinfo {author} {\bibfnamefont {D.~V.}\ \bibnamefont {Nazipov}}, \bibinfo {author} {\bibfnamefont {A.~E.}\ \bibnamefont {Nikiforov}},\ and\ \bibinfo {author} {\bibfnamefont {V.~A.}\ \bibnamefont {Chernyshev}},\ }\bibfield  {title} {\bibinfo {title} {Exchange interaction in pyrochlore vanadates {Lu}$_{\textrm{2}}${V}$_{\textrm{2}}${O}$_{\textrm{7}}$ and {Y}$_{\textrm{2}}${V}$_{\textrm{2}}${O}$_{\textrm{7}}$: {Ab} initio approach},\ }\href {https://doi.org/10.1134/S1063783416100255} {\bibfield  {journal} {\bibinfo  {journal} {Physics of the Solid State}\ }\textbf {\bibinfo {volume} {58}},\ \bibinfo {pages} {1989} (\bibinfo {year} {2016}{\natexlab{a}})}\BibitemShut {NoStop}%
\bibitem [{\citenamefont {Nazipov}\ \emph {et~al.}(2016{\natexlab{b}})\citenamefont {Nazipov}, \citenamefont {Nikiforov},\ and\ \citenamefont {Chernyshev}}]{Nazipov2016a}%
  \BibitemOpen
  \bibfield  {author} {\bibinfo {author} {\bibfnamefont {D.~V.}\ \bibnamefont {Nazipov}}, \bibinfo {author} {\bibfnamefont {A.~E.}\ \bibnamefont {Nikiforov}},\ and\ \bibinfo {author} {\bibfnamefont {V.~A.}\ \bibnamefont {Chernyshev}},\ }\bibfield  {title} {\bibinfo {title} {Structure, lattice dynamics, and exchange interaction in {Lu}$_{\textrm{2}}${V}$_{\textrm{2}}${O}$_{\textrm{7}}$, {Y}$_{\textrm{2}}${V}$_{\textrm{2}}${O}$_{\textrm{7}}$: an ab initio approach},\ }\href {https://doi.org/10.1134/S0030400X16100179} {\bibfield  {journal} {\bibinfo  {journal} {Optics and Spectroscopy}\ }\textbf {\bibinfo {volume} {121}},\ \bibinfo {pages} {544} (\bibinfo {year} {2016}{\natexlab{b}})}\BibitemShut {NoStop}%
\bibitem [{\citenamefont {Mena}\ \emph {et~al.}(2014)\citenamefont {Mena}, \citenamefont {Perry}, \citenamefont {Perring}, \citenamefont {Le}, \citenamefont {Guerrero}, \citenamefont {Storni}, \citenamefont {Adroja}, \citenamefont {R{\"u}egg},\ and\ \citenamefont {McMorrow}}]{Mena2014}%
  \BibitemOpen
  \bibfield  {author} {\bibinfo {author} {\bibfnamefont {M.}~\bibnamefont {Mena}}, \bibinfo {author} {\bibfnamefont {R.~S.}\ \bibnamefont {Perry}}, \bibinfo {author} {\bibfnamefont {T.~G.}\ \bibnamefont {Perring}}, \bibinfo {author} {\bibfnamefont {M.~D.}\ \bibnamefont {Le}}, \bibinfo {author} {\bibfnamefont {S.}~\bibnamefont {Guerrero}}, \bibinfo {author} {\bibfnamefont {M.}~\bibnamefont {Storni}}, \bibinfo {author} {\bibfnamefont {D.~T.}\ \bibnamefont {Adroja}}, \bibinfo {author} {\bibfnamefont {C.}~\bibnamefont {R{\"u}egg}},\ and\ \bibinfo {author} {\bibfnamefont {D.~F.}\ \bibnamefont {McMorrow}},\ }\bibfield  {title} {\bibinfo {title} {{Spin-Wave} {Spectrum} of the {Quantum} {Ferromagnet} on the {Pyrochlore} {Lattice} {Lu}$_{\textrm{2}}${V}$_{\textrm{2}}${O}$_{\textrm{7}}$},\ }\href {https://doi.org/10.1103/PhysRevLett.113.047202} {\bibfield  {journal} {\bibinfo  {journal} {Phys. Rev. Lett.}\ }\textbf {\bibinfo {volume} {113}},\ \bibinfo {pages} {047202} (\bibinfo {year} {2014})}\BibitemShut {NoStop}%
\bibitem [{\citenamefont {Onose}\ \emph {et~al.}(2010)\citenamefont {Onose}, \citenamefont {Ideue}, \citenamefont {Katsura}, \citenamefont {Shiomi}, \citenamefont {Nagaosa},\ and\ \citenamefont {Tokura}}]{onose2010}%
  \BibitemOpen
  \bibfield  {author} {\bibinfo {author} {\bibfnamefont {Y.}~\bibnamefont {Onose}}, \bibinfo {author} {\bibfnamefont {T.}~\bibnamefont {Ideue}}, \bibinfo {author} {\bibfnamefont {H.}~\bibnamefont {Katsura}}, \bibinfo {author} {\bibfnamefont {Y.}~\bibnamefont {Shiomi}}, \bibinfo {author} {\bibfnamefont {N.}~\bibnamefont {Nagaosa}},\ and\ \bibinfo {author} {\bibfnamefont {Y.}~\bibnamefont {Tokura}},\ }\bibfield  {title} {\bibinfo {title} {Observation of the {{Magnon Hall Effect}}},\ }\href {https://doi.org/10.1126/science.1188260} {\bibfield  {journal} {\bibinfo  {journal} {Science}\ }\textbf {\bibinfo {volume} {329}},\ \bibinfo {pages} {297} (\bibinfo {year} {2010})}\BibitemShut {NoStop}%
\bibitem [{Note91()}]{Note91}%
  \BibitemOpen
  \bibinfo {note} {Interestingly, the breathing pyrochlore (BP) material Ba$_3$Yb$_2$Zn$_5$O$_{11}$ seems well-described by the Hamiltonian of Eq.~\protect \eqref {eq:hamiltonian} over the small tetrahedra of the BP structure~\cite {rau2016b}.}\BibitemShut {Stop}%
\bibitem [{\citenamefont {Alzate-Cardona}\ \emph {et~al.}(2019)\citenamefont {Alzate-Cardona}, \citenamefont {Sabogal-Suárez}, \citenamefont {Evans},\ and\ \citenamefont {Restrepo-Parra}}]{ALZATE-CARDONA2019}%
  \BibitemOpen
  \bibfield  {author} {\bibinfo {author} {\bibfnamefont {J.~D.}\ \bibnamefont {Alzate-Cardona}}, \bibinfo {author} {\bibfnamefont {D.}~\bibnamefont {Sabogal-Suárez}}, \bibinfo {author} {\bibfnamefont {R.~F.~L.}\ \bibnamefont {Evans}},\ and\ \bibinfo {author} {\bibfnamefont {E.}~\bibnamefont {Restrepo-Parra}},\ }\bibfield  {title} {\bibinfo {title} {Optimal phase space sampling for {Monte} {Carlo} simulations of {Heisenberg} spin systems},\ }\href {https://doi.org/10.1088/1361-648X/aaf852} {\bibfield  {journal} {\bibinfo  {journal} {J. Phys. Condens. Matter}\ }\textbf {\bibinfo {volume} {31}},\ \bibinfo {pages} {095802} (\bibinfo {year} {2019})}\BibitemShut {NoStop}%
\bibitem [{\citenamefont {Binder}\ and\ \citenamefont {Heermann}(2010)}]{BINDER2010}%
  \BibitemOpen
  \bibfield  {author} {\bibinfo {author} {\bibfnamefont {K.}~\bibnamefont {Binder}}\ and\ \bibinfo {author} {\bibfnamefont {D.~W.}\ \bibnamefont {Heermann}},\ }\href@noop {} {\emph {\bibinfo {title} {Monte {Carlo} {Simulation} in {Statistical} {Physics}: {An} {Introduction}}}}\ (\bibinfo  {publisher} {Springer Science \& Business Media},\ \bibinfo {year} {2010})\BibitemShut {NoStop}%
\bibitem [{\citenamefont {Creutz}(1987)}]{CREUTZ1987}%
  \BibitemOpen
  \bibfield  {author} {\bibinfo {author} {\bibfnamefont {M.}~\bibnamefont {Creutz}},\ }\bibfield  {title} {\bibinfo {title} {Overrelaxation and {Monte} {Carlo} simulation},\ }\href {https://doi.org/10.1103/PhysRevD.36.515} {\bibfield  {journal} {\bibinfo  {journal} {Phys. Rev. D}\ }\textbf {\bibinfo {volume} {36}},\ \bibinfo {pages} {515} (\bibinfo {year} {1987})}\BibitemShut {NoStop}%
\bibitem [{\citenamefont {Bouchaud}\ and\ \citenamefont {Zérah}(1993)}]{bouchaud1993}%
  \BibitemOpen
  \bibfield  {author} {\bibinfo {author} {\bibfnamefont {J.~P.}\ \bibnamefont {Bouchaud}}\ and\ \bibinfo {author} {\bibfnamefont {P.~G.}\ \bibnamefont {Zérah}},\ }\bibfield  {title} {\bibinfo {title} {Dipolar ferromagnetism: {A} {Monte} {Carlo} study},\ }\href {https://doi.org/10.1103/PhysRevB.47.9095} {\bibfield  {journal} {\bibinfo  {journal} {Phys. Rev. B}\ }\textbf {\bibinfo {volume} {47}},\ \bibinfo {pages} {9095} (\bibinfo {year} {1993})}\BibitemShut {NoStop}%
\bibitem [{\citenamefont {Chaikin}\ and\ \citenamefont {Lubensky}(2000)}]{CHAIKIN2000}%
  \BibitemOpen
  \bibfield  {author} {\bibinfo {author} {\bibfnamefont {P.~M.}\ \bibnamefont {Chaikin}}\ and\ \bibinfo {author} {\bibfnamefont {T.~C.}\ \bibnamefont {Lubensky}},\ }\href@noop {} {\emph {\bibinfo {title} {Principles of {Condensed} {Matter} {Physics}}}}\ (\bibinfo  {publisher} {Cambridge University Press},\ \bibinfo {year} {2000})\BibitemShut {NoStop}%
\bibitem [{\citenamefont {Holstein}\ and\ \citenamefont {Primakoff}(1940)}]{Holstein1940}%
  \BibitemOpen
  \bibfield  {author} {\bibinfo {author} {\bibfnamefont {T.}~\bibnamefont {Holstein}}\ and\ \bibinfo {author} {\bibfnamefont {H.}~\bibnamefont {Primakoff}},\ }\bibfield  {title} {\bibinfo {title} {Field {Dependence} of the {Intrinsic} {Domain} {Magnetization} of a {Ferromagnet}},\ }\href {https://doi.org/10.1103/PhysRev.58.1098} {\bibfield  {journal} {\bibinfo  {journal} {Phys. Rev.}\ }\textbf {\bibinfo {volume} {58}},\ \bibinfo {pages} {1098} (\bibinfo {year} {1940})}\BibitemShut {NoStop}%
\bibitem [{\citenamefont {Rau}\ \emph {et~al.}(2019)\citenamefont {Rau}, \citenamefont {Moessner},\ and\ \citenamefont {McClarty}}]{Rau2019a}%
  \BibitemOpen
  \bibfield  {author} {\bibinfo {author} {\bibfnamefont {J.~G.}\ \bibnamefont {Rau}}, \bibinfo {author} {\bibfnamefont {R.}~\bibnamefont {Moessner}},\ and\ \bibinfo {author} {\bibfnamefont {P.~A.}\ \bibnamefont {McClarty}},\ }\bibfield  {title} {\bibinfo {title} {Magnon interactions in the frustrated pyrochlore ferromagnet {Yb}$_{\textrm{2}}${Ti}$_{\textrm{2}}${O}$_{\textrm{7}}$},\ }\href {https://doi.org/10.1103/PhysRevB.100.104423} {\bibfield  {journal} {\bibinfo  {journal} {Phys. Rev. B}\ }\textbf {\bibinfo {volume} {100}},\ \bibinfo {pages} {104423} (\bibinfo {year} {2019})}\BibitemShut {NoStop}%
\bibitem [{\citenamefont {Mourigal}\ \emph {et~al.}(2013)\citenamefont {Mourigal}, \citenamefont {Fuhrman}, \citenamefont {Chernyshev},\ and\ \citenamefont {Zhitomirsky}}]{MOURIGAL2013}%
  \BibitemOpen
  \bibfield  {author} {\bibinfo {author} {\bibfnamefont {M.}~\bibnamefont {Mourigal}}, \bibinfo {author} {\bibfnamefont {W.~T.}\ \bibnamefont {Fuhrman}}, \bibinfo {author} {\bibfnamefont {A.~L.}\ \bibnamefont {Chernyshev}},\ and\ \bibinfo {author} {\bibfnamefont {M.~E.}\ \bibnamefont {Zhitomirsky}},\ }\bibfield  {title} {\bibinfo {title} {Dynamical structure factor of the triangular-lattice antiferromagnet},\ }\href {https://doi.org/10.1103/PhysRevB.88.094407} {\bibfield  {journal} {\bibinfo  {journal} {Phys. Rev. B}\ }\textbf {\bibinfo {volume} {88}},\ \bibinfo {pages} {094407} (\bibinfo {year} {2013})}\BibitemShut {NoStop}%
\bibitem [{\citenamefont {Mahan}(2000)}]{mahan2000}%
  \BibitemOpen
  \bibfield  {author} {\bibinfo {author} {\bibfnamefont {G.~D.}\ \bibnamefont {Mahan}},\ }\href@noop {} {\emph {\bibinfo {title} {{Many-Particle} {Physics}}}},\ \bibinfo {edition} {3rd}\ ed.\ (\bibinfo  {publisher} {Springer},\ \bibinfo {address} {New York, NY},\ \bibinfo {year} {2000})\BibitemShut {NoStop}%
\bibitem [{Note4()}]{Note4}%
  \BibitemOpen
  \bibinfo {note} {Note that we also considered the possibility that $\vu *{m}$ is not along one of the high-symmetry directions, but this does not change any of our results.}\BibitemShut {Stop}%
\bibitem [{Note5()}]{Note5}%
  \BibitemOpen
  \bibinfo {note} {Similar results where obtained for the entropy difference between the $\expval {111}$ and the $\expval {100}$ magnetization directions, i.e. $\protect \mathcal {S}_{\expval {111}}-\protect \mathcal {S}_{\expval {100}}>0$ (not shown).}\BibitemShut {Stop}%
\bibitem [{Note103()}]{Note103}%
  \BibitemOpen
  \bibinfo {note} {In a system where there are an extensive number of quadratic zero modes, there will be a thermal contribution to the internal energy, see for example Ref.~\cite {chalker1992}.}\BibitemShut {Stop}%
\bibitem [{\citenamefont {Chalker}\ \emph {et~al.}(1992)\citenamefont {Chalker}, \citenamefont {Holdsworth},\ and\ \citenamefont {Shender}}]{chalker1992}%
  \BibitemOpen
  \bibfield  {author} {\bibinfo {author} {\bibfnamefont {J.~T.}\ \bibnamefont {Chalker}}, \bibinfo {author} {\bibfnamefont {P.~C.~W.}\ \bibnamefont {Holdsworth}},\ and\ \bibinfo {author} {\bibfnamefont {E.~F.}\ \bibnamefont {Shender}},\ }\bibfield  {title} {\bibinfo {title} {Hidden order in a frustrated system: {Properties} of the {Heisenberg} {Kagom\'e} antiferromagnet},\ }\href {https://doi.org/10.1103/PhysRevLett.68.855} {\bibfield  {journal} {\bibinfo  {journal} {Phys. Rev. Lett.}\ }\textbf {\bibinfo {volume} {68}},\ \bibinfo {pages} {855} (\bibinfo {year} {1992})}\BibitemShut {NoStop}%
\bibitem [{\citenamefont {Maryasin}\ \emph {et~al.}(2016)\citenamefont {Maryasin}, \citenamefont {Zhitomirsky},\ and\ \citenamefont {Moessner}}]{Maryasin2016}%
  \BibitemOpen
  \bibfield  {author} {\bibinfo {author} {\bibfnamefont {V.~S.}\ \bibnamefont {Maryasin}}, \bibinfo {author} {\bibfnamefont {M.~E.}\ \bibnamefont {Zhitomirsky}},\ and\ \bibinfo {author} {\bibfnamefont {R.}~\bibnamefont {Moessner}},\ }\bibfield  {title} {\bibinfo {title} {Low-field behavior of an $xy$ pyrochlore antiferromagnet: Emergent clock anisotropies},\ }\href {https://doi.org/10.1103/PhysRevB.93.100406} {\bibfield  {journal} {\bibinfo  {journal} {Phys. Rev. B}\ }\textbf {\bibinfo {volume} {93}},\ \bibinfo {pages} {100406} (\bibinfo {year} {2016})}\BibitemShut {NoStop}%
\bibitem [{Note6()}]{Note6}%
  \BibitemOpen
  \bibinfo {note} {Using a number of numerical and analytical methods, Ref.~\cite {lozano-gomez2023} found evidence that the $D/J=2$ point for a spin-$\protect \frac {1}{2}$ system may be a quantum spin liquid, described by the combination of a rank-1 and rank-2 emergent gauge field, which corresponds to a triple point in the classical phase diagram where two long-range ordered quadrupolar phases meet with a spin ice state. In the classical $S=\infty $ limit, evidence is compelling that the point $D/J=2$ is a classical spin liquid~\cite {lozano-gomez2023}}\BibitemShut {NoStop}%
\bibitem [{\citenamefont {Weinberg}(1972)}]{WEINBERG1972}%
  \BibitemOpen
  \bibfield  {author} {\bibinfo {author} {\bibfnamefont {S.}~\bibnamefont {Weinberg}},\ }\bibfield  {title} {\bibinfo {title} {Approximate {Symmetries} and {Pseudo}-{Goldstone} {Bosons}},\ }\href {https://doi.org/10.1103/PhysRevLett.29.1698} {\bibfield  {journal} {\bibinfo  {journal} {Phys. Rev. Lett.}\ }\textbf {\bibinfo {volume} {29}},\ \bibinfo {pages} {1698} (\bibinfo {year} {1972})}\BibitemShut {NoStop}%
\bibitem [{\citenamefont {Burgess}(2000)}]{BURGESS2000}%
  \BibitemOpen
  \bibfield  {author} {\bibinfo {author} {\bibfnamefont {C.}~\bibnamefont {Burgess}},\ }\bibfield  {title} {\bibinfo {title} {Goldstone and pseudo-{Goldstone} bosons in nuclear, particle and condensed-matter physics},\ }\href {https://doi.org/10.1016/S0370-1573(99)00111-8} {\bibfield  {journal} {\bibinfo  {journal} {Phys. Rep.}\ }\textbf {\bibinfo {volume} {330}},\ \bibinfo {pages} {193} (\bibinfo {year} {2000})}\BibitemShut {NoStop}%
\bibitem [{\citenamefont {Aharony}(1973)}]{AHARONY1973}%
  \BibitemOpen
  \bibfield  {author} {\bibinfo {author} {\bibfnamefont {A.}~\bibnamefont {Aharony}},\ }\bibfield  {title} {\bibinfo {title} {Critical {Behavior} of {Anisotropic} {Cubic} {Systems}},\ }\href {https://doi.org/10.1103/PhysRevB.8.4270} {\bibfield  {journal} {\bibinfo  {journal} {Phys. Rev. B}\ }\textbf {\bibinfo {volume} {8}},\ \bibinfo {pages} {4270} (\bibinfo {year} {1973})}\BibitemShut {NoStop}%
\bibitem [{\citenamefont {Ferer}\ \emph {et~al.}(1981)\citenamefont {Ferer}, \citenamefont {Van~Dyke},\ and\ \citenamefont {Camp}}]{ferer1981}%
  \BibitemOpen
  \bibfield  {author} {\bibinfo {author} {\bibfnamefont {M.}~\bibnamefont {Ferer}}, \bibinfo {author} {\bibfnamefont {J.~P.}\ \bibnamefont {Van~Dyke}},\ and\ \bibinfo {author} {\bibfnamefont {W.~J.}\ \bibnamefont {Camp}},\ }\bibfield  {title} {\bibinfo {title} {Effect of a cubic crystal field on the critical behavior of a {{3D}} model with {{Heisenberg}} exchange coupling: {{A}} high-temperature series investigation},\ }\href {https://doi.org/10.1103/PhysRevB.23.2367} {\bibfield  {journal} {\bibinfo  {journal} {Phys. Rev. B}\ }\textbf {\bibinfo {volume} {23}},\ \bibinfo {pages} {2367} (\bibinfo {year} {1981})}\BibitemShut {NoStop}%
\bibitem [{\citenamefont {Newman}\ and\ \citenamefont {Riedel}(1982)}]{newman1982}%
  \BibitemOpen
  \bibfield  {author} {\bibinfo {author} {\bibfnamefont {K.~E.}\ \bibnamefont {Newman}}\ and\ \bibinfo {author} {\bibfnamefont {E.~K.}\ \bibnamefont {Riedel}},\ }\bibfield  {title} {\bibinfo {title} {Cubic {{$N$}}-vector model and randomly dilute {{Ising}} model in general dimensions},\ }\href {https://doi.org/10.1103/PhysRevB.25.264} {\bibfield  {journal} {\bibinfo  {journal} {Phys. Rev. B}\ }\textbf {\bibinfo {volume} {25}},\ \bibinfo {pages} {264} (\bibinfo {year} {1982})}\BibitemShut {NoStop}%
\bibitem [{\citenamefont {Manuel~Carmona}\ \emph {et~al.}(2000)\citenamefont {Manuel~Carmona}, \citenamefont {Pelissetto},\ and\ \citenamefont {Vicari}}]{manuelcarmona2000}%
  \BibitemOpen
  \bibfield  {author} {\bibinfo {author} {\bibfnamefont {J.}~\bibnamefont {Manuel~Carmona}}, \bibinfo {author} {\bibfnamefont {A.}~\bibnamefont {Pelissetto}},\ and\ \bibinfo {author} {\bibfnamefont {E.}~\bibnamefont {Vicari}},\ }\bibfield  {title} {\bibinfo {title} {{$N$}-component {{Ginzburg-Landau Hamiltonian}} with cubic anisotropy: {{A}} six-loop study},\ }\href {https://doi.org/10.1103/PhysRevB.61.15136} {\bibfield  {journal} {\bibinfo  {journal} {Phys. Rev. B}\ }\textbf {\bibinfo {volume} {61}},\ \bibinfo {pages} {15136} (\bibinfo {year} {2000})}\BibitemShut {NoStop}%
\bibitem [{\citenamefont {Adzhemyan}\ \emph {et~al.}(2019)\citenamefont {Adzhemyan}, \citenamefont {Ivanova}, \citenamefont {Kompaniets}, \citenamefont {Kudlis},\ and\ \citenamefont {Sokolov}}]{ADZHEMYAN2019}%
  \BibitemOpen
  \bibfield  {author} {\bibinfo {author} {\bibfnamefont {L.~T.}\ \bibnamefont {Adzhemyan}}, \bibinfo {author} {\bibfnamefont {E.~V.}\ \bibnamefont {Ivanova}}, \bibinfo {author} {\bibfnamefont {M.~V.}\ \bibnamefont {Kompaniets}}, \bibinfo {author} {\bibfnamefont {A.}~\bibnamefont {Kudlis}},\ and\ \bibinfo {author} {\bibfnamefont {A.~I.}\ \bibnamefont {Sokolov}},\ }\bibfield  {title} {\bibinfo {title} {Six-loop $\varepsilon$ expansion study of three-dimensional $n$-vector model with cubic anisotropy},\ }\href {https://doi.org/10.1016/j.nuclphysb.2019.02.001} {\bibfield  {journal} {\bibinfo  {journal} {Nuclear Physics B}\ }\textbf {\bibinfo {volume} {940}},\ \bibinfo {pages} {332} (\bibinfo {year} {2019})}\BibitemShut {NoStop}%
\bibitem [{\citenamefont {Wallace}(1973)}]{wallace1973}%
  \BibitemOpen
  \bibfield  {author} {\bibinfo {author} {\bibfnamefont {D.~J.}\ \bibnamefont {Wallace}},\ }\bibfield  {title} {\bibinfo {title} {Critical behaviour of anisotropic cubic systems},\ }\href {https://doi.org/10.1088/0022-3719/6/8/007} {\bibfield  {journal} {\bibinfo  {journal} {J. Phys. C: Solid State Phys.}\ }\textbf {\bibinfo {volume} {6}},\ \bibinfo {pages} {1390} (\bibinfo {year} {1973})}\BibitemShut {NoStop}%
\bibitem [{\citenamefont {Ketley}\ and\ \citenamefont {Wallace}(1973)}]{ketley1973}%
  \BibitemOpen
  \bibfield  {author} {\bibinfo {author} {\bibfnamefont {I.~J.}\ \bibnamefont {Ketley}}\ and\ \bibinfo {author} {\bibfnamefont {D.~J.}\ \bibnamefont {Wallace}},\ }\bibfield  {title} {\bibinfo {title} {A modified epsilon expansion for a {{Hamiltonian}} with cubic point-group symmetry},\ }\href {https://doi.org/10.1088/0305-4470/6/11/006} {\bibfield  {journal} {\bibinfo  {journal} {J. Phys. A}\ }\textbf {\bibinfo {volume} {6}},\ \bibinfo {pages} {1667} (\bibinfo {year} {1973})}\BibitemShut {NoStop}%
\bibitem [{\citenamefont {Pelissetto}\ and\ \citenamefont {Vicari}(2002)}]{pelissetto2002}%
  \BibitemOpen
  \bibfield  {author} {\bibinfo {author} {\bibfnamefont {A.}~\bibnamefont {Pelissetto}}\ and\ \bibinfo {author} {\bibfnamefont {E.}~\bibnamefont {Vicari}},\ }\bibfield  {title} {\bibinfo {title} {{Critical} {Phenomena} and {Renormalization-Group} {Theory}},\ }\href {https://doi.org/10.1016/S0370-1573(02)00219-3} {\bibfield  {journal} {\bibinfo  {journal} {Phys. Rep.}\ }\textbf {\bibinfo {volume} {368}},\ \bibinfo {pages} {549} (\bibinfo {year} {2002})}\BibitemShut {NoStop}%
\bibitem [{Note7()}]{Note7}%
  \BibitemOpen
  \bibinfo {note} {For the classical model, the instability discussed for the spin-$\protect \frac {1}{2}$ does not occur for any $|D|/|J|<2$}\BibitemShut {NoStop}%
\bibitem [{Note8()}]{Note8}%
  \BibitemOpen
  \bibinfo {note} {The precise determination of the extension of the $\expval {100}$ phase as the temperature drops below $T_c$ is challenged by the strong thermal fluctuations near criticality as well as the weak thermal selection associated to this phase, as well as the $\expval {111}$ (see Fig.~\ref {fig:CLTE_entropic_weight}) as $|D|/J \rightarrow 0^+$. As such, the identification of a clearer $\expval {100}/\expval {111}$ phase boundary in this region of the phase diagram would require a significantly more extensive numerical study while also considering larger system sizes.}\BibitemShut {Stop}%
\bibitem [{Note126()}]{Note126}%
  \BibitemOpen
  \bibinfo {note} {In these distributions of the magnetization direction, the highest intensity appears near the point $\cos (\theta ) = 1/\protect \sqrt {3}, \phi = \pi /4$, suggesting a $\expval {111}$ orientation.}\BibitemShut {Stop}%
\bibitem [{\citenamefont {Ernzerhof}(1994)}]{Ernzerhof1994}%
  \BibitemOpen
  \bibfield  {author} {\bibinfo {author} {\bibfnamefont {M.}~\bibnamefont {Ernzerhof}},\ }\bibfield  {title} {\bibinfo {title} {Taylor-series expansion of density functionals},\ }\href {https://doi.org/10.1103/PhysRevA.50.4593} {\bibfield  {journal} {\bibinfo  {journal} {Phys. Rev. A}\ }\textbf {\bibinfo {volume} {50}},\ \bibinfo {pages} {4593} (\bibinfo {year} {1994})}\BibitemShut {NoStop}%
\end{thebibliography}%
\end{document}